\documentclass[twocolumn]{aastex631}
\usepackage{amsmath}
\usepackage[T1]{fontenc}

\shorttitle{Planet-disk Interaction in Modest MRI Turbulence with a Disk Wind I}
\shortauthors{Aoyama and Bai}

\graphicspath{{./}{figures/}}
\newcommand{\bvec}[1]{\mbox{\boldmath $#1$}}
\newcommand{\etaA}{\eta_\mathrm{A}}
\newcommand{\etaO}{\eta_\mathrm{O}}
\newcommand{\al}{$\alpha$}
\newcommand{\AM}{Am}
\newcommand{\AMzero}{Am_\mathrm{disk}}
\newcommand{\AMatm}{Am_\mathrm{atm}}

\newcommand{\Mth}{M_\mathrm{th}}
\newcommand{\Macc}{\dot{M}_\mathrm{acc}}
\newcommand{\RP}{R_\mathrm{P}}
\newcommand{\rH}{r_\mathrm{H}}
\newcommand{\MP}{M_\mathrm{P}}
\newcommand{\cs}{c_\mathrm{s}}
\newcommand{\OmegaP}{\Omega_\mathrm{P}}
\newcommand{\st}{\sin{\theta}}

\newcommand{\RA}{R_\mathrm{A}}
\newcommand{\OmegaK}{\Omega_\mathrm{K}}
\newcommand{\vK}{v_\mathrm{K}}
\newcommand{\vP}{v_\mathrm{P}}
\newcommand{\SigmaP}{\Sigma_\mathrm{P}}
\newcommand{\tauP}{\tau_\mathrm{P}}

\newcommand{\taupass}{\tau_\mathrm{pass}}
\newcommand{\taulib}{\tau_\mathrm{lib}}

\newcommand{\dMwind}{\dot{M}_\mathrm{wind}}
\newcommand{\alphaMx}{\alpha^\mathrm{Max}}
\newcommand{\alphaRe}{\alpha^\mathrm{Rey}}
\newcommand{\GP}{\gamma_\mathrm{P}}

\newcommand{\GMx}{\gamma_\mathrm{Max}}
\newcommand{\Gwind}{\gamma_\mathrm{wind}}
\newcommand{\Gwave}{\gamma_\mathrm{wave}}
\newcommand{\Gvis}{\gamma_\mathrm{vis}}
\newcommand{\GA}{\gamma_\mathrm{A}}
\newcommand{\Gd}{\gamma_\mathrm{dep}}
\newcommand{\GPt}{\Gamma_\mathrm{P}}

\newcommand{\JMx}{J_\mathrm{Max}}
\newcommand{\Jwind}{J_\mathrm{wind}}
\newcommand{\Jwave}{J_\mathrm{wave}}
\newcommand{\Jvis}{J_\mathrm{vis}}
\newcommand{\JMHD}{J_\mathrm{MHD}}
\newcommand{\JA}{J_\mathrm{A}}
\newcommand{\av}[1]{\left\langle #1 \right\rangle}

\newcommand{\sint}{\sin{\theta}}

\newcommand{\TthetaD}{\theta_\mathrm{t'}}
\newcommand{\hdisk}{h_\mathrm{disk}}
\newcommand{\hatm}{h_\mathrm{atm}}
\newcommand{\idtheta}{\int_{\theta_\mathrm{u}}^{\theta_\mathrm{d}} d\theta}
\newcommand{\idphi}{\int_{0}^{2\pi} d\phi}
\newcommand{\rhoi}{\rho_\mathrm{ini}}
\newcommand{\Bmi}{B_{z0\mathrm{,mid}}}
\newcommand{\Bzm}{B_{z\mathrm{,mid}}}
\newcommand{\Alf}{Alfv\'en}
\newcommand{\pa}{\partial}
\newcommand{\mb}{\boldsymbol}
\newcommand{\dd}{\mathrm{d}}
\def\Flh{0.45\hsize}
\def\Fh{0.48\hsize}
\def\Ft{0.32\hsize}

\begin{document}
\title{Three-dimensional Global Simulations of Type-II Planet-disk Interaction with a Magnetized Disk Wind: I. Magnetic Flux Concentration and Gap Properties}

\author[0000-0003-0568-9225]{Yuhiko Aoyama}
\affiliation{Institute for Advanced Study, Tsinghua University, Beijing 100084, China}
\affiliation{Kavli Institute for Astronomy and Astrophysics, Peking University, Beijing 100084, China}
\email{yaoyama@pku.edu.cn}

\author[0000-0001-6906-9549]{Xue-Ning Bai}
\affiliation{Institute for Advanced Study, Tsinghua University, Beijing 100084, China}
\affiliation{Department of Astronomy, Tsinghua University, Beijing 100084, China}
\email{xbai@tsinghua.edu.cn}

\begin{abstract}
Giant planets embedded in protoplanetary disks (PPDs) can create annulus density gaps around their orbits in the type-II regime, potentially responsible for the ubiquity of annular substructures observed in PPDs. Despite of substantial amount of works studying type-II planet migration and gap properties, they are almost exclusively conducted under the viscous accretion disk framework. However, recent studies have established magnetized disk winds as the primary driving disk accretion and evolution, which can co-exist with turbulence from the magneto-rotational instability (MRI) in the outer PPDs.
We conduct a series of 3D global non-ideal magneto-hydrodynamic (MHD) simulations of type-II planet-disk interaction applicable to the outer PPDs. Our simulations properly resolve the MRI turbulence and accommodate the MHD disk wind.
We found that the planet triggers the poloidal magnetic flux concentration around its orbit. The concentrated magnetic flux strongly enhances angular momentum removal in the gap, which is along the inclined poloidal field through a strong outflow emanating from the disk surface outward of the planet gap. The resulting planet-induced gap shape is more similar to an inviscid disk, while being much deeper, which can be understood from a simple inhomogeneous wind torque prescription. The corotation region is characterized by a fast trans-sonic accretion flow that is asymmetric in azimuth about the planet and lacking the horseshoe turns, and the meridional flow is weakened.
The torque acting on the planet generally drives inward migration, though the migration rate can be affected by the presence of neighboring gaps through stochastic, planet-free magnetic flux concentration.
\end{abstract}


\section{Introduction} \label{sec:intro}

Recent observations of protoplanetary disk (PPD) have ubiquitously found the presence of substructures, which are mainly in the form of rings and gaps \citep[see the review of][]{Andrews+2020}.
These ring-like substructures form when a massive planet is embedded in and gravitationally interacts with the PPD \citep{Dong+2015,Bae+2017}, and at least some of these substructures suggest the presence of unseen planets, especially when combined with the kinematic signatures, such as velocity kinks \citep[e.g.,][]{Perez+2015,Pinte+2018, Pinte+2019,Speedie+Dong2022}. On the other hand, rings and gaps can also form from internal physical processes in PPDs, such as snow lines \citep[e.g.,][]{Saito+Sirono2011,Zhang+2015,Okuzumi+2016,Hu+2019}, dust-gas interaction \citep{Tominaga+2019}, and in particular, magnetic processes \citep[e.g.,][]{Suriano+2018,Riols+2020a,Cui+Bai2021}.

Sufficiently massive planets open gaps in PPDs \citep[e.g.,][]{Lin+Papaloizou1986}, which is referred as type-II planet-disk interaction. This process has been well studied in 2D pure-hydrodynamic simulations \citep[e.g.,][]{Artymowicz+Lubow1994,Kley1999,Varniere+2004,Tanaka+2022}.
Based on the numerical results, the gap depth and width are studied as functions of planetary mass and PPD properties \citep[e.g., viscosity,][]{Crida+2006,Duffell+MacFadyen2013,Fung+2014,Duffell2015,Kanagawa+2015,Kanagawa+2017,Ginzburg+Sari2018,Duffell2020}. The 3D hydrodynamic simulations largely confirm these results \citep{Kley+2001,Bate+2003,Fung+Chiang2016,Dong+Fung2017b}, and these theoretical predictions of gap profile have been used to predict the planetary mass from the observations of planet-induced gaps \citep{Rosotti+2016,Asensio-Torres+2021}.

Gap formation, regardless of induced by a massive planet or not, requires redistribution of mass, and hence angular momentum in PPDs, which in turn leads to type-II planet migration. While planet-disk interaction naturally provides the gravitational torque that leads to this redistribution, the aforementioned planet-free mechanisms suggest that internal disk processes can yield similar outcomes. Therefore, it is essential to study planet-disk interaction under more realistic disk conditions. Such studies will also allow us to reassess the important topics such as planet migration \citep{Kley+Nelson2012}, kinematic signatures \citep{Pinte+2022}, and potentially towards the formation and dynamics of circumplanetary disks \citep[e.g.,][]{Kley1999,Suzulagyi+2014}.

Conventionally, disk angular momentum transport is thought to be due to turbulence, presumably due to the MRI \citep{Balbus+Hawley1998} or pure hydrodynamic instabilities \citep{Lyra+Umurhan2019}. Such turbulence is usually treated as a viscosity and parameterized by the dimensionless Shakura-Sunyaev $\alpha$ parameter \citep{Shakura+Sunyaev1973}. Most studies of planet-disk interaction are conducted under this framework of $\alpha$ disk model, where the gap-opening torque by planet gravity and is balanced by the viscous torque to fill in the gap \citep{Lin+Papaloizou1993}.

However, recent studies have shown that angular momentum transport in PPDs is likely dominated by magnetized disk winds \citep{Blandford+Payne1982,Bai+Stone2013b}. This is primarily because PPDs are poorly ionized, where the gas and magnetic fields are only weakly coupled, described by three non-ideal MHD effects \citep[e.g.,][]{Fleming+Stone2003,Ilgner+Nelson2008,Oishi+MacLow2009,Bai+Stone2011,Bai2013,Gressel+2015,Bethune+2017,Bai2017}. These effects generally suppress or damp the MRI, making it less effective in driving disk accretion \citep{Gammie1996,Wardle2007,Bai2011a}. While pure hydrodynamic instabilities can also drive turbulence, they are similarly considered insufficient in driving accretion \citep[see the recent review of ][]{Lyra+Umurhan2019}. Magnetized disk winds are more efficient ways of transporting angular momentum in thin disks \citep[e.g.,][]{Wardle2007,Bai+Goodman2009}. Launching magnetized disk winds requires the presence of large-scale poloidal magnetic flux threading the disk, and the rate of wind-driven angular momentum transport is directly correlated with the density of the poloidal flux \citep[e.g.,][]{Bai+Stone2017,Lesur2021,Tabone+2022}. Note that both the MRI and the hydrodynamic instabilities can co-exist with the disk wind, with the wind dominating angular momentum transport \citep{Cui+Bai2020,Cui+Bai2021,Cui+Bai2022}. The difference in the governing mechanism of angular momentum transport naturally calls for a re-investigation of planet-disk interaction under more realistic physical scenarios.

In addition, under the magnetized disk wind scenario, it has been found that the poloidal flux threading the disks tend to spontaneously and stochastically concentrate into thin magnetic flux sheets \citep{Bai+Stone2014}, which naturally lead to the formation of disk substructures known as zonal flows \citep{Johansen+2009,Simon+Armitage2014}. Regions with flux concentration get depleted in density, resembling planet-induced gaps \citep[e.g.,][]{Bethune+2017,Suriano+2018,Suriano+2019,Riols+Lesur2019,Riols+2020a,Cui+Bai2021}. Therefore, the process of wind-driven accretion is intrinsically stochastic and inhomogeneous.

The paradigm shift towards wind-driven angular momentum transport in PPDs has motivated planet-disk interaction studies that incorporate magnetic fields and wind-driven accretion. Early global simulations conducted under ideal MHD only incorporated mean toroidal magnetic flux without launching disk winds \citep{Nelson+Papaloizou2003,Papaloizou+2004,Winters+2003b,Baruteau+2011}. More recent local shearing-box MHD simulations have included a net poloidal magnetic flux, and found that the magnetic flux gets concentrated into the planet-induced gap, making the gap deeper and wider due to enhanced MRI turbulence within the gap \citep{Zhu+2013,Carballido+2016,Carballido+2017}. However, besides being local, these studies are unstratified without wind-driven angular momentum transport, and were conducted in the ideal MHD regime. 
Very recently, several groups carried out 2D hydrodynamic simulations with prescribed torques mimicking wind-driven accretion \citep{Kimmig+2020,Elbakyan+2022}, finding that wind-driven advection plays a very different role than viscous diffusion.
On the other hand, planet-induced gap strongly modifies the disk structure, while such approach adopt prescriptions based on winds from unperturbed disks.

In this paper, we conduct the first global MHD simulations of planet-disk interaction that self-consistently captures the launching of magnetized disk winds. Our simulations target the most observable outer regions of PPDs, and incorporate ambipolar diffusion as the dominant non-ideal MHD effect. Our simulations also feature sufficiently high numerical resolution in the disk region to properly resolve the MRI, while having sufficiently large simulation domain spanning two orders of magnitude in radius and opening all the way to the polar region to accommodate wind launching and propagation.
This will enable us to self-consistently study in this paper the magnetic field structures around the planet, reveal the torque balance maintaining the gap, and then evaluate how the gap structure and planet migration differ from the conventional understandings under the $\alpha$-disk framework. 

This paper is organized as follows. In \S~\ref{sec:method}, we describe the numerical methods and simulation setup. The numerical results including the disk and gap structures and magnetic flux concentration phenomena are shown in \S~\ref{sec:Disk}. In \S~\ref{sec:flow}, we discuss the flow structure around the planet-induced gap. We further examine in \S~\ref{sec:GOM} the torque balance that determines the gap structure and discuss how our results differ from that of previous studies in the viscous disk framework. We then discuss the migration torque acting on the planet in \S~\ref{sec:D_mig} and the implications of our results in \S~\ref{sec:Discussion}. Finally, we conclude in \S~\ref{sec:conclusion}.

\section{Numerical Method} \label{sec:method}
In this section, we describe the setup of our 3D non-ideal MHD simulations of planet-disk interaction. The background disk simultaneously possesses a magnetized wind launched from the surface and the MRI turbulence over the bulk disk, which we refer to as a ``windy'' disk. For comparison, we also conduct 2D global pure hydrodynamic simulations with constant-$\alpha$ viscosity as a reference.

\subsection{Basic equations}
We use Athena++, a grid-based, higher-order Godunov MHD code \citep{Athena++2020} to solve the following 
fluid equations 
in conservative form
\begin{align}
    \label{eq:con}
    &\frac{\partial \rho}{\partial t} +\nabla \cdot (\rho \bvec{v}) = 0,\\
    &\frac{\partial \rho \bvec{v}}{\partial t} + \nabla\cdot  \mathsf{M} = - \rho \nabla \Phi,
    \label{eq:mom}\\
    &\frac{\partial \bvec{B}}{\partial t} = \nabla \times(\bvec{v} \times \bvec{B} - \bvec{E}),
\end{align}
where $\rho, \Phi$ are the gas density and gravitational potential,
$\bvec{v}$ and $\bvec{B}$ are vectors of gas velocity and magnetic field,
$\mathsf{M}= \rho\bvec{v}\bvec{v} - \bvec{BB} + P^* \mathsf{I}$ is the stress tensor where $\mathsf{I}$ is the rank two unit tensor and $P^*=P+B^2/2$ is the total pressure with $P$ being the thermal pressure,
and
$\bvec{E}= \etaO \bvec{J}+ \etaA \bvec{J}_\perp$ is the non-ideal electric field in the rest fluid frame, where $\etaO$ and $\etaA$ are the Ohmic and ambipolar diffusivities, respectively, and $\bvec{J}=\nabla\times \bvec{B}$ is the current density vector,
$\bvec{J}_\perp$ is the current density vector perpendicular to $\bvec{B}$, respectively.
In the code, the units for $\bvec{B}$ is chosen such that the magnetic pressure is $B^2/2$ and hence magnetic permeability is 1.

We adopt a locally isothermal equation of state with $P=\rho  \cs^2$, where $\cs$ is the isothermal sound speed to be specified in Section \ref{sec:M_disk}. In doing so, we are free from the buoyancy resonance \citep[e.g.,][]{Bae+2021} which allows us to focus on MHD effects, and this condition is likely satisfied in the outer PPDs \cite[e.g.][]{Lin+Youdin2015,Pfeil+2019}. On the other hand, the system is subject to the vertical shear instability \citep[e.g.,][]{Nelson+2013}, which can coexist with the MRI as a source of disk turbulence \citep{Cui+Bai2022}. In our simulation setup ($\AM=1$ and 3, see Section \ref{sec:M_disk}), the MRI generally dominates.

We primarily focus on the outer regions of PPDs, where ambipolar diffusion (AD) is the dominant non-ideal MHD effect \citep{Wardle2007,Bai2011a}. The ambipolar diffusivity $\eta_A$ is characterized by the dimensionless Elsasser number defined as
\begin{equation}
    \AM     = \frac{v^2_\mathrm{A}}{\etaA \OmegaK}
\end{equation}
where $v_\mathrm{A}=\sqrt{B^2/\rho}$ is the Alfv\'en speed, $\OmegaK=\sqrt{GM_* R^{-3}}$ is the Keplerian angular velocity, $M_*$ is the central star mass, and $R=r\sint$ is the cylindrical radius. Note that in the absence of abundant small grains, $\eta_A\propto B^2$, so that $\AM$ is independent of field strength \citep{Bai2011b}. We further add some Ohmic resistivity $\eta_O$ near the inner region to stabilize the inner boundary (see Section \ref{sec:M_disk}).

Our simulations are conducted in a frame corotating with the planet, and describe the implementation of rotating frame appropriate source terms for angular momentum conservation in Appendix~\ref{sec:rot}. On the other hand, in the presentation of this paper, we consider $v_\phi$ as in lab-frame for convenience unless otherwise noted.

For 3D and 2D simulations, we solve these equations in the spherical polar coordinates ($r$, $\theta$, $\phi$) and cylindrical coordinate ($R$, $\phi$), respectively, centered at the central star. Note that we also use the 3D cylindrical coordinate ($R$, $\phi$, $z$) in analyzing the results of 3D simulations, where $z=r\cos{\theta}$.

\subsection{Gravitational potential}
The gravitational potential in the rotating frame is given as \citep[e.g.,][]{Fung+Chiang2016}
\begin{align}
\label{eq:GPot}
    \Phi = -GM_*  \left\{ \frac{1}{r}+\frac{q}{\sqrt{|\bvec{r}-\bvec{R}_\mathrm{P}|^2 + r_\mathrm{s}^2}}
    - \frac{q R \cos{\phi}}{\RP^2}
    \right\},
\end{align}
where $q=\MP/M_*$ is the planetary mass ratio to the central star where $\MP$ is the planetary mass, $R= r\st$ is the cylindrical radius, $\RP$ is the planet orbital radius, and $r_\mathrm{s}$ is the softening radius.
The third term in Eq.~(\ref{eq:GPot}) is the indirect potential due to fixing the star at the center of the simulation domain. 
We set $r_\mathrm{s}$ to be twice the cell size in $\phi$-direction at the planet.
As a first study, we fix the planetary orbit with circular motion, while we also note that the gap structure can be affected by planet migration \citep[e.g.,][]{Meru+2019,Kanagawa+2020} and orbital eccentricity \cite[e.g.,][]{Sanchez-Salcedo+2023}.

The planet gravity is introduced after a time of $t_\mathrm{P0}$ (see the Table~\ref{tab:param}). To minimize the influence of sudden mass addition, we make the planet mass linearly increase to the final mass at the rate of
\begin{equation}
    \dot{M}_\mathrm{P} = \frac{\Mth}{5 \tauP},
\end{equation}
where 
\begin{equation}
    \Mth= M_* (H_\mathrm{P}/\RP)^3
\end{equation}
is the thermal mass, which is used as a criteria whether planet can open a density gap at its orbit,
$H_\mathrm{P}$ is the disk scale height at the planetary orbit (see next subsection), $\tauP = 2\pi/\OmegaP$ is the planetary orbital period, and $\OmegaP$ is the planetary angular velocity.
This growth rate is comparable or longer than that in \citet{Fung+2014}.
For reference, the planetary Hill radius is defined as
\begin{equation}
    \rH= [\MP/(3M_*)]^{1/3}\RP \ ,
\end{equation}
and hence for $\MP=3\Mth$, $r_H=H_P$.

\subsection{Disk model for initial condition}
\label{sec:M_disk}
In our setup, the disk temperature is characterized by $c_s$, which we specify as a function of $R$ and is represented by the local disk scale height $H$ and disk aspect ratio $h$
\begin{equation}
    H=c_s/\Omega_K\ ,\quad h\equiv H/R\ .
\end{equation}
In this work, we fix the aspect ratio of the bulk disk $\hdisk\equiv H_{\rm disk}/R$ to be $0.1$ (constant, and without causing ambiguity, we drop the subscript ``disk"). Then, the vertical density profile in hydrostatic equilibrium is given by \citep[e.g.,][]{Gressel+2015}
\begin{align}
    \rhoi=\rho_0 \left(\frac{R}{r_0}\right)^{-q_\rho} \exp \left\{ \frac{GM}{c_\mathrm{s}^2}\left(\frac{1}{r}-\frac{1}{R}\right)
    \right\},
\end{align}
where we use $r_0=1$, $\rho_0=1$, and $q_\rho=2.25$ for the fiducial runs (see Table~\ref{tab:param}). We also set a density floor value of $10^{-8} (r/r_0)^{-q_\rho}$ in the code units. The initial surface density $\Sigma_0 \approx \sqrt{2\pi}H_{\rm disk}\rhoi\propto r^{-1.25}$. This is also adopted in 2D simulations.
To achieve force balance, the $\phi$-direction velocity is initially set to
\begin{align}
    v_\phi = \sqrt{ \left(\frac{GM}{R}\right) 
    \left\{ \frac{R}{r}
    -h_{\rm disk}^2(q_\rho+1) \right\} } - R\OmegaP
\end{align}
in the corotating frame with the planet.
The velocities in other directions $v_r$ and $v_\theta$ are initialized with random values up to $2.5\%$ of the local sound speed.

At high altitudes, the disk atmosphere is tenuous but is expected to be hot due to the UV and X-ray heating \citep[e.g.,][]{Glassgold+2004}. To mimic such a hot atmosphere, we allow the disk aspect ratio $h$ to transition from $h_{\rm disk}$ to some higher value $h_{\rm atm}$ in the following functional form: 
\begin{align}
    h =& \exp \left\{ \frac{\ln(\hdisk)+\ln(\hatm)}{2}
    \right. \nonumber\\ & \left.
    +\frac{\ln(\hdisk)-\ln(\hatm)}{2} \tanh{\left[ 2\left( \frac{R\cos{\theta} - z_\mathrm{t}}{H}  \right) \right] }
    \right\},
    \label{eq:htrans}
\end{align}
and we choose $\hatm=0.4$ and $z_\mathrm{t}=4H$ is the transition height. Similar approaches have been adopted in our earlier works \cite[e.g.,][]{Bai+Stone2017,Cui+Bai2021}.
We note that although hydrostatic equilibrium is not initially maintained at the disk upper layer, the subsequent wind launching process will quickly override any imbalance, and the outcome is largely independent of the detailed treatment here.

The poloidal magnetic field is initialized as
\begin{align}
  \boldsymbol{B} = \nabla \times (A_\phi \hat{\phi}),
\end{align}
where the vector potential is given in \citet{Zanni+2007,Bai+Stone2017} as
\begin{align}
    A_\phi = \frac{2 B_{z0} R}{3-q_\rho} \left(\frac{R}{r_0}\right)^{-\frac{q_\rho+1}{2}} 
    \left\{ 1+ (m \tan{\theta})^{-2}
    \right\}^{-\frac{5}{8}},
\end{align}
where $B_{z0}$ is the initial poloidal field at the midplane of the $r$-inner boundary ($r=r_0=1$) and
$m=1$ is the geometry parameter. We set $B_{z0}=\sqrt{2\beta_0 \rho_0 c_\mathrm{s}^2}$ so that the entire disk is magnetized with the same plasma beta ($\beta_0$, ratio of initial gas pressure to magnetic pressure), and we choose $\beta_0=10^{4}$.
Hence, the initial magnetic flux at the midplane is $\Bmi=\sqrt{2\beta_0 \rhoi h \vK}$, where $\vK=R\OmegaK$ is the Keplerian velocity.

Following earlier works \citep{Bai2011a,Bai2011b,Simon+2013b,Bai+Stone2017,Riols+Lesur2019,Cui+Bai2021}, we set the ambipolar Elsasser number to be constant of order of unity over the bulk disk column, while the value of $\AM$ transitions to much higher values in the disk atmosphere to mimic higher ionization due to the penetration of far-UV radiation \citep{Perez-Becker+Chiang2011}.
Since the physics are similar to the heating of atmosphere, we use the same functional form as Eq.~(\ref{eq:htrans}) for the transition in $Am$ value. 
We set the $\AMzero=1$ or 3 and $\AMatm=100$ in the bulk disk and the atmosphere, respectively. The midplane $\AM$ of 1 or 3 leads to a moderately strong MRI turbulence that overwhelms the potential influence of the VSI \citep{Cui+Bai2022}. 
Although gap formation would modify the ionization profile and hence the $\AM$ values in the vicinity of the planet orbit, we note that the previous studies on transitional disks infer that $\AM$ is still of order unity even inside the deep density cavity \citep{Wang+Goodman2017,Martel+Lesur2022}.
As the first study of planet-disk interaction with both MRI turbulence and MHD disk wind, this simplified treatment allows us to focus on purely MHD effects, and we defer more self-consistent treatment of ionization chemistry to a future study.
It is beyond the scope of this work to explore broader ranges of $\AM$ values and/or more self-consistent chemistry \citep[e.g.][]{Bai2017}.
In this study, we treat $\AMzero=3$ as the fiducial, which is on the high side but is also in line with the realization that cosmic-ray traveling along magnetic fields can enhance the ionization rate in the disk outer region \citep{Fujii+2022}.
Additionally, following \S~2.3 in \citet{Cui+Bai2021}, we apply Ohmic resistivity in the inner region ($r<2$) to help stabilize the inner boundary.

\subsection{Numerical settings}

For the main 3D simulations, the simulation domain spans from $r=1$ to $100$ in code units with logarithmic grid spacing, and the $\theta$ and $\phi$ domains cover full $\pi$ and $2\pi$, respectively, with the polar boundary condition \citep{Zhu+Stone2018}.
The planet orbit is fixed at $R_p=6$ in code units\footnote{It is known that global MHD simulations of disk winds tend to be affected by inner boundary conditions (e.g. \citealp{Cui+Bai2021}). Despite of rendering the simulations more expensive (per planet orbit), we choose a relatively large $R_p$ in order to leave sufficient dynamical range to minimize this influence.}, and in the corotating frame, its coordinate is $(r,\theta,\phi)=(6,\pi/2,0)$. We note that the large radial extent relative to the planetary orbit and the $\theta$ extent to both poles are essential to properly accommodate the MHD disk wind.
Similar to \citet{Cui+Bai2021}, the $\theta$ grid size increases from the midplane to the pole by a constant factor, and the pole to midplane grid size ratio is four. We use three levels of static mesh refinement on top of the root grid, which has 112, 48, and 96 grid cells in $r$-, $\theta$-, and $\phi$-directions, respectively. 
The finest grid ranges from $2.5<r<12$, $1.5<\theta<1.64$, and full 2$\pi$ in phi-direction, respectively.
In the $\theta$-direction, the finest grid cell resolves the disk scale height by $\sim26$ cells. At the planet orbit ($R=6$), the cell size is $(\Delta r,r \Delta \theta, r\Delta \phi)=(0.030,0.023,0.049)$ in code units.

At both radial boundaries, $\rho$ and $T$ are copied from the nearest grid cell to the ghost zones with extrapolating following the initial radial gradient. 
At the outer boundary, the $v_r$ and $v_\theta$ are copied from the nearest cell to the ghost cell, while for $v_r<0$, $v_r$ is forced to zero to prevent inflow. We also copy $v_\phi$ with extrapolation according to $r^{-1/2}$.
At the inner boundary, both $v_r$ and $v_\theta$ are fixed to zero, and $v_\phi$ is fixed as initial, which was found useful to stabilize the inner boundary region \citep[see also][]{Bai+Stone2017}. 
Magnetic field in the ghost zone is copied from the nearest grid cell assuming $B_r\propto r^{-2}$ and $B_\phi \propto r^{-1}$, with $B_\theta$ unchanged. Besides, the electric field in $\phi$-direction is forced to zero at the radial inner boundary to anchor the initial magnetic flux in the inner boundary.

In the 2D simulations, with cylindrical coordinates $(R,\phi)$, no magnetic field is included, and we include the standard $\alpha$ viscosity $\nu=\alpha\cs H$. Other settings are same as the 3D simulation described above.

\begin{table}[t]
    \centering
    \begin{tabular}{ c   c c c c c c}
    \multicolumn{7}{c}{3D MHD simulations}\\ \hline
    Run & $\AMzero$ & $q_{\rho}$  & $\MP$ & $t_\mathrm{P0} /\tauP$ & $t_\mathrm{Pf}/\tauP$ & $t_\mathrm{e}/\tauP$\\
    \hline
    Mt1Am3 &   3 & 2.25 & $1\Mth$ & 3.4 & 8.4 & 120\\
    Mt3Am3 &   3 & 2.25 & $3\Mth$ & 3.4 & 18.4 & 140\\
    Mt5Am3 &   3 & 2.25 & $5\Mth$ & 3.4 & 28.4 & 140\\
    Mt3Am1 &   1 & 2.0  & $3\Mth$ & 40  & 55   & 200\\ \hline\hline
    \multicolumn{7}{c}{2D viscous simulations}\\ \hline
    Run &   $\alpha (10^{-3})$ & $q_{\rho}$  & $\MP$ & $t_\mathrm{P0} /\tauP$ & $t_\mathrm{Pf}/\tauP$ & $t_\mathrm{e}/\tauP$\\
    \hline
    Mt1$\alpha$6 &  {$6$} & 2.25 & $1\Mth$ & 3.4 & 8.4 & 120\\
    Mt3$\alpha$6 &  {$6$} & 2.25 & $3\Mth$ & 3.4 & 18.4 & 120\\
    Mt5$\alpha$6 &  {$6$} & 2.25 & $5\Mth$ & 3.4 & 28.4 & 120\\
    Mt1$\alpha$0 &  {0} & 2.25 & $1\Mth$ & 3.4 & 8.4 & 120\\
    Mt3$\alpha$0 &  {0} & 2.25 & $3\Mth$ & 3.4 & 18.4 & 120\\
    Mt5$\alpha$0 &  {0} & 2.25 & $5\Mth$ & 3.4 & 28.4 & 120\\
    Mt3$\alpha$3A & {$3$} & 2.0 & $3\Mth$ & 40 & 55 & 200\\
    Mt3$\alpha$0A & {0} & 2.0 & $3\Mth$ & 40 & 55 & 200\\
    \hline
    \end{tabular}    
    \caption{List of simulation runs and parameters. The initial plasma $\beta=10^4$ for poloidal magnetic fields in all MHD runs.}
    \label{tab:param}
\end{table}

The main parameters for the simulations are listed in Table~\ref{tab:param}, where $t_\mathrm{P0}$, $t_\mathrm{Pf}$, and $t_\mathrm{e}$ are the times when the planet is introduced, when the planet reaches the final mass, and when the simulation stops, respectively.
In this paper, we consider three different planet masses, with $\MP=1\Mth$, $3\Mth$ and $5\Mth$, respectively, and take the case with $\MP=3\Mth$ and $\AM_0=3$ as fiducial. Our presentation of simulation results mostly focus on the $Am=3$ simulations, and we mainly summarize the results from the $Am=1$ simulations in Appendix \ref{sec:Am1} which are consistent with our fiducial runs. Given that we anticipate spontaneous formation of ring-like substructures even without planets \citep{Cui+Bai2021}, we choose planet masses that are expected to be largely enough to compete with the planet-free gap-formation mechanisms.

\subsection{Diagnostics}
\label{sec:M_ana}
Before we present simulation results, we introduce and define certain concepts and quantities helpful for diagnostics.
The simulation domain is divided into the disk and wind regions which coincide with the temperature and $\AM$ transition \citep{Bai+Stone2013b,Gressel+2015}. Although the characteristic height for the transition is set to $z=4H$, the temperature and $\AM$ transitions start at a bit lower height as following Eq.~\ref{eq:htrans}. 
Judging from the vertical profile of the velocities (see also figure~\ref{fig:vertical_vB}), we define the disk zone as $|z|<3.5H$. Hence, the disk surface density is defined as 
\begin{equation}
    \Sigma = \frac{1}{2\pi}\int_{0}^{2\pi} d\phi \int_{-3.5H}^{3.5H} dz ~ \rho.
\end{equation}
Note that the disk boundary is constant at $\theta= \TthetaD \equiv \arctan(3.5h)$ because the disk aspect ratio is assumed to be constant.

To illustrate the evolution of poloidal magnetic flux, we use linearly equally spaced contour for the poloidal magnetic flux function defined as
\begin{align}
    \Phi_B(r,\theta) = \int_0^{\theta} r d\theta' \int_0^{2\pi} r d\phi \sin{\theta'} B_r(r,\theta',\phi)
\end{align}
We also define the mean vertical magnetic field in the midplane region
\begin{equation}
    \Bzm = \frac{1}{2 H} \int_{-H}^{H} dz ~ B_z\ ,
\end{equation}
which is more convenient when considering, e.g., correlations with disk surface density profiles.
Since the poloidal magnetic field is nearly vertical at the midplane but bends radially outward at high altitudes (see figure~\ref{fig:rzslice}), we average only within $1H$ rather than within the whole disk ($3.5H$). 

The disk structure including the planet-induced gap is governed by angular momentum transport, which is more naturally analyzed in spherical-polar coordinates, and the equation for angular momentum conservation can be expressed as
\begin{align}
    &\frac{\partial \left( \rho j\right)}{\partial t} 
    + \frac{1}{r^2} \frac{\partial}{\partial r}  \left(r^2 \rho v_r j 
    + r^3 \sint T^\mathrm{Max}_{r\phi}\right)
    \nonumber \\
    +& \frac{1}{\sint} \frac{\partial }{\partial \theta} \left( \sin{\theta} \rho v_\theta j 
    +r \sin^2{\theta} T^\mathrm{Max}_{\theta \phi} \right)
    \nonumber\\
    +& \frac{\partial}{\partial \phi}\left( \rho v_\phi j 
    + r\sint T^\mathrm{Max}_{\phi \phi} + r\sint P^* \right)
    \nonumber \\
    =& -r\sint \rho \frac{\partial \Phi}{\partial \phi}
    \label{eq:AM_sph}
\end{align}
where $j= r\sint v_\phi$ is the specific angular momentum in the $\phi$-direction (in the lab frame), and $T^\mathrm{Max}_{ij} = -B_i B_j$ is the $ij$th component of the Maxwell stress tensor.

We can define the standard disk integral as
\begin{equation}
    \iint ds\equiv \idtheta \idphi ~ r^2 \sint\ ,
\end{equation}
where $\theta_\mathrm{d,u} = \pi/2\pm \TthetaD$ (note $\TthetaD=0.35$ in our case). In particular, we can define the (spherical-shell-integrated) surface density and accretion rate as
\begin{equation}
    \Sigma_\mathrm{sph} = \frac{1}{2\pi r}\iint ds \rho\ ,\ 
    \Macc = -\iint ds ~ \rho v_r\ .
\end{equation}
We can further define the disk average of a quantity $A$, denoted by angle bracket as
\begin{equation}
    \langle A\rangle\equiv\iint ds A\bigg/\iint ds\ .
\end{equation}
In our analysis, unless otherwise noted, we further average the data over $130<t/\tauP<140$ in time, and $\pm 0.25~H$ in $r$ to enhance statistics, while excluding the region where $|\bvec{r}-\bvec{\RP}|<\rH$.
In particular, we define
\begin{equation}
    \langle\rho v_r\rangle\equiv-\Macc\bigg/\iint ds\ ,\ 
    \langle j\rangle \equiv \iint ds j \bigg/\iint ds\ ,
\end{equation}
and local deviations from these averaged quantities can be denoted by $\delta(\rho v_r)$ and $\delta j$. 

By integrating over $\phi$- and $\theta$-directions, and by utilizing the continuity equation, Eq.~(\ref{eq:AM_sph}) can be cast into 
\begin{align}
    \label{eq:AM_decom}
    2 \pi r \Sigma_{\rm sph}\frac{\partial \langle j\rangle}{\partial t} \approx 
    \GA + \GP + \Gwind
    -\frac{\partial}{\partial r}\left(\Jwave + \JMx \right)
\end{align}
where
\begin{align}
    \label{eq:JA}
    \gamma_A &= \Macc\frac{\partial \langle{j}\rangle}{\partial r}, \\
    \label{eq:GP}
    \GP &= - \iint ds  \left( r\sint \rho \frac{\partial \Phi}{\partial \phi} \right),\\    
    \label{eq:Gwind}
    \Gwind&= -\int d\phi \left[r\sint \left( \rho v_\theta ~\delta j + r\sint T^\mathrm{Max}_{\theta \phi}\right)\right]_{\theta_\mathrm{u}}^{\theta_\mathrm{d}}, \\
    \label{eq:Jwave}
    \Jwave &= \iint ds \left(r \sint T^\mathrm{Rey}_{r \phi} \right),\\
    \label{eq:JMx}
    \JMx&= \iint  ds \left( r\sint ~ T^\mathrm{Max}_{r \phi}\right),
\end{align}
In the above, $\GP$ and $\Gwind$ are the torque densities due to the planet gravity and MHD wind, $\Jwave$ and $\JMx$ are the angular momentum flux due to the Reynolds and Maxwell stresses, respectively.

In addition, we define the cumulative torque $\Gamma$, which has a same dimension as the angular momentum flux, as
\begin{align}
\label{eq:GPt}
\GPt &= -\int_{\RP}^{r} dr' \GP(r'),\\
\Gamma_\mathrm{wind} &= -\int_{\RP}^{r} dr' \Gwind(r').
\label{eq:Gwt}
\end{align}
Note that $-\{\GPt(r_\mathrm{out}) -\GPt(r_\mathrm{in})\}$ is the torque acting on the planet as the backreaction ($r_{\rm in}$, $r_{\rm out}$ are the radius of the inner and outer radial boundary), so-called the planet migration torque, where $r_\mathrm{in}$ and $r_\mathrm{out}$ are the radius of the inner and outer boundary of the disk.

We have already defined $T^{\rm Max}$, and the Reynolds stress is defined as
\begin{equation}\label{eq:Rey}
   T^\mathrm{Rey}_{r\phi} \equiv \delta(\rho v_r) \delta v_\phi
   =\delta(\rho v_r)\frac{\delta j}{r\sint}\ .
\end{equation}
Note that the Reynolds stress has a contribution from the MHD-driven flows (e.g., the MRI turbulence), as well as the planetary density wave/shock, although it is not straightforward to separate them.
The remaining term $\gamma_A$ is the torque density associated with bulk accretion, which should be understood as $\Macc$ being the response to the action of the other four terms. 

The stresses are related to the Shakura-Sunyaev $\alpha$ values \citep{Shakura+Sunyaev1973},
which are given, when averaged over the disk, as
\citep[e.g.,][]{Balbus+Hawley1998}
\begin{align}
    \label{eq:alpha_Re}
    \alphaRe&= \av{ T^\mathrm{Rey}_{r \phi}} /  \av{P}, \\
    \alphaMx&= \av{T^\mathrm{Max}_{r\phi}} / \av{P},
    \label{eq:alpha_Mx}
\end{align}
and their sum gives the total $\alpha$ value.

There are a few points worth further clarification. First,
while $\Gwave$ comprises of the Reynolds stress, it is largely dominated by the planetary density wave (here the spiral shock) at least in the vicinity of the planet. Following \citet{Kanagawa+2017}, we refer to this term as the ``wave" term representing the angular momentum flux carried by the density wave/shock.
Second, while the MRI turbulence produces both the Reynolds and Maxwell stresses, the total stress and hence $\alpha$ is dominated by the Maxwell stress by a factor of a few (e.g., \citealp{Bai+Stone2011}). Thus, we consider $\JMx$ as the proxy for total viscous flux of angular momentum (see also \S~\ref{sec:alpha}). 
Third, we analyze angular momentum transport in spherical polar coordinates, which is more rigorous than in cylindrical coordinates, despite of being less optimal for wind diagnostics since the direction of the wind deviates substantially from the $\theta$-direction.
Here, the {\it local} wind mass-loss rate can be defined as 
\begin{equation}
    \frac{d\dMwind}{d\ln r} = \int d\phi [r^2 \sin{\theta} \rho v_\theta]_{\theta_\mathrm{u}}^{\theta_\mathrm{d}}\ .
\end{equation}
Finally, we note that in deriving Equation (\ref{eq:AM_decom}), we have already accounted for the angular momentum loss associated with bulk mass loss $\dMwind\langle j\rangle$, and hence in $\Gwind$, the first term only accounts for the deviation $\delta j$ instead of the total $j$.
Also note that we have omitted a term of $\iint ds \partial(\rho \delta j)/(\partial t)$ from Eq.~(\ref{eq:AM_decom}), which is expected to be small at least in the quasi-steady state that we have achieved.

Similarly to the above diagnostics in 3D spherical coordinate, the torque in the 2D viscous simulations is categorized into $\GP$, $\GA$, $\JA$, $\Jwave$, and $\Jvis$. Here, there is no $\Gwind$, and the $\GMx$ is replaced by the viscous torque density defined as \citep[e.g.,][]{Kanagawa+2015}
\begin{align}
    \Gvis = \frac{d}{dR} \left(-2\pi R^3 \nu \Sigma \frac{d\Omega}{dR} \right).
\end{align}

The direct contribution from the planet arises from both the $\GP$ term of planet gravity, together with the $\Jwave$ term representing wave propagation.
The net torque deposition rate can be approximately expressed as \citep[e.g.,][]{Kanagawa+2017} 
\begin{equation}
    \Gd\equiv
    \GP -\frac{\partial \Jwave}{\partial r},
\end{equation}
which is associated with the dissipation of the density waves and directly leads to planet-induced substructure formation. 

Without external forces, fluid parcels in the planet vicinity experience horseshoe orbits over the radial extent of $\sim\rH$ \citep[e.g.,][]{Masset+2006,Paardekooper+Papaloizou2009a}, and this region is thus called the corotation region or horseshoe region.
For convenience, we just refer to $|R-\RP|<\rH$ as the corotation region in this work.

\section{General gas dynamics and disk structure}
\label{sec:Disk}
\begin{figure*}
    \centering
    \begin{minipage}{\Fh}
        \centering
        \includegraphics[width=\hsize]{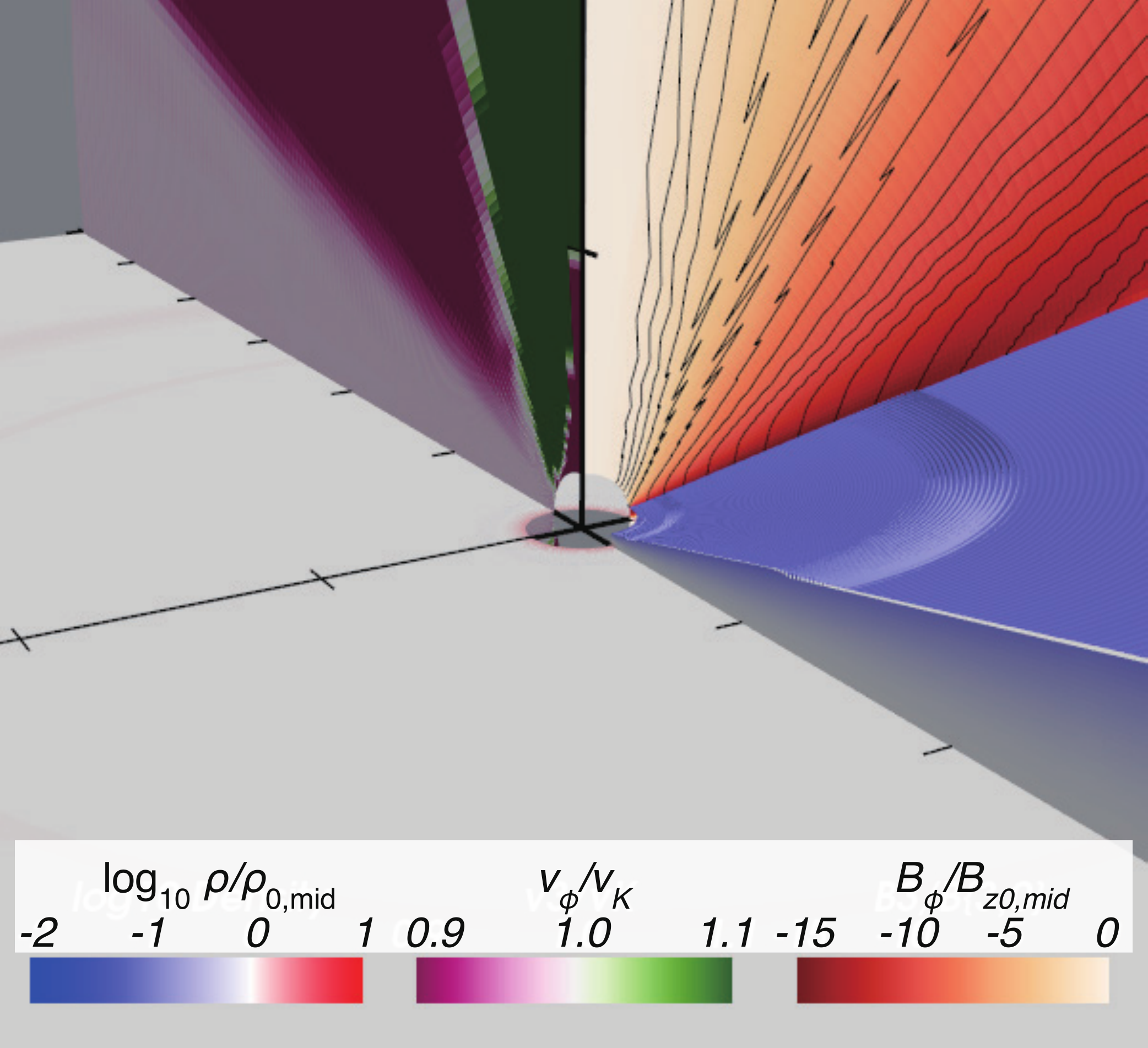}
        {(a) $t=3~\tauP$}
    \end{minipage}
    \hspace{0.015\hsize}
    \begin{minipage}{\Fh}
        \centering
        \includegraphics[width=\hsize]{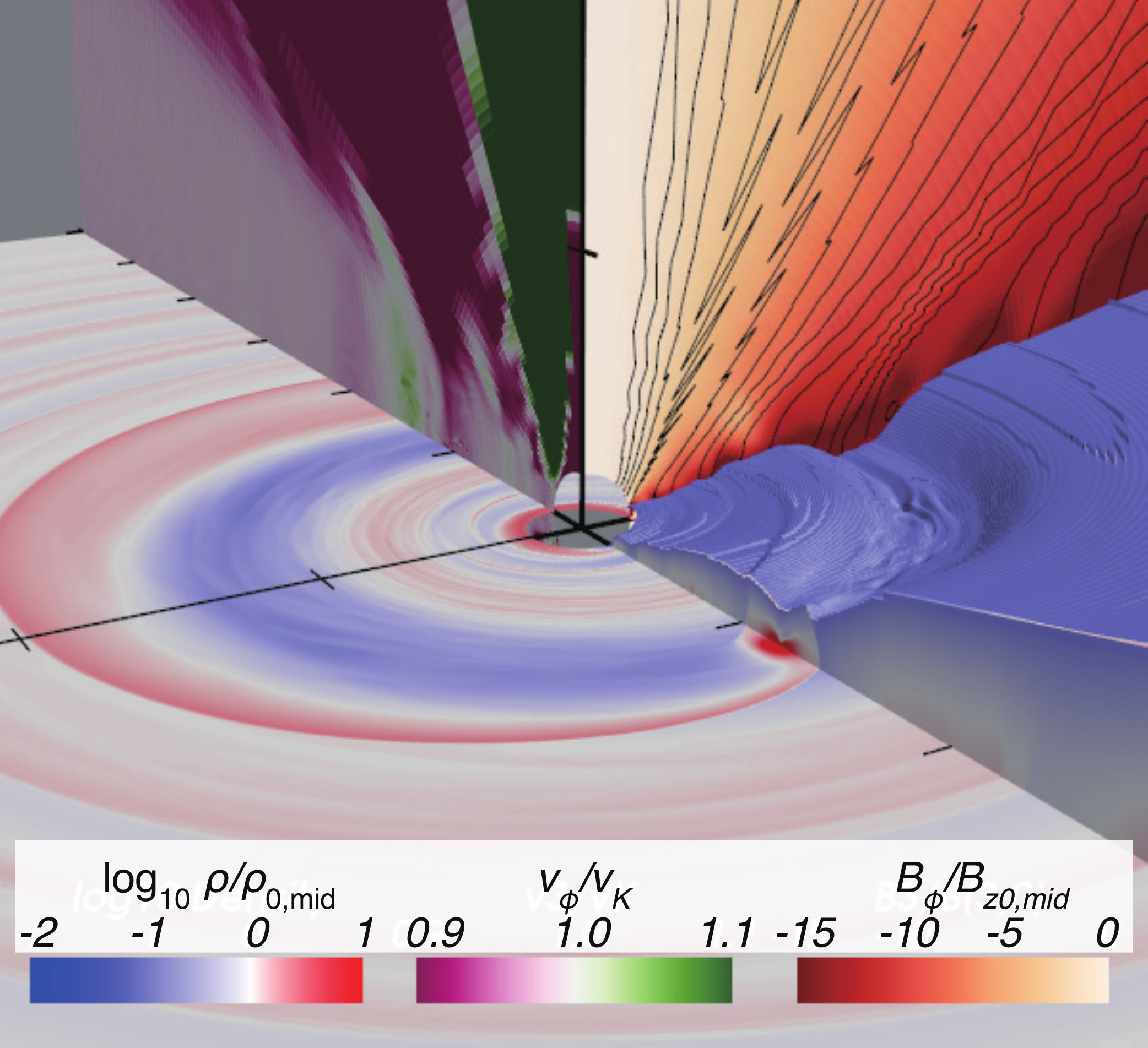}
        {(b) $t=30~\tauP$}
    \end{minipage}\\
    \vspace{0.02\hsize}
    \begin{minipage}{\Fh}
        \centering
        \includegraphics[width=\hsize]{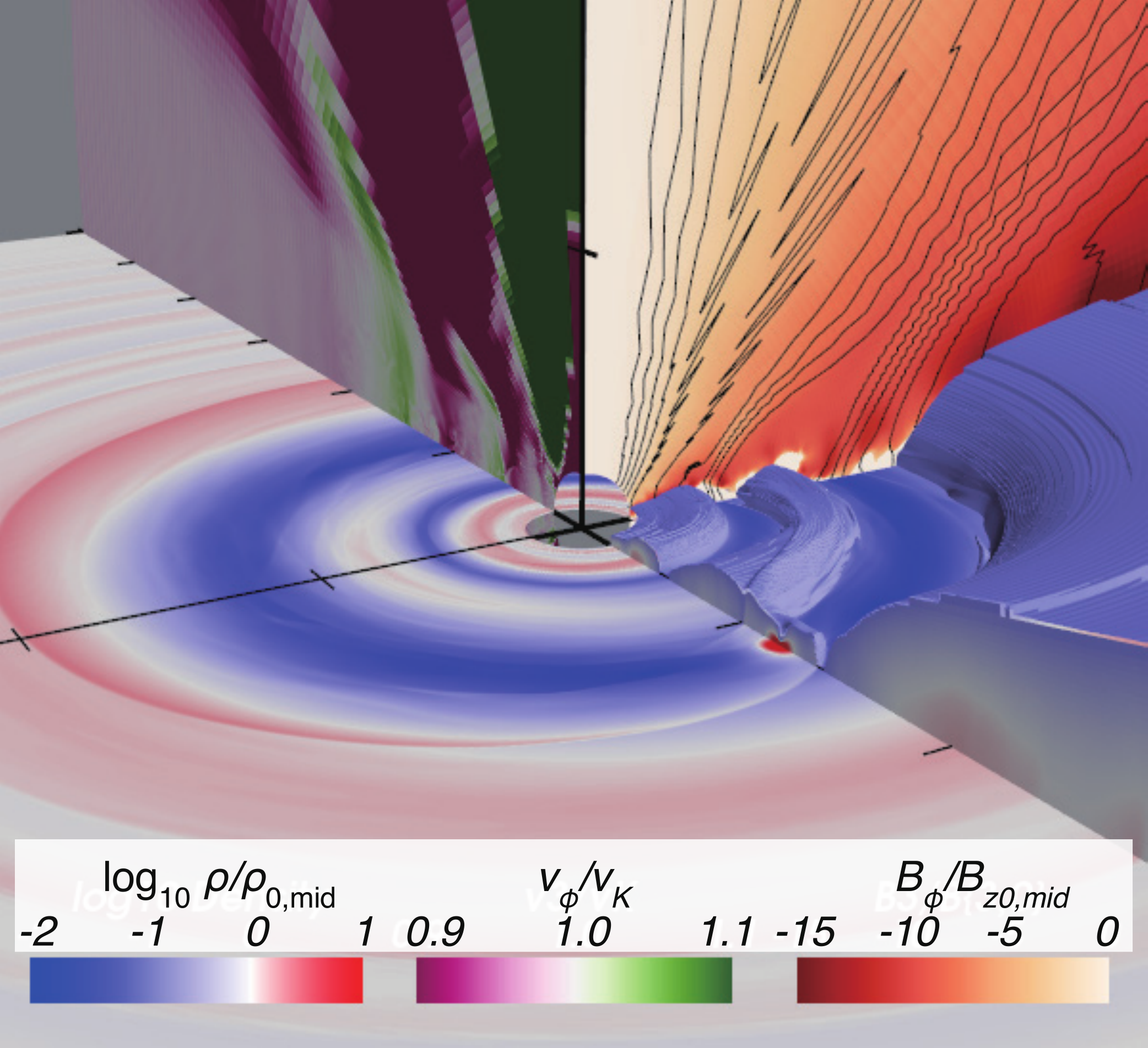}
        {(c) $t=100~\tauP$}
    \end{minipage}
    \hspace{0.015\hsize}
    \begin{minipage}{\Fh}
        \centering
        \includegraphics[width=\hsize]{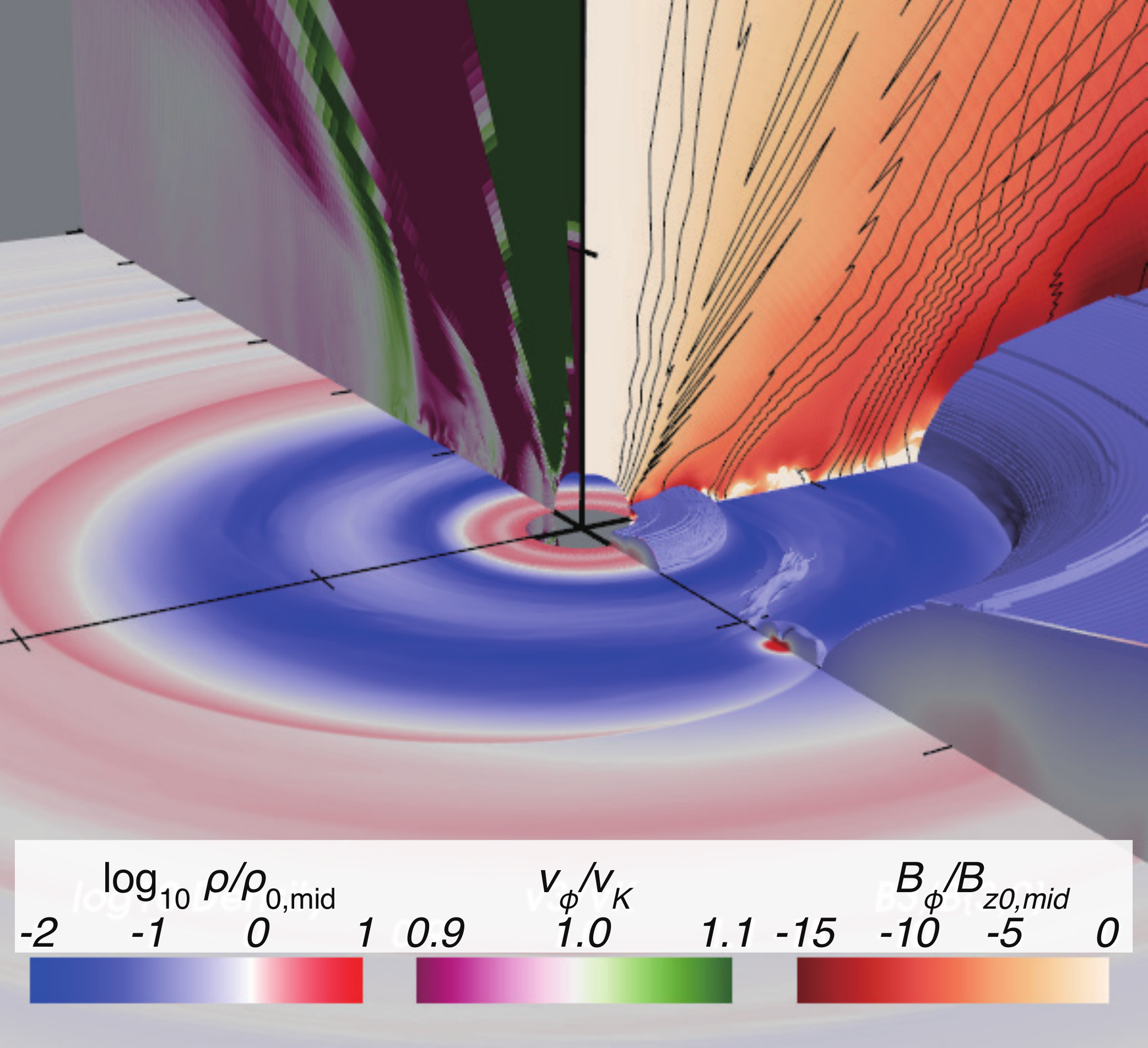}
        {(d) $t=140~\tauP$}
    \end{minipage}
    \caption{Snapshots of simulation results at $t=3$, 30, 100, and $140~\tauP$ in the fiducial run Mt1Am3. In each panel, the bottom slice shows the midplane density, the left and right vertical walls show $\phi$-averaged $v_\phi/v_K$ and $B_\phi/B_{z,0}$, respectively, and the blue iso-surface marks a constant density of $\rho=\rho_\mathrm{0,mid} ~\mathrm{e}^{-2H}~$, where $\rho_\mathrm{0,mid}= \rho_0 R^{-q_\rho}$. The black lines on the right wall show the contours of equi-spaced azimuthally-averaged poloidal magnetic fluxes. Along the three axes ($x=R\sin\phi$, $y=R\cos\phi$, z), the black ticks have a uniform spacing of 5 in code units.
    }
    \label{fig:3D}
\end{figure*}

In this section, we discuss, primarily at the phenomenological level, how the presence of a planet modifies the overall gas dynamics and disk structure. We focus on the $\AM=3$ simulations with different planet masses, where the background gas dynamics are characterized by MHD wind with modestly strong MRI turbulence.
We note that the background gas dynamics without the planet is very similar to that in \citet{Cui+Bai2021}, and we refer to that work for further information.

\subsection{Overview of the fiducial case}
\label{sec:fid}
Figure~\ref{fig:3D} shows the major MHD quantities in our fiducial case (Mt3Am3). The panels from (a) to (d) are the snapshot at $t=3$, 30, 100, and $140~\tauP$, respectively.
Since we introduce the planet at $t=3.4~\tauP$, there is no planet and no significant structure in panel (a). Note that this snapshot is common with the three runs with $\AM=3$, and the density feature in this snapshot is part of the initial relaxation process.
The disk is subject to both the MHD-disk wind and the MRI turbulence. While the MRI is not yet well developed at the planetary orbit, the disk wind is established and starts to drive the disk accretion. 

The planet reaches to the final mass of $3~\Mth$ at $t=28.4~\tauP$. In the panel (b) of $t=30~\tauP$, the planet excites the density wave (spiral shock) and starts to open a density gap around its orbit.
Interestingly, we see that poloidal magnetic flux gets concentrated in the gap region, which we identify as a major characteristic of planet-disk interaction in the presence of MHD winds. We note that this phenomenon was also identified in earlier local shearing-box simulations \citep{Zhu+2013,Carballido+2016,Carballido+2017}, nevertheless, these simulations were unstratified thus do not launch disk winds, and they were conducted under ideal MHD conditions. 
Also note that, at this time, the MRI turbulence is fully developed at the planetary orbit. 

In panel (c) at $t=100~\tauP$, the planetary gap gets wider and deeper. In addition, a planet-free gap appears inside the planetary orbit. This phenomenon has been found in previous simulations of the MRI turbulence and/or disk winds without a planet \citep{Bai+Stone2014, Bai2015, Bethune+2017, Suriano+2018, Suriano+2019, Riols+Lesur2019, Riols+2020a, Cui+Bai2021}, where magnetic flux spontaneously concentrate into quasi-axisymmetric flux sheets, leading to the formation of ring-like substructures. This process is expected to be stochastic, and applies for a wide range of $\AM$ values \citep{Cui+Bai2021}.
Both the planetary and the planet-free gaps get wider over time, and by the time of $t=140~\tauP$, they start to overlap with each other, namely the density between the two gaps is getting depleted over time relative to the initial (white), and the two gaps are on a merging trajectory (see Figure \ref{fig:GapProf} in the next subsection). Due to substantial computational cost, we terminate our simulation at this time. Also note that in practice, it will likely take $10^3$ planetary orbits for the gap profile to reach a steady state as found in pure hydrodynamic simulations \citep[e.g.,][]{Fung+Chiang2016}.

\begin{figure*}
    \centering
    \includegraphics[width=\hsize]{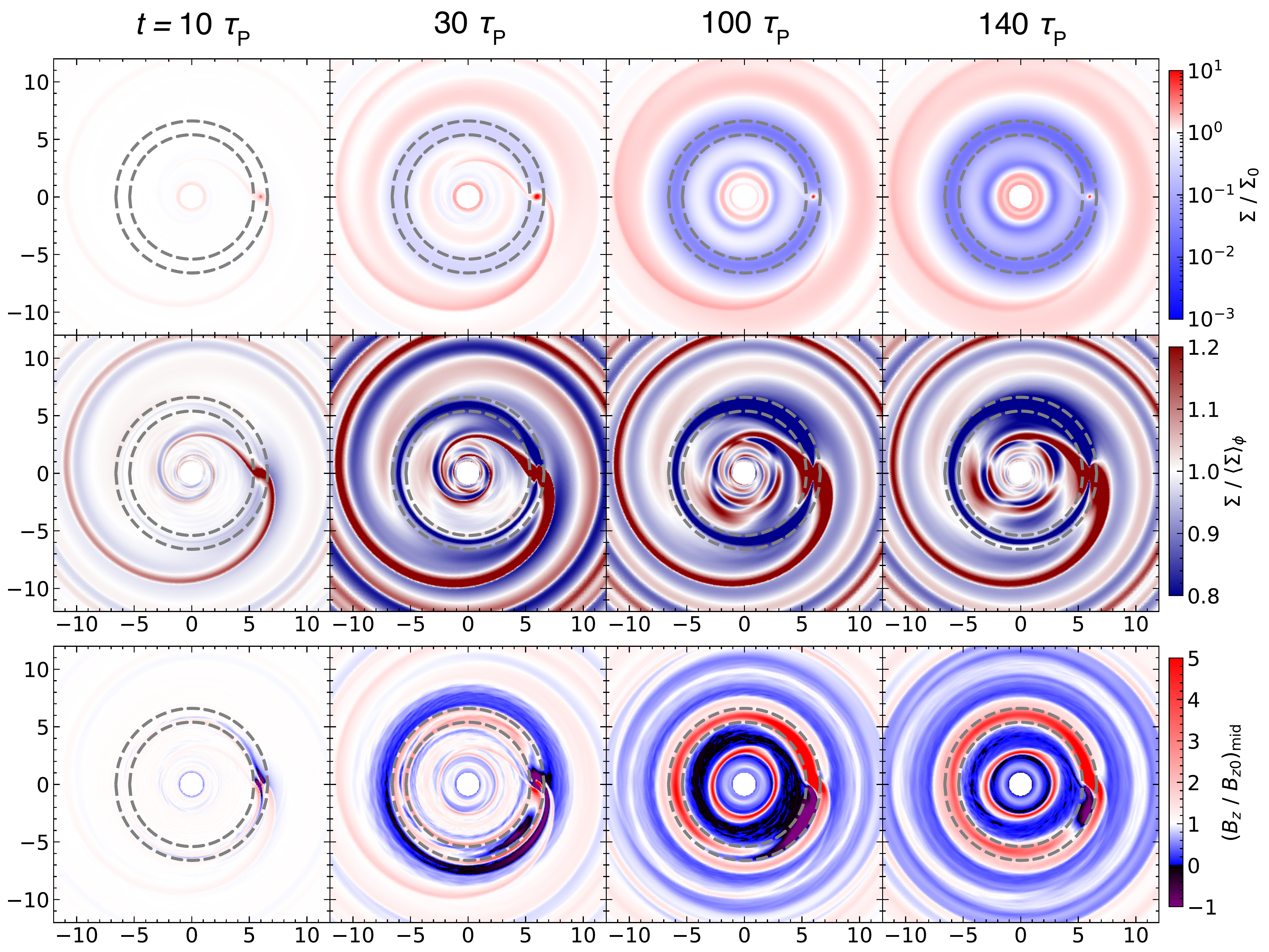}
    \caption{$R$-$\phi$ plots of surface density (top) and vertical magnetic field averaged within $|z|<H$ (bottom) normalized by the initial values and surface density normalized by the average at the orbit (middle), at $t=10$, $30$, $100$, and $140~\tauP$ from left to right. The time average is taken over $10~\tauP$ before the quoted values.
    The grey dashed circles correspond to $R=\RP\pm\rH$.
    }
    \label{fig:Rphi}
\end{figure*}

To further illustrate the gap opening process and magnetic flux concentration, we show in figure~\ref{fig:Rphi} the snapshots of $\Sigma$ (upper) and $\Bzm$ (bottom) at $t=10$, $30$, $100$ and $140~\tauP$.
The planetary gap ($R=\RP(=6)$) is clearly associated with the spiral density waves/shocks. For the vertical magnetic flux threading the disk, we see that in the inner planet-free gap, magnetic flux largely concentrates in the gap region, consistent with previous findings \citep{Cui+Bai2021}. On the other hand, magnetic flux distribution around the planet-induced gap is clearly asymmetric, akin to the spiral density waves/shocks. Moreover, a patch of negative magnetic flux is present at the trailing side of the spiral shock near the planet.
We will discuss the magnetic field concentration in \S~\ref{sec:MFC}. Its effects on the disk turbulence and gap opening will be discussed in \S~\ref{sec:alpha} and \S~\ref{sec:GOM}, respectively.

To highlight the density waves/shocks, we further show in the middle panels of Figure \ref{fig:Rphi} (and also in Figure \ref{fig:Rphi_15}) the azimuthal surface density variations $\Sigma/\av{\Sigma}_\phi$. Besides the primary planet density waves/shocks already visible in the top panels, we can identify a secondary spiral interior of the planet orbit, especially at early time of $t=30\tauP$. Secondary spirals have been found in previous hydrodynamic simulations of planet-disk interaction in inviscid/low-viscosity disks \citep{Dong+2015,Zhu+2015,Bae+2017}, and is a result of constructive interference of a set of spiral modes \cite{BaeZhu2018}.
They can also open additional gaps in low viscosity ($\alpha\lesssim10^{-3}$) disks \citep{Bae+2017}. While they might contribute to the initial formation of the planet-free gap in our simulations, we note that the effective viscosity in our simulations is higher, and that the amplitudes the secondary spirals in our simulations are relatively low ($\lesssim20\%$). We also observe that at later time, the secondary spiral appears to break apart towards disk interior upon encountering the planet-free gap, and quantifying their role in our simulation is thus not straightforward. We thus refrain from further discussion of secondary spirals, and regardless of the onset of the planet-free gap, they are clearly sustained by magnetic flux concentration in our simulations.

\subsection{Structure of planet-induced density gap}
\label{sec:gap_structure}

\begin{figure*}
    \centering
    \includegraphics[width=\hsize]{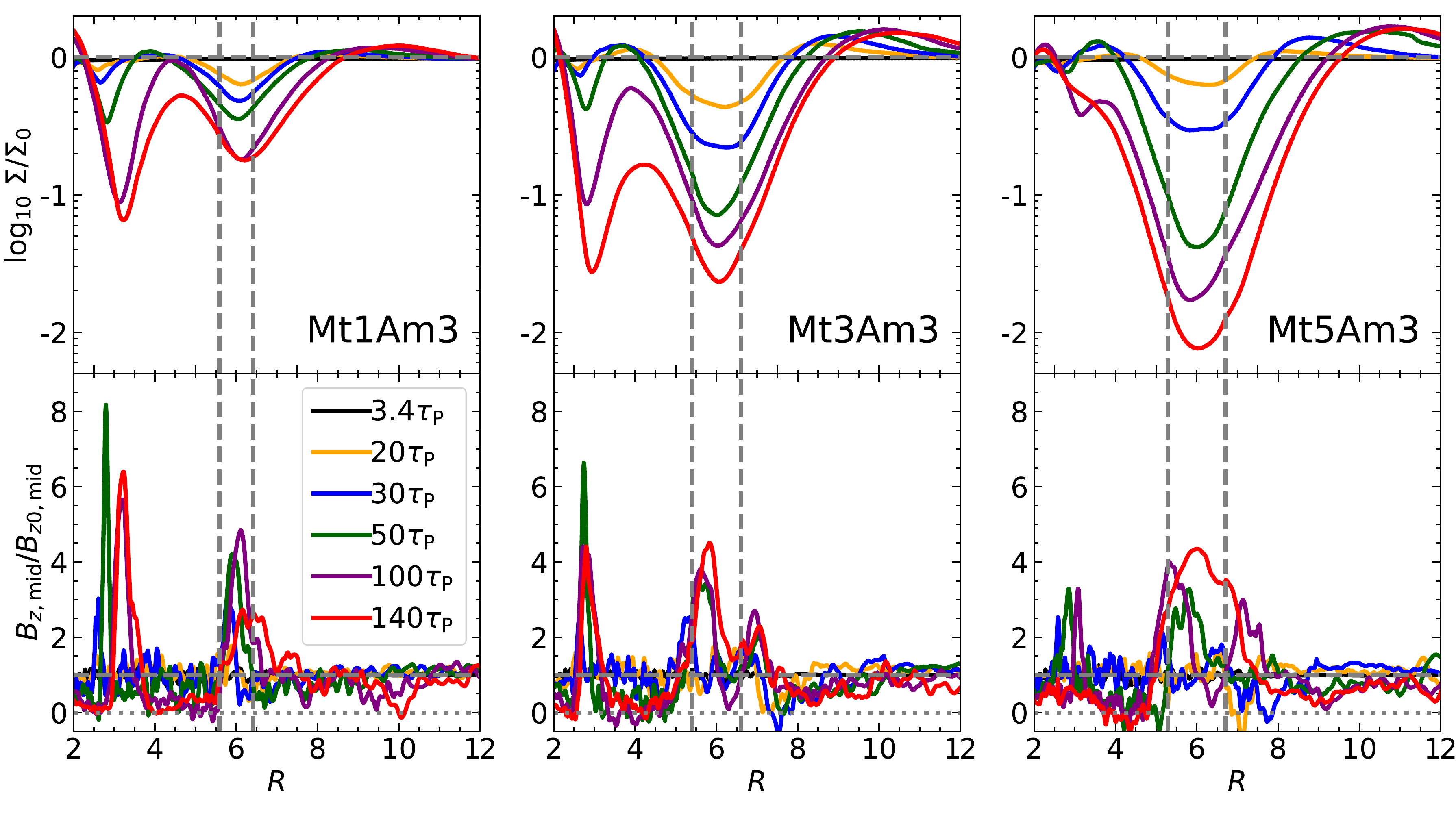}
    \caption{Normalized radial profiles of surface density (upper) and vertical magnetic field within $|z|<H$ (lower), at $t=3.4$ (black), 20 (orange), 30 (blue), 50 (green), 100 (purple), and $140~\tauP$ (red), respectively. From left to right, each panel corresponds to run Mt1Am3, Mt3Am3, and Mt5Am3, respectively. The vertical grey dashed lines mark $R=\RP\pm\rH$, and the horizontal dashed and dotted lines correspond to $\Sigma=\Sigma_0$, $\Bzm=\Bmi$, and $\Bzm=0$, respectively.
    }
    \label{fig:GapProf}
\end{figure*}

In this subsection, we study the structure of the planet-induced density gap in detail. Firstly, we show in the top panels of figure~\ref{fig:GapProf} the surface density profiles of the three runs Mt1Am3, Mt3Am3, and Mt5Am3 at different snapshots, normalized by the initial surface density profile. As discussed before, two gaps are developed. The one peaking at $R\sim\RP$ is induced by planet, and the other one at $R\sim3$ is induced by magnetic flux concentration. As the two gaps are relatively close, they get overlapped with time, and in the case of run Mt5Am3, they get fully merged at $t=140~\tauP$.

\begin{figure}
    \centering
    \includegraphics[width=8.55cm]{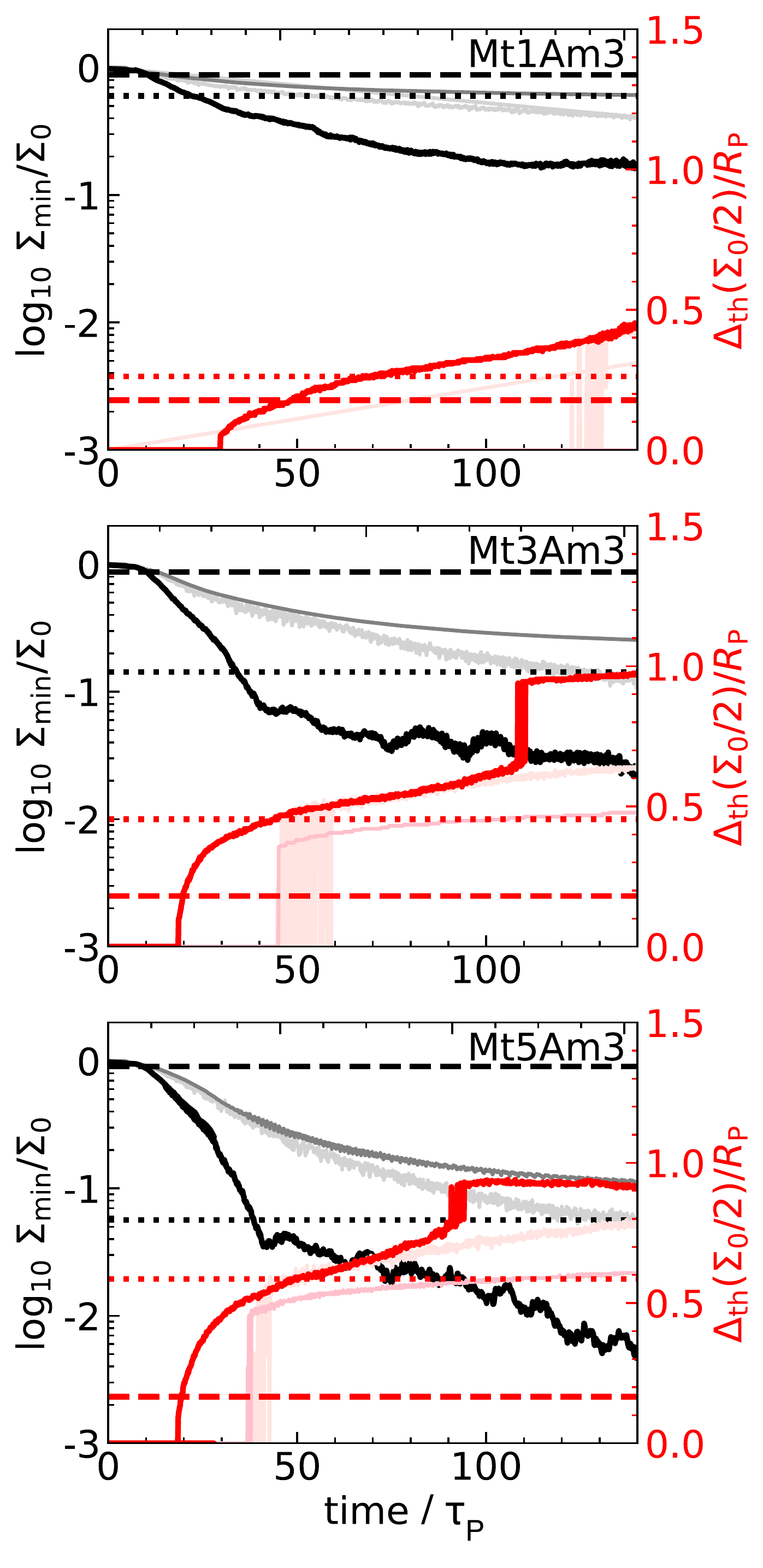}
    \caption{Temporal evolution of the minimum surface density (solid black) and full width at $\Sigma=\Sigma_0/2$ (solid red) of the gap in runs Mt1Am3, Mt3Am3, and Mt5Am3 (from left to right). Solid lines with medium and light colors correspond to the viscous (Mt3\al6) and inviscid case (Mt3\al0), respectively.
    The horizontal lines show the predicted gap structure in viscous accretion disk \citep{Kanagawa+2016} for 
    $\alpha=6\times10^{-3}$ (dotted, approximately the planet-free value) and the measured $\alpha$ at $|R-\RP| < H$ and $t=140~\tauP$ (dashed, $\alpha=0.03$, 0.2, and 0.9 for Mt1Am3, Mt3Am3, and Mt5Am3).
    The gap is much deeper and modestly wider than that under the viscous accretion disk.
    }
    \label{fig:gap_time}
  \end{figure}

To quantify the gap properties, we define the gap depth as the ratio of the local minimum to the initial surface density profile, and gap width being the full width at half of the initial surface density, normalized by the planet orbital radius. Figure~\ref{fig:gap_time} shows the temporal evolution of the depth (black solid) and width (red solid) of the planet-induced gap in the three runs.
The gap width gradually get wider with time but shows a jump at $t\sim110$ and $\sim 90 \tauP$ for run Mt3Am3 and Mt5Am3, respectively. It corresponds to when the $\Sigma$ between the planet-induced and planet-free gaps gets smaller than $\Sigma_0/2$. After the jump, the width in Mt3Am3 almost stays constant while the radial gap structure of $\Sigma$ is still evolving. On the contrary, the width in Mt5Am3 gets slightly narrower due to the smoothing after the two gaps get fully merged (see the purple and red lines in figure~\ref{fig:GapProf}).
In run Mt1Am3, the jump has not yet occurred, but $\Sigma$ between two gaps is getting lower towards the end of the simulation.

The width and depth of planet-induced gaps have been widely studied in previous works \citep[e.g.,][]{Crida+2006,Duffell+MacFadyen2013,Fung+2014,Duffell2015,Kanagawa+2015,Kanagawa+2016,Ginzburg+Sari2018,Duffell2020}, almost exclusively under the framework of viscous disks. To compare our results from these studies, in figure~\ref{fig:gap_time}, we overplot the depth (grey) and width (pink) in the viscous disks. 
Hence, for comparison, we also compare our results with the semi-analytical fit for the gap depth and width in $\alpha$-viscous disks by \citet{Kanagawa+2016}
\begin{align}
    \frac{\Sigma_\mathrm{min}}{\Sigma_0} &= \frac{1}{1+0.04K}, \label{eq:Kanagawa_SDm}\\
    \frac{\Delta_\mathrm{gap}(\Sigma_\mathrm{th})}{\RP} &= \left( \frac{\Sigma_\mathrm{th}}{2\Sigma_0} + 0.16 \right) K'^{1/4},\label{eq:Kanagawa_W}
\end{align}
where
\begin{align}
    K &= \left( \frac{M_\mathrm{P}}{M_*} \right)^2 h_\mathrm{P}^{-5} \alpha^{-1}, \label{eq:Kanagawa_K}
    \\
    K' &= \left( \frac{M_\mathrm{P}}{M_*} \right)^2 h_\mathrm{P}^{-3} \alpha^{-1},\label{eq:Kanagawa_Kd}
\end{align}
$\Sigma_\mathrm{th}=\Sigma_0/2$ is the threshold surface density,
and $h_\mathrm{P}$ is the disk aspect ratio at the planetary orbit. 
In our runs, the turbulent viscosity is effectively evaluated as $\alpha=6\times10^{-3}$ far from the plane ($R\sim 20$; see \S~\ref{sec:alpha}).
By substituting this $\alpha$ and $h_\mathrm{P}=0.1$ into Eqs.~(\ref{eq:Kanagawa_SDm})-(\ref{eq:Kanagawa_Kd}), the gap depth and width in the viscous accretion disk are predicted to be $(\Sigma_\mathrm{min}/\Sigma_0, \Delta_\mathrm{gap}(\Sigma_0/2)/\RP)=(6.0\times10^{-1}, 0.26)$, $(1.4\times10^{-1}, 0.45)$, and $(5.6\times10^{-2}, 0.59)$ for $\MP=1$, 3, and $5\Mth$, respectively. They are shown as the dotted lines in figure~\ref{fig:gap_time}.
At the end of our simulations, the gap profile is wider and deeper than the prediction of Eqs.~(\ref{eq:Kanagawa_SDm}) and (\ref{eq:Kanagawa_W}) by factors of (4, 1.7), (6, 1.5), and (10, 1.3) for the three runs (here the width are measured just before the jump for the latter two runs Mt3Am3 and Mt5Am3 for fair comparison).
Note that our viscous simulations of Mt1\al6, Mt3\al6, and Mt5\al6 have not yet reach to the full steady state by the end of simulation and hence the depth is shallower than the prediction.

As we shall see in Section~\ref{sec:alpha}, the $\alpha$ values in our simulations are not spatially constant, and become much higher in the gap region due to magnetic flux concentration. If we adopt $\alpha$ in the gap region (averaged within $|R-\RP|< \rH$) for runs Mt1Am3, Mt3Am3, and Mt5AM, their values measured  at $t=140~\tauP$ are found to be $3\times10^{-2}$, $2\times10^{-1}$, and $9\times10^{-1}$, respectively.
Applying the same semi-analytical formula, we find the predicted gaps would be much shallower and narrower (thus deviate much more substantially from simulation results), as shown in dashed lines in figure~\ref{fig:gap_time}.

In addition, we compare the gap depth and width with those in the inviscid disk (pale grey and pale pink in figure~\ref{fig:gap_time}).
Even comparing with the inviscid disk, the gap depth is deeper by a factor of 3, 5, and 10, respectively. On the contrary, interestingly, the gap width evolves similarly in the wind and inviscid disks, at least before the jump due to gap merging occurs.

Overall, we conclude that for planet mass $\gtrsim$ thermal mass, the planet-induced gap in the MHD-wind-driven disk is much deeper and modestly wider than that in a viscous disk.
Rather, in terms of the gap width, the windy disk is more similar to the inviscid disk, while the gap depth is still much deeper in the windy disk.
This result is due to a combination of the nature of wind-driven accretion and magnetic flux concentration, as will be detailed in Section~\ref{sec:GOM}.

\subsection{Magnetic flux concentration}
\label{sec:MFC}
\begin{figure*}
    \centering
    \includegraphics[width=\hsize]{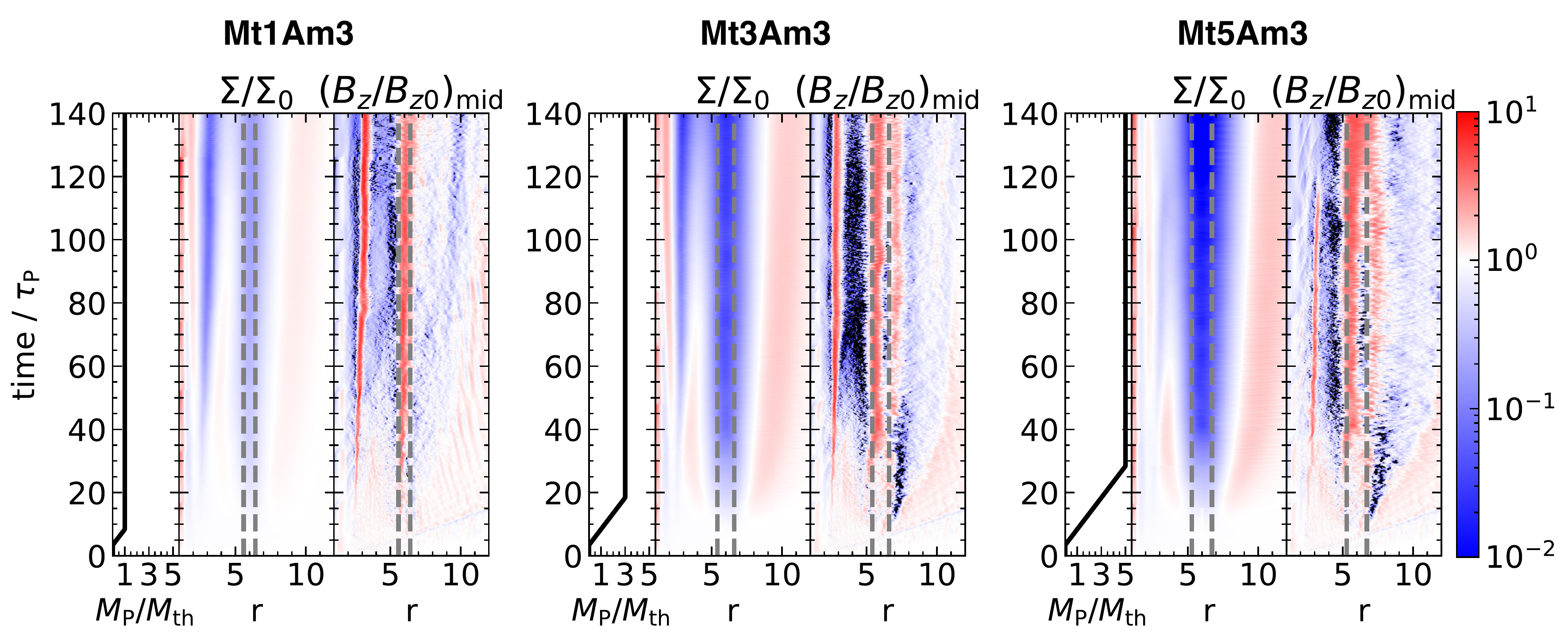}
    \caption{Illustration of the time evolution for runs Mt1Am3 (left), Mt3Am3 (middle), and Mt5Am3 (right). Each of the three panels show the temporal evolution of planetary mass (left), surface density $\Sigma$ (center), and mean vertical magnetic field at the midplane $\Bzm$ (right), where $\Sigma$ and $\Bzm$ are normalized to initial values.
    The planet is located at $r=6$, and the gray dashed line shows $r=\RP\pm\rH$. The black color corresponds to negative sign.
    }
    \label{fig:trSB}
\end{figure*}

We have already mentioned that our simulations form a planet-free gap due to magnetic flux concentration at $R\sim3$, together with the planet-induced gap, which also shows magnetic flux concentration. To better comprehend how these structures come into being, we show in figure~\ref{fig:trSB} the temporal evolution of the disk surface density $\Sigma$ and mean vertical magnetic field at the midplane $\Bzm$, for all three runs with $Am=3$.

We start by discussing the gap around the planet orbit. Magnetic flux concentration in the planet gap is one of the most significant new findings in our study. 
It is the main reason that leads to the formation of wider and deeper gaps, as we analyze in more detail in Section \ref{sec:GOM}. How magnetic flux spontaneously concentrates to the gap at first place, on the other hand, is a much more complex question, which generally involves the complex interplay between the gas dynamics (leading to flux advection), and non-ideal MHD and (turbulent) dissipative processes (leading to flux diffusion), yet both of which are dependent upon the magnetic flux configuration itself that drives wind and turbulence. Despite numerous attempts \citep[e.g.,][]{Lubow+1994,Guilet+Ogilvie2013,Okuzumi+2014,Riols+2020a}, most of these works rely on phenomenological assumptions of advection/diffusion that lack dynamical feedback, while simulation results also appear to depend on implementation details \citep[e.g.,][]{Bai+Stone2017,Lesur2021}. Clearly, flux concentration in planet-induced gap reflects the back-reaction of gravitational influence of the planet on magnetic flux transport, which further complicates the picture.
Here, we focus our discussion mainly at phenomenological level. Further analysis will be presented in Appendix~\ref{sec:M_MFC}, where we show that flux concentration in planet-induced gaps firstly occurs at high altitudes and is likely associated with vertical flows above the planet Hill sphere.

\begin{figure}
    \centering
    \includegraphics[width=8.55cm]{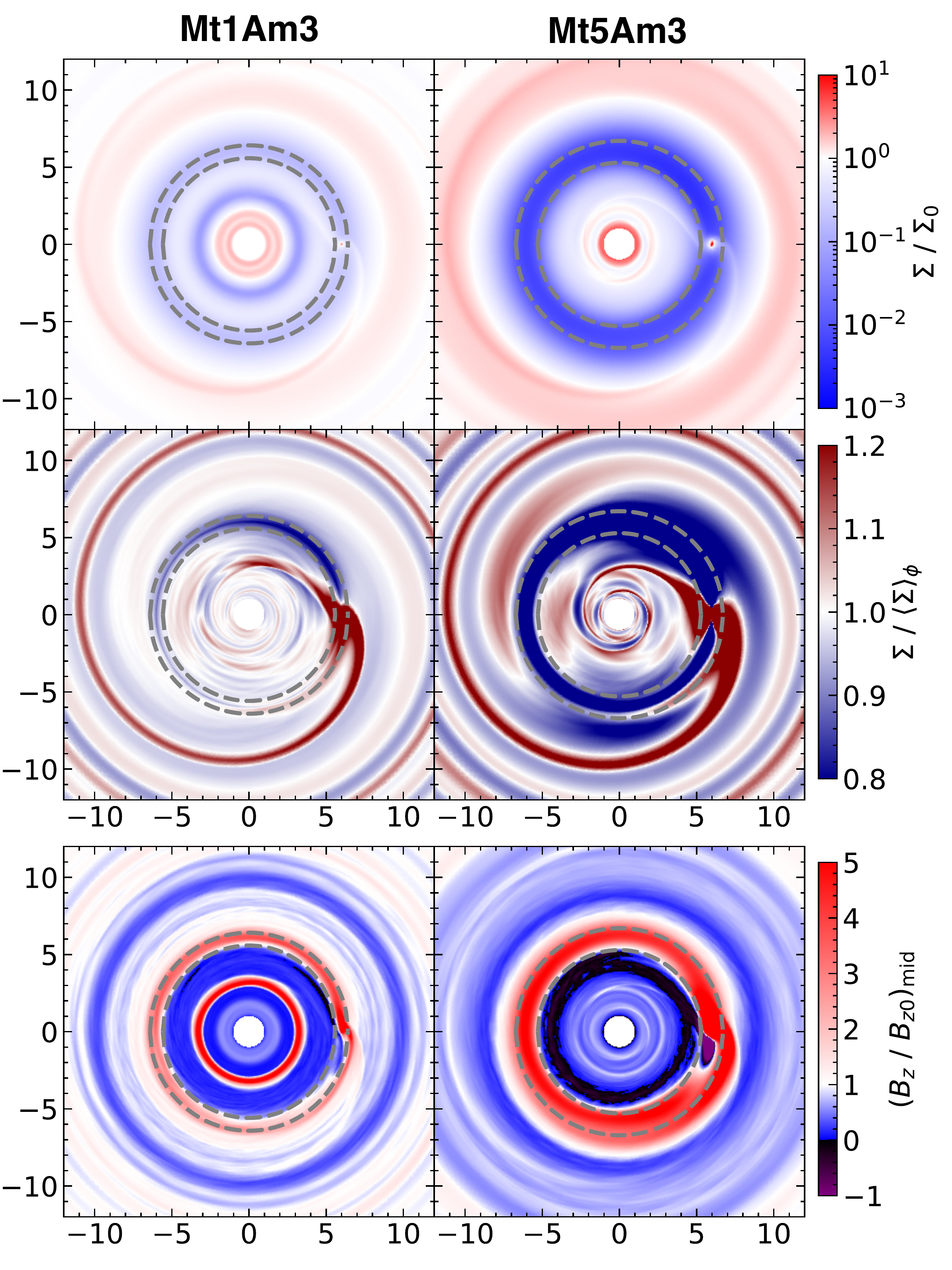}
    \caption{Same as figure~\ref{fig:Rphi} but at $t=140~\tauP$ for Mt1Am3 (left) and Mt5Am3 (right), respectively.
    }
    \label{fig:Rphi_15}
\end{figure}

There are several noticeable features in magnetic flux concentration around the planet gap. First, this process requires the presence of net poloidal flux. By contrast, earlier simulations with toroidal magnetic flux (but zero net poloidal flux) showed that toroidal magnetic flux gets expelled from the gap region \citep{Nelson+Papaloizou2003,Winters+2003b,Papaloizou+2004}. 
Second, the radial profile of $\Bzm$ is not necessarily centrally peaked, but can exhibit double peaks at higher planet mass, as seen in figures~\ref{fig:GapProf} and ~\ref{fig:trSB}. We note that similar findings were reported in local simulations by \citet{Carballido+2016} despite of being in the ideal MHD regime without stratification. 
This double-peak structure is related to the asymmetric distribution of $B_z$ in the $R-\phi$ plane as seen in figure \ref{fig:Rphi} for run Mt3Am3, as well as shown in figure \ref{fig:Rphi_15} for the other two runs.
We also find that the double-peak profile merge into a single peak at higher altitudes, as can also be seen in figures~\ref{fig:3D} and \ref{fig:rzslice}.
Third, the level of magnetic flux concentration is relatively strong, with $\Bzm$ reaching a factor of $\sim2\text{--}4$ the background level within the $R_H$ from the planetary orbit. This requires gathering magnetic flux from a wide range of neighboring regions, especially as the corotation region (and the gap) becomes wider for more massive planets.

The planet-free magnetic flux concentration has been reported in various MHD simulations of PPDs without planet \citep[e.g.,][]{Bai+Stone2014, Bai2015, Bethune+2017, Suriano+2018, Suriano+2019, Riols+Lesur2019,Riols+2020a,Cui+Bai2021}.
In the presence of the MRI turbulence, the location and evolution of magnetic flux concentration is stochastic and generally unpredictable.
We clarify that the location of the planet-free gap in our three fiducial runs are largely the same because they are restarted from the same initial condition. The specific location of $R\approx3$ is likely the outcome of the stochasticity from our initial condition.\footnote{We cannot rule out that the planet influences the formation of the planet-free gap due to secondary spirals \citep{Bae+2017} and subsequent magnetic flux concentration, especially as the planet is inserted relatively early. On the other hand, the $R=3$ gap is not present in our simulation with $Am=1$ presented in Appendix \ref{sec:Am1}, where the planet is inserted at much later time (and note that the ``gap" further inward is too close to the inner boundary to be trustworthy).} Also note that there are additional flux concentration beyond the radial range of figure~\ref{fig:trSB} at larger radii.
The planet-free $\Bzm$ concentration moves slightly outward in the Mt1Am3 and Mt5Am3 cases, but stays at the initial location in the Mt3Am3 case. We anticipate such migration to be less predictable.
On the other hand, the amount of magnetic flux trapped at $R\sim3$ appears to anti-correlate with the planet mass. Being in the vicinity of the planet orbit, reflecting that more flux gets ``attracted" to the more prominent planet gap for more massive planets.

We further look at the spatial distribution of the magnetic flux in figures~\ref{fig:Rphi} and \ref{fig:Rphi_15}, in the $R-\phi$ plane. 
In the planet-free gap, the distribution of $\Bzm$ is largely axisymmetric. However, in the planet gap, the distribution is clearly asymmetric, and magnetic flux concentration traces the spiral density shocks in the leading and trailing side in the vicinity of the planet.
The positive concentration of $\Bzm$ has two arms starting from the planet; the inner (outer) arm azimuthally extends within (to the outer side) the corotation region ($|R-\RP|<\rH$) towards the leading (tailing) side. On the other hand, the tailing side of the gap has a localized region with negative $\Bzm$.
As a result, the two separated $\Bzm$ peaks in the radial profile is a continuous distribution in the $r$-$\phi$ plane.
We also observe that the double peaks in run Mt5Am3 turn to a single peak at $t\gtrsim120~\tauP$ as the gap merges with the planet-free gap and gather more flux from there.

\subsection{Turbulence and stress}
\label{sec:alpha}
\begin{figure*}
    \centering
    \includegraphics[width=\Ft]{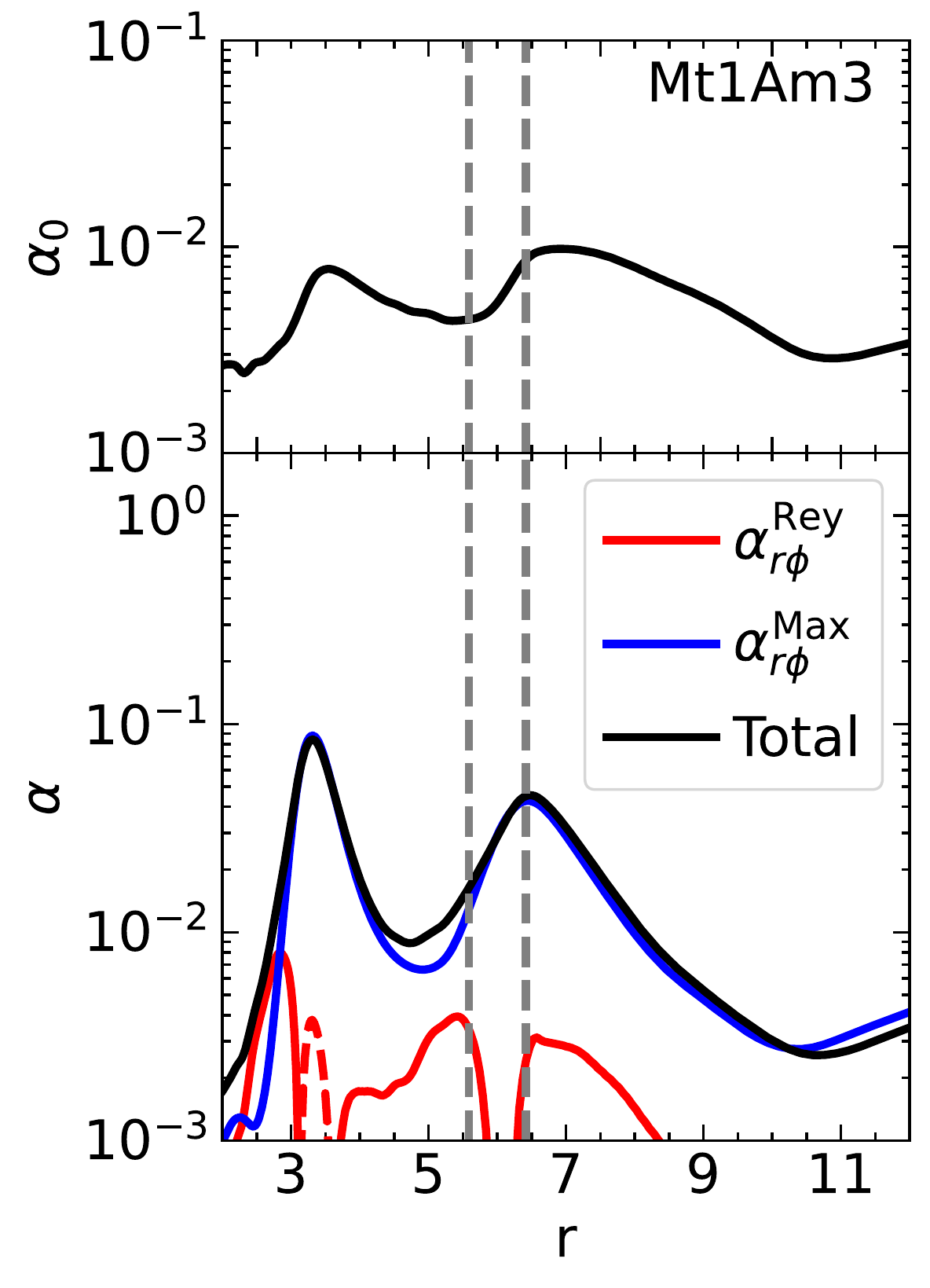}
    \includegraphics[width=\Ft]{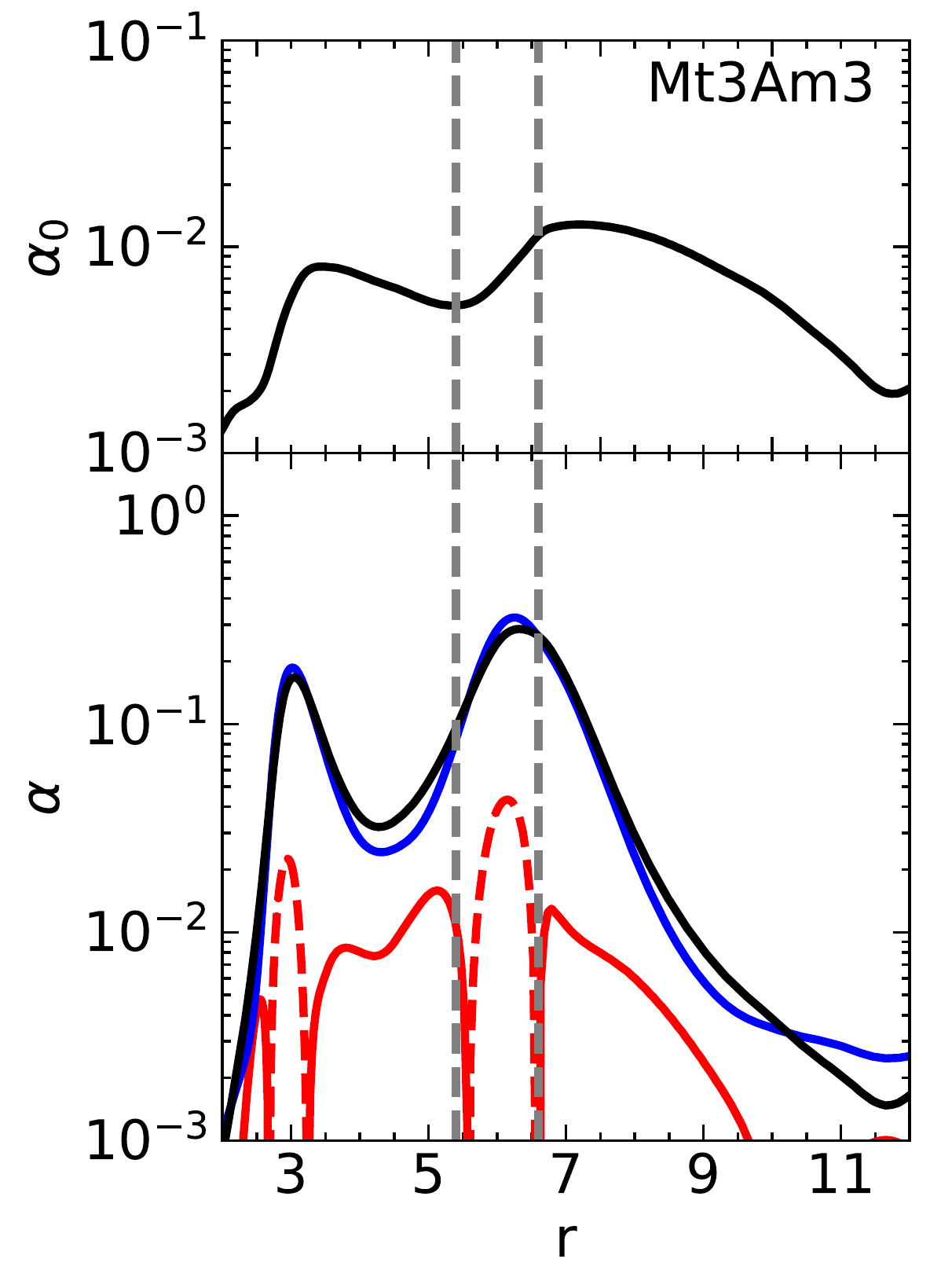}
    \includegraphics[width=\Ft]{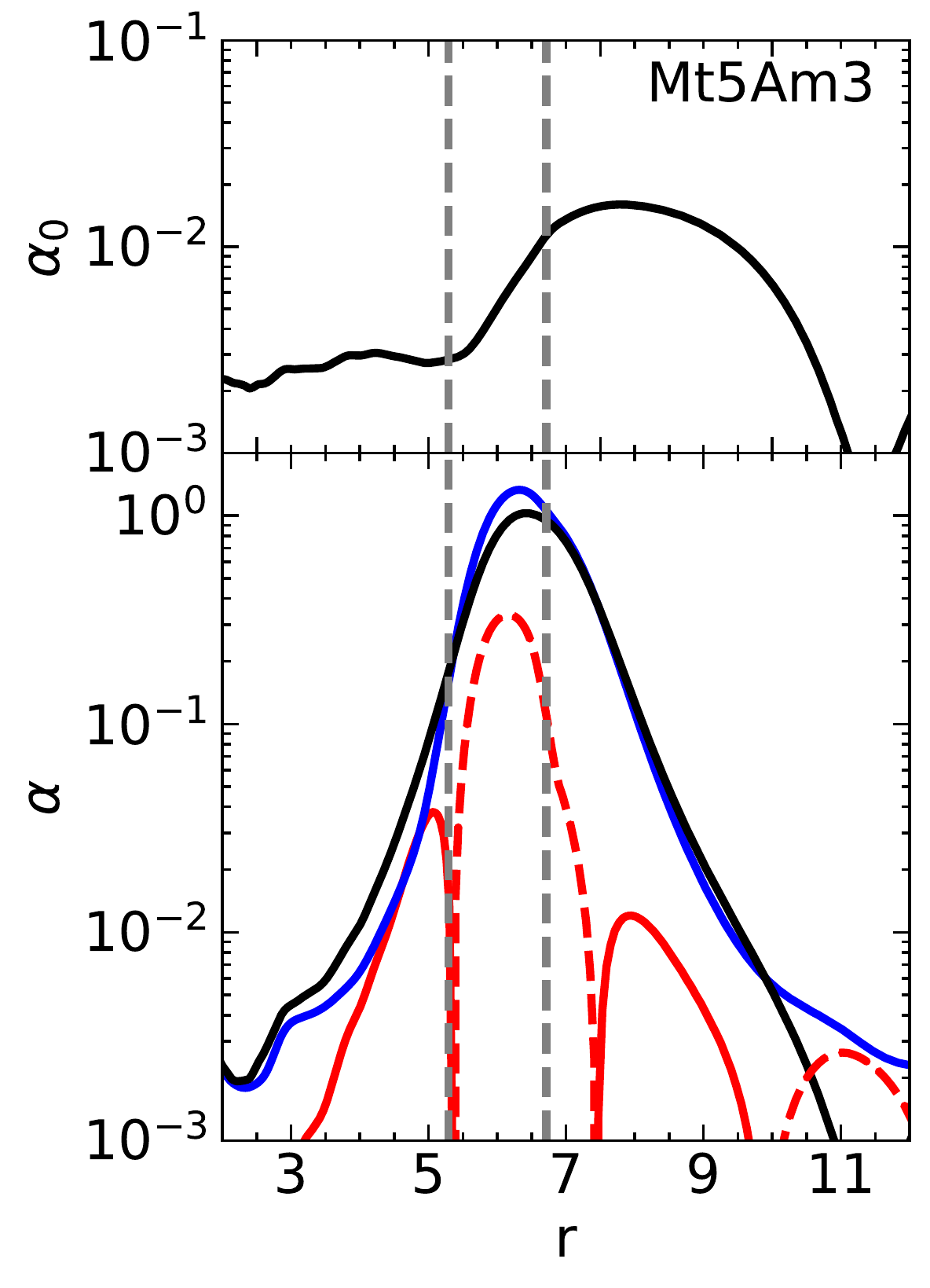}
    \caption{Bottom: the radial profiles of the total $\alpha$ (black) and the $\alpha$ caused by Reynolds stress (red) and Maxwell stress (blue), for the case of Mt1Am3 (left), Mt3Am3 (center), and Mt5Am3 (right), respectively. The solid and dashed lines corresponds to the positive and negative values, respectively.
    Top: total stress normalized by the initial pressure, $\alpha_0$.
    The time average is taken over $130<t/\tauP<140$. 
    The grey dashed lines mark $r=\RP\pm\rH$.
    }
    \label{fig:alp}
\end{figure*}

It is well known that the strength of the MRI turbulence, together with the Maxwell and Reynolds stresses, directly correlates with the amount of net vertical magnetic flux threading the disk \citep[e.g.,][]{Hawley+1995, Bai+Stone2011}. The spontaneous flux concentration thus implies that the level of turbulence in the disk is inhomogeneous.

Figure \ref{fig:alp} shows the radial distribution of the stresses for the three runs with $Am=3$.
As the $\alpha$ values correspond to stresses normalized by pressure, radial variations in surface density (and hence pressure) yields additional variations in $\alpha$ values. To facilitates the analysis, we show the $\alpha$ values normalized by both the {\it initial} pressure profile (top) and the concurrent pressure profile (bottom).
Without the planet, and/or sufficiently far away from the planet, we see that $\alphaMx$ is on average greater than $\alphaRe$ by a factor of several, and we quote $\alphaMx \sim 6\times 10^{-3}$ for $\AM=3$ as a reference value. Below, we focus on the influence of the planet on the $\alpha$ profiles.

The radial profile of $\alpha^{\rm Max}$ (and hence $\alpha$) largely reflects magnetic flux concentration, which peaks at the radii where magnetic flux concentrates at $R=\RP$ as well as in the planet-free gaps. With both density depletion and magnetic flux concentration, effective magnetization is greatly enhanced in the gap region, leading to a very large $\alpha\gtrsim0.1$. This value increases with increasing planet mass, reflecting deeper density gap.
Similar behavior was found in previous unstratified sharing-box simulations with net poloidal magnetic flux
\citep{Zhu+2013,Carballido+2016,Carballido+2017}. These works also found that the radial profile of the Maxwell stress (not normalized by pressure) is largely flat across the gap.
In other words, the larger $\alphaMx$ is mainly due to the lower density within the gap.
In our simulation, as shown in the upper panels of figure~\ref{fig:alp}, the Maxwell stress in fact increases from the inner to the outer sides of the gap. This is mainly because poloidal fields are inclined radially outwards at higher altitudes accompanying wind launching, which greatly enhances $\alphaMx$ at outer radii. This fact will be further discussed in Section \ref{sec:AMT_MHD}.
On the other hand, we observe that for the Reynolds stress, $\alphaRe$ becomes negative around the planet orbit, as well as in the planet-free gap, which can be partly attributed to the strong accretion flow (see Section \ref{sec:Flow_AF_DR}) in the midplane region in addition to the density waves/shocks.\footnote{The Reynolds stress has major contributions from the density wave/shock exhibiting strong azimuthal variations. We find that another major contribution arises from the highly non-uniform distribution of the accretion flow concentrated in the midplane region with large accretion velocities (see Section \ref{sec:Flow_AF_DR}) that yields a negative $\delta(\rho v_r)$. Given our definition of the Reynolds stress (see Equation \ref{eq:Rey}), the midplane region also tends to share a small excess of (positive) $\delta j$ thanks to the $\sin\theta$ factor in $j$. The combined effect thus promotes a negative Reynolds stress in the midplane.}

\begin{figure}
    \centering
    \includegraphics[width=8.55cm]{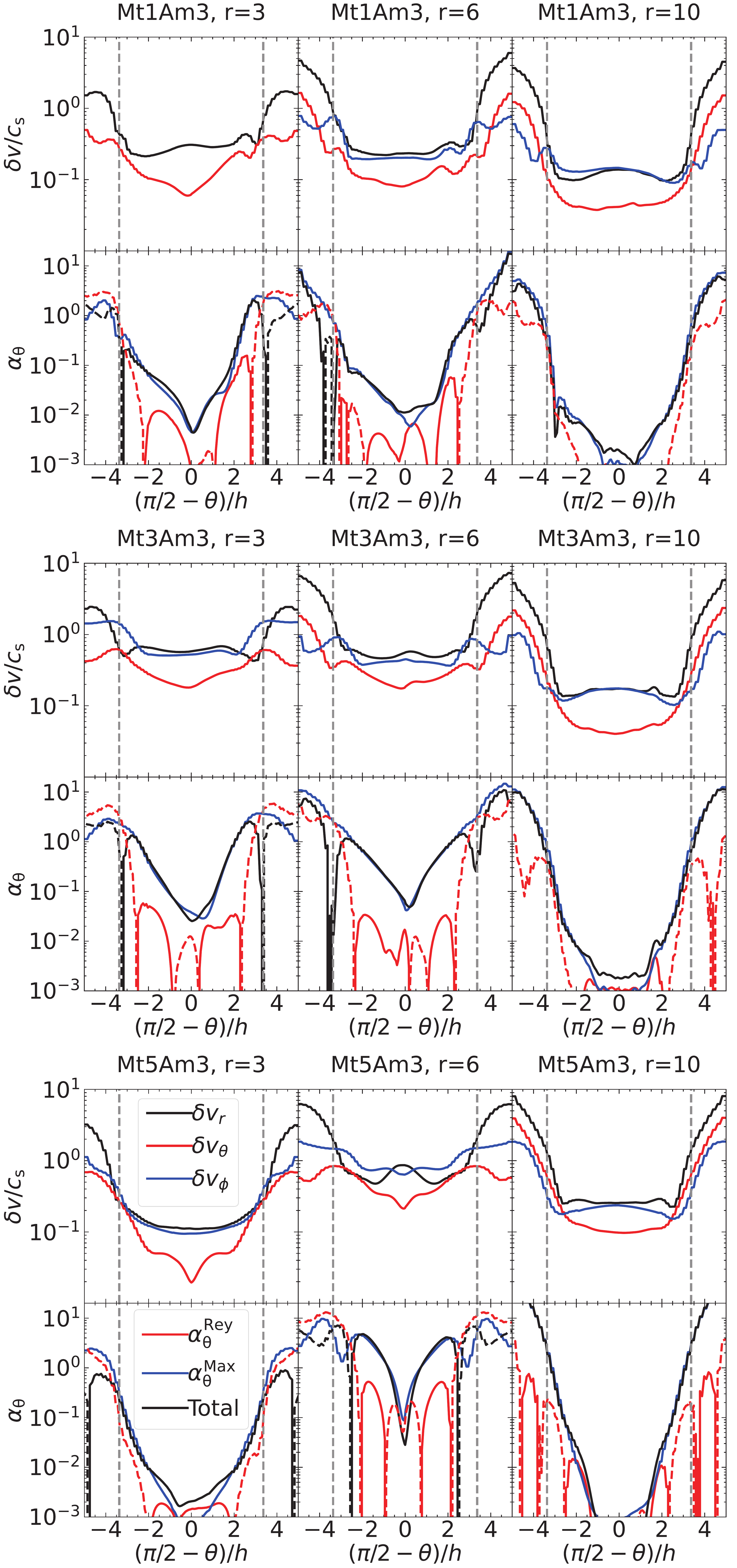}
    \caption{
    Velocity fluctuation ($\delta v$, upper) and the $r$-$\phi$ component of the stress tensor normalized by the local pressure ($\alpha_\theta$, lower) for runs Mt1Am3 (top), Mt3Am3 (middle) and Mt5Am3 (bottom) at $r=3$ (left), $6$ (center) and 10 (right). The solid and dashed correspond to the positive and negative values, respectively. The velocity mean $\av{v}$ and velocity fluctuation $\delta \bvec{v} \equiv \sqrt{ (\bvec{v}-\av{\bvec{v}})^2 }$ are averaged over the fiducial range, i.e., $130<t/\tauP<140$ in time, $\pm 0.25~H$ in $r$, and full $2\pi$ in $\phi$.
    The grey dashed lines mark the disk surface $|z|=3.5H$.
    }
    \label{fig:dvalpha}
\end{figure}

To further look into turbulence and stresses, we show in figure~\ref{fig:dvalpha} the vertical profiles of the velocity fluctuations (upper) and $\alpha_\theta \equiv T_{r\phi}/P$ (lower) at $r=3$ (left), $6$ (center) and $10$ (right), for all three runs with $Am=3$, where the location $r=10$ is chosen to be well outside of the planet-induced density gap.
There are several noticeable features. First, the value of $\alphaMx_\theta$ increases from the midplane to disk surface, consistent with standard expectations (e.g., \citealp{Simon+2013b,Bai2015}), and it largely dominates the total stress at all heights and both radial locations. Second, the vertical profile of the Reynolds stress can be oscillatory, with amplitudes comparable to the Maxwell stress. This can be related to the periodic passing of the density shocks, though its net contribution to angular momentum transport is reduced through vertical averaging. 
At highest altitude, the Reynolds stress is dominated by the wind. Third, there are strong velocity fluctuations. At $r=10$, the level of velocity fluctuations in all cases are similar and are stronger than the planet-free case reported in \citet{Cui+Bai2021} (see their figure~11), by a factor of $\gtrsim3$, and the dominant component is $v_r$. We anticipate that the main contribution is from the spiral shocks rather than the MRI turbulence, given that the physical conditions remains similar to the planet-free case at that radius. The gap region is characterized by even stronger velocity fluctuations, with $\delta v$ increasing towards higher planet mass, and reaching $\sim c_s$ for $M_p\gtrsim3M_t$. In fact, the flow structure in the planet orbital region is very complex, which we address further in the next section.
The planet-free gap at $r=3$ (for $\MP\leq3\Mth$) also has strong velocity fluctuations with a positive correlation to planetary mass, which also indicates that the planet-induced density waves/shocks, including the secondaries, can enhance turbulence in these strongly magnetized planet-free gaps not far from the planet orbit. As the planet-free gap already merges with the planetary gap in the $\MP=5\Mth$ run, the magnetization at $r=3$ is weaker, thus showing weaker velocity fluctuations there.

\section{Flow around the planetary orbit}
\label{sec:flow}

\begin{figure}
    \begin{minipage}[]{8.55cm}
    \centering    
    \includegraphics[width=\hsize]{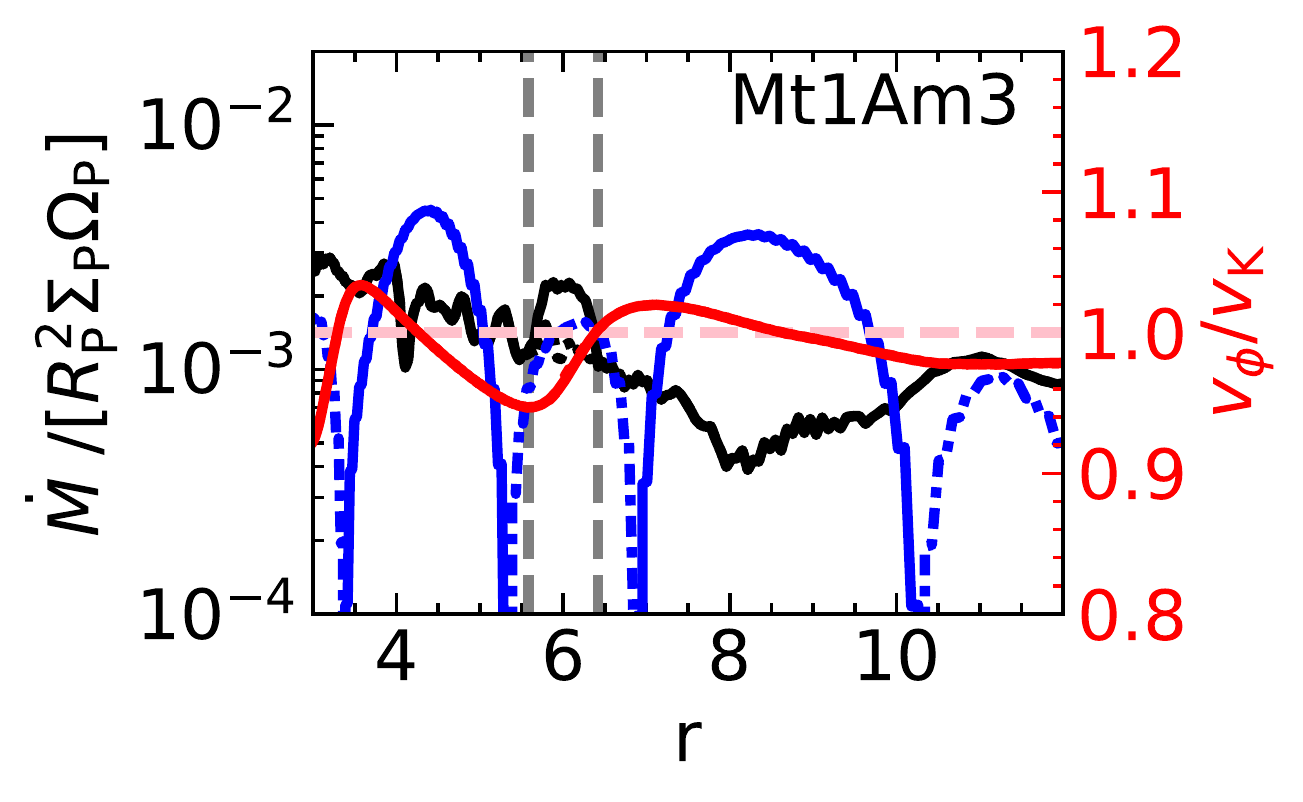}
    \includegraphics[width=\hsize]{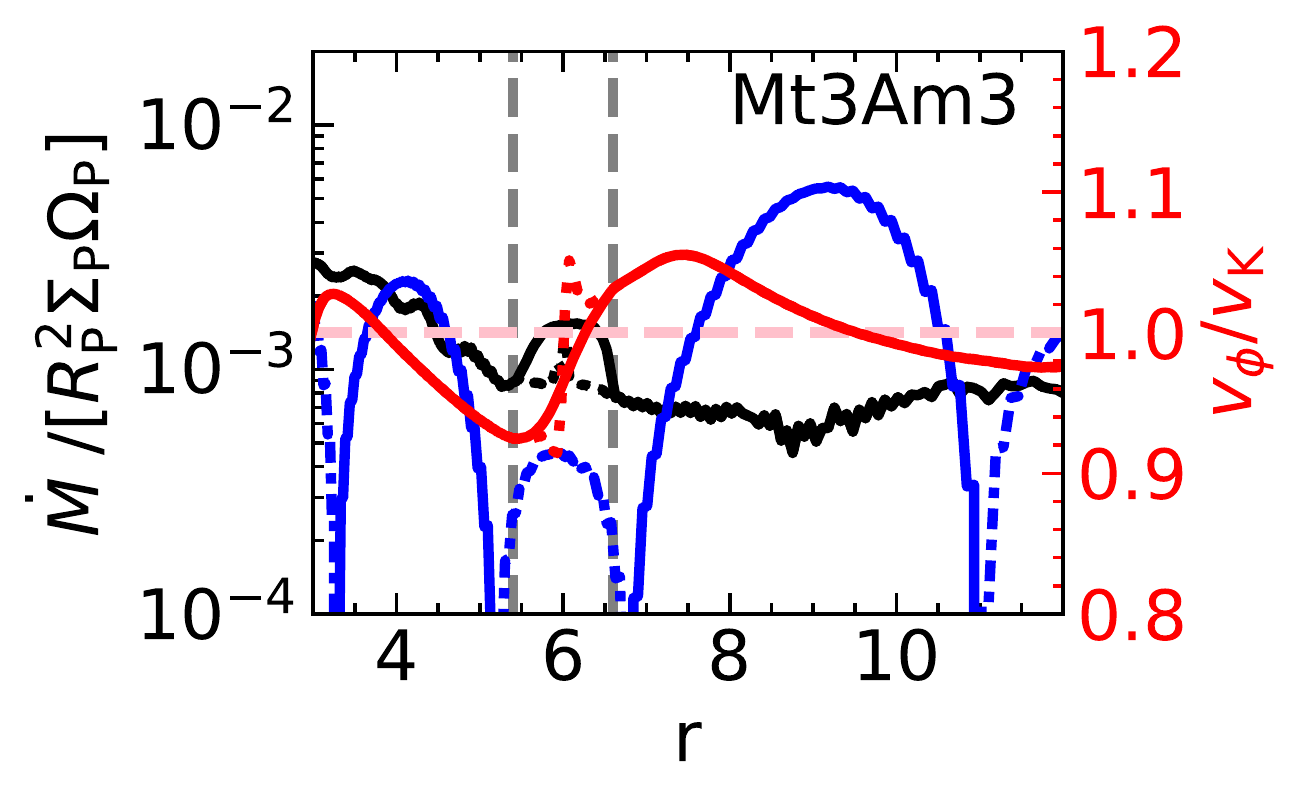}
    \includegraphics[width=\hsize]{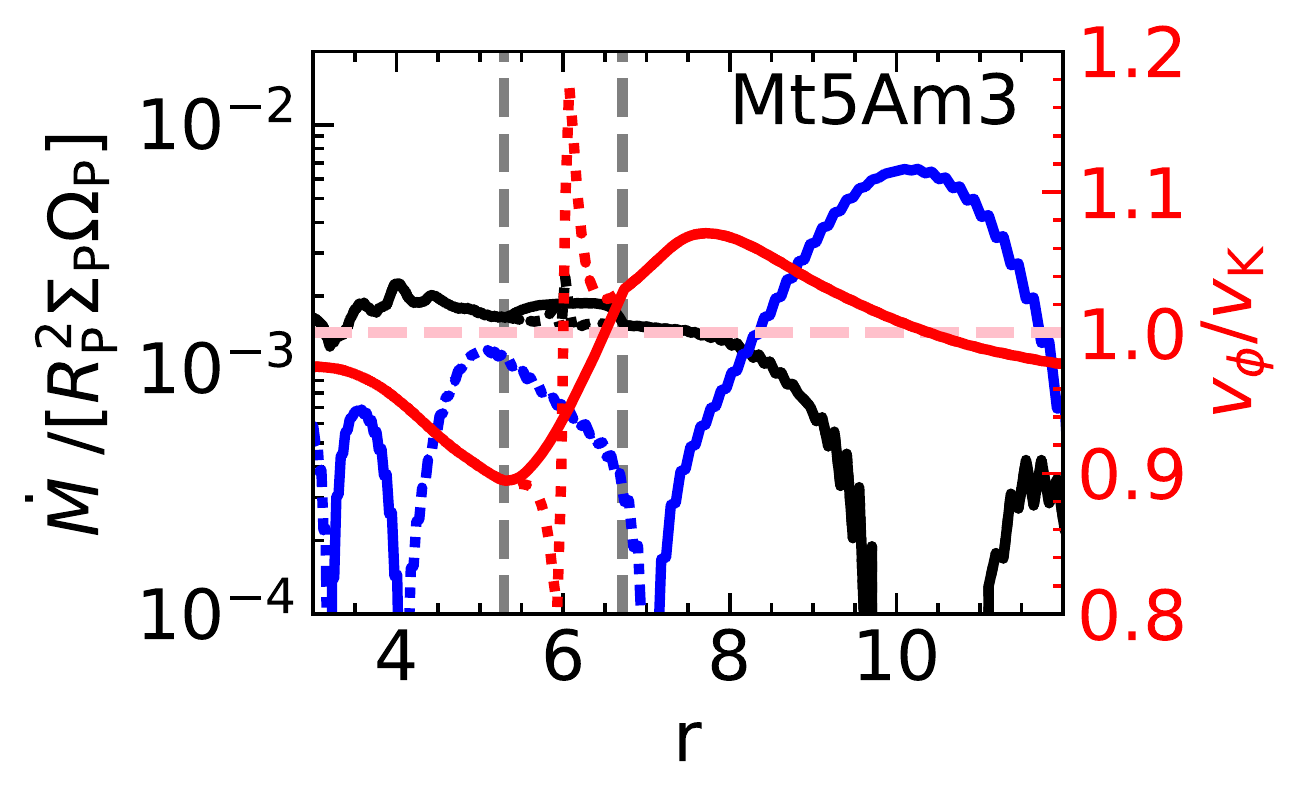}
    \end{minipage}
    \caption{
    Disk mass accretion rate ($\Macc$, black), wind mass-loss rate ($\Delta \dot{M}_\mathrm{wind}$, blue), and density-weighted averaged rotating velocity (red) for runs Mt1Am3 (top), Mt3Am3 (middle), and Mt5Am3 (bottom), respectively, where $\Delta \dMwind= \partial \dMwind/(\partial \ln{r})$. The vertical integration is taken along the spherical $\theta$-direction.
    The black and red dotted line shows the results without excluding the planet Hill sphere ($|\bvec{r}-\bvec{\RP}|<\rH$), and the blue dashed line corresponds to the negative mass-loss, namely, vertical inflow through the disk surface.
    The vertical grey and horizontal pink dashed lines mark $r=\RP\pm\rH$ and $v_\phi=v_K$, respectively.
    }
    \label{fig:Mdot}
\end{figure}

\begin{figure}
    \centering
    \includegraphics[width=8.55cm]{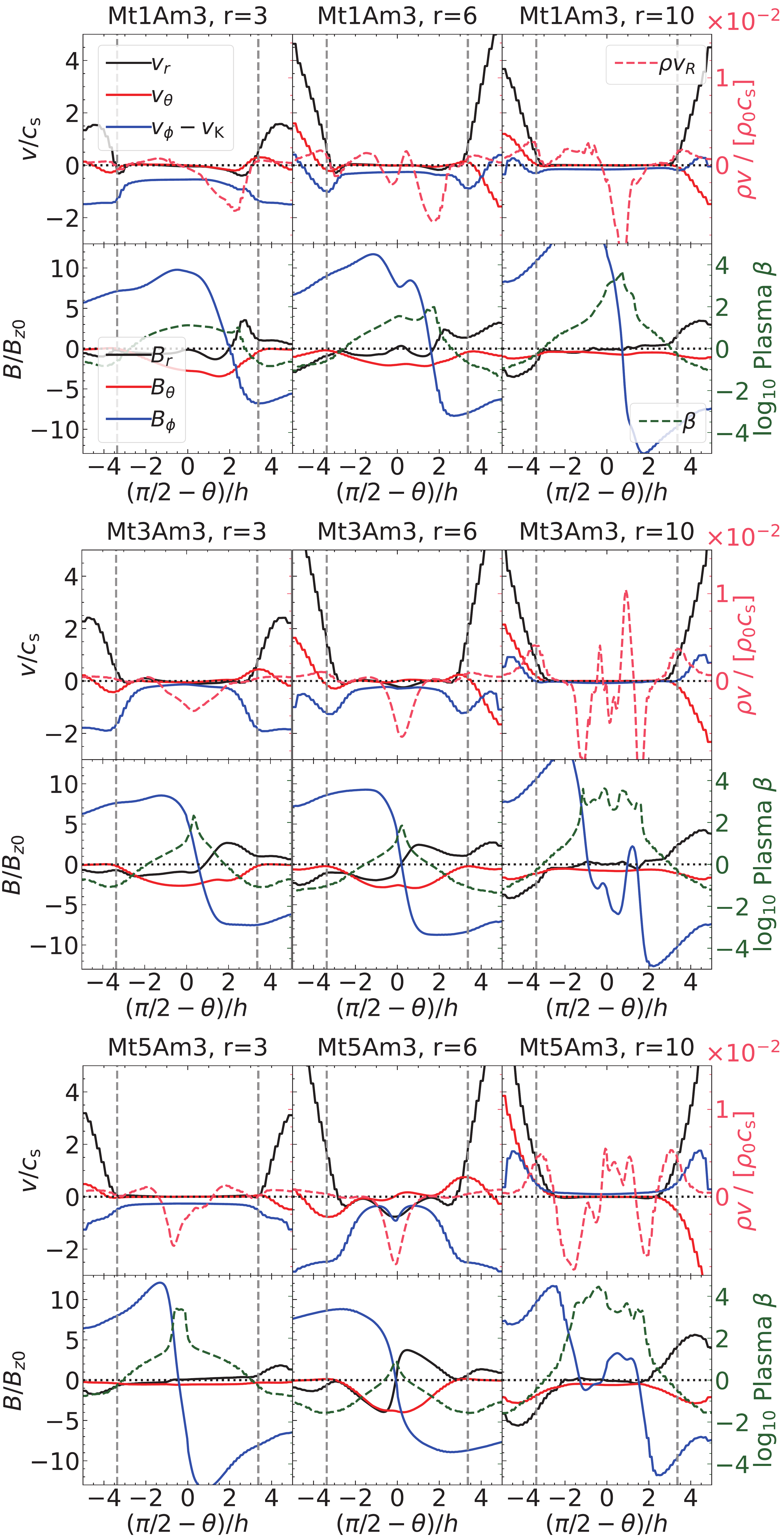}
    \caption{
    Vertical profile of the mean velocity (upper) and mean magnetic field (lower) for runs Mt1Am3 (top), Mt3Am3 (middle), and Mt5Am3 (bottom) at $r=3$ (left), $6$ (center), and $10$ (right). In addition, the pink and green dashed lines correspond to radial momentum and plasma $\beta$, respectively. The vertical grey dashed lines correspond to the disk surface ($z=3.5~H$). 
    }
    \label{fig:vertical_vB}
\end{figure}

There are several aspects of flow structures around the planet orbit, including sub/super-Keplerian rotation, the accretion flow, wind and outflow, as well as the potential for meridional flows around the planet. These structures are also closely connected to magnetic fields, which we discuss in this section. In addition, we discuss vortex formation and evolution due to the Rossby wave instability at the gap outer edges in Appendix~\ref{sec:vortex}.

\subsection{The mean accretion flow and disk rotation}
\label{sec:Flow_AF_DR}
Figure~\ref{fig:Mdot} shows 
the radial profiles of accretion rate and rotational velocity (density weighted over the bulk disk). Without the planet, given the radial density gradient and $h=0.1$, we anticipate $|v_\phi-\vK|$ to be $0.98~\vK$. With the planet, we see that the disk rotating velocity is super- and sub-Keplerian outside and inside the planet, respectively. The deviation from the local Keplerian velocity increases with the planetary mass. The maximum and minimum $v_\phi/\vK$ are measured as $(1.02, 0.95)$, $(1.06,0.92)$, and $(1.07, 0.89)$ for runs Mt1Am3, Mt3Am3, and Mt5Am3, respectively.
We also note the sub-/super-Keplerian modulation due to the planet-free gap. The variation in $v_\phi$ is at a similar amplitude as the planet-induced gap. This suggests that such variations are insufficient to distinguish between gap formation mechanisms.

The accretion rate profile is not flat in our simulations, implying that a steady state is not reached. This is not a major concern for us because firstly, in wind-driven accretion, the accretion rate profile is largely set by magnetic flux distribution (e.g., \citealp{Bai+2016,Lesur2021}), which is not known a priori\footnote{Our initial radial profile of surface density is relatively steep, leading to relatively large amount of magnetic flux in the inner disk, and hence higher accretion rate in the inner than outer disk regions.}. In addition, long-term disk evolution is subject to the not-so-predictable nature of magnetic flux evolution, which we are not in a position to address in this work. Therefore, we mainly focus on the main stage of gap formation, with more detailed analysis in Section \ref{sec:GOM}. We speculate the gap profile may undergo some long-term secular evolution, though this is beyond the scope of this work.

Focusing on the planet orbital region, the mass accretion rate is on the order of $1$--$2\times10^{-3}~[2\pi \RP^2 \SigmaP\tauP^{-1}]$, where $\SigmaP=\Sigma_0(\RP)$, and remains at a similar level as the accretion rates at neighboring radii. With the surface density depleted by a factor of ten to hundreds (see figure~\ref{fig:gap_time}), the mean accretion velocity increases by a similar factor to cancel out density depletion.
While there are accretion rate bumps in the planet orbital region, they are related to the exclusion of the planetary Hill sphere. Further incorporating the Hill sphere indicates roughly constant accretion flow across the planetary gap region (dotted black lines in figure~\ref{fig:Mdot}). Note that without a mass sink at the planet, there is relatively large mass accumulation within the circumplanetary region, giving substantial mass weighting in the measurements, as also reflected in the $v_\phi$ profiles. We will study the dynamics of this region with further mesh refinement and sink prescriptions in a follow up work.

\begin{figure*}
    \centering
    \includegraphics[width=\hsize]{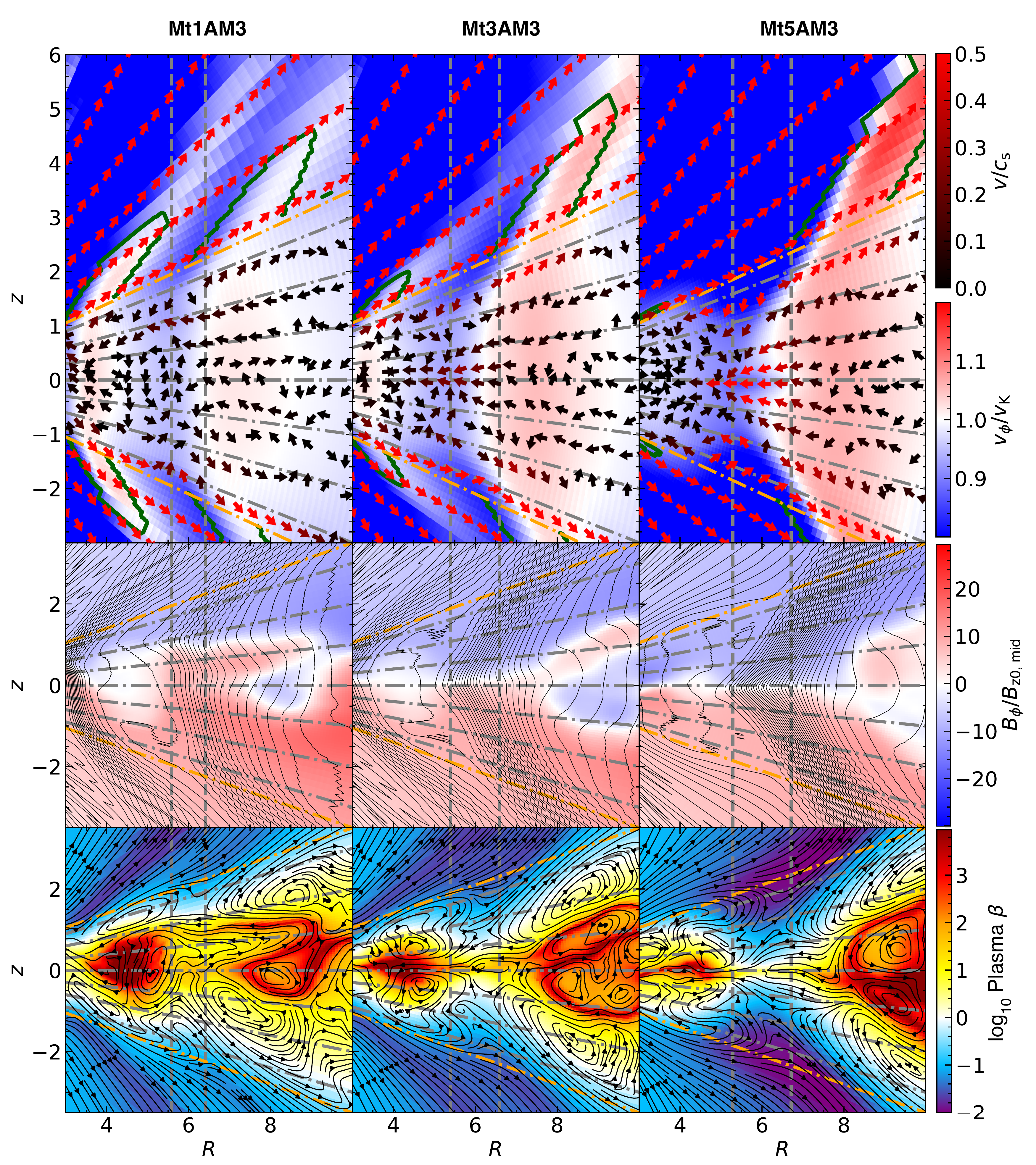}
    \caption{
    The $R-z$ slices of the disk in Mt1Am3, Mt3Am3, and Mt5Am3 from left to right, respectively.
    Top: poloidal velocity (arrow) and disk-rotation velocity (background color).
    Middle: isocontour of the poloidal magnetic flux function $\Phi_\mathrm{B}$ and toroidal magnetic field (background color).
    Bottom: momentum streamline (black lines) and plasma $\beta$ (background color).
    All quantities are averaged in $0<\phi<2\pi$ and $130<t/\tauP<140$. 
    The vertical and horizontal grey dashed lines show $R=\RP\pm\rH$ and $z/H=$-3, -2, -1, 0, 1, 2, 3 from bottom to top, respectively. The orange dash-dotted and green lines correspond to the disk surface ($z/H=3.5$) and \Alf\ surface where the poloidal velocity equals to the poloidal \Alf\ velocity.
    Note that $\vK$ is based on cylindrical coordinates, and $\Bmi$ is the initial vertical field in the midplane.
    }
    \label{fig:rzslice}
\end{figure*}

Next, we look into more details of the global flow structure and its relation to the magnetic field configuration. Figure~\ref{fig:vertical_vB} shows the vertical profiles of mean velocity and magnetic fields at the planetary gap (r=6), at some further distance away from the gap ($r=10$), and at the planet-free gap ($r=3$), and figure~\ref{fig:rzslice} shows the azimuthally-averaged configurations of the flow and magnetic field properties. At the gaps with magnetic flux concentration ($r=3$ and 6), we see that a strong inflow appears at the midplane (run Mt3Am3 and Mt5Am3) or in the disk upper layer (run Mt1Am3) to achieve the mass accretion rate. For wind-driven accretion, the accretion mass flux is directly proportional to the vertical gradient of toroidal field, and is the strongest when $B_\phi$ changes sign (e.g., \citealp{Bai+Stone2013b}). This can be clearly seen by comparing the 2nd and 3rd rows in figure~\ref{fig:rzslice}, as well as in figure~\ref{fig:vertical_vB}. We note that the location where $B_\phi$ flips in the presence of the MRI turbulence is generally stochastic and evolves over time \citep{Cui+Bai2021}, whereas in runs Mt3Am3 and Mt5Am3, the strongly magnetized nature in the gap region (with plasma $\beta\lesssim1$) makes toroidal field to flip in the midplane, confining the accretion flow there (see also \citealp{Martel+Lesur2022}). 
In this case, the accretion velocity becomes on the order of the sound speed.

\subsection{The outflow}\label{sssec:outflow}
The segregation of magnetic flux due to magnetic flux concentration into the planet gap also strongly affect the disk wind properties. It is generally expected that the wind is launched from the disk surface around the height (here $|z|\sim3.5H$) where the flow approaches the ideal MHD regime \citep{Bai+Stone2013b,Gressel+2015}, above which various conservation laws along poloidal field lines apply. As we see from figure~\ref{fig:rzslice}, this is still largely the case, except that the outflow from the planetary gap region can be launched from lower height at $z\sim2$--$3H$. This is likely because of the strongly magnetized nature there, where the Lorentz force starts to dominate at lower heights.

The outflow launched from the planetary gap is asymmetric. As poloidal magnetic fields bend radially outward from the bulk disk, the magnetic flux threading the disk gap region, when approaching the wind base, is already outside of the planet radius. Therefore, most of the wind mass flux arises from the outer gap edge, as can be clearly identified in figure~\ref{fig:rzslice}. This means that mass loss and angular momentum extraction from the wind is strongly concentrated on the outer side of the gap.

Thanks to the asymmetry, the wind from the planetary gap region is launched with super-Keplerian rotation. Together with the strong magnetic flux concentration, the wind efficiency is strongly enhanced \citep{Bai+2016}. We also show the location of the Alfv\'en surface in the top panel of figure~\ref{fig:rzslice}, where the poloidal wind speed matches the poloidal Alfv\'en velocity. Along a poloidal field line, the ratio of the Alfv\'en radii to wind launching radii ($\RA/R_{0w}$, in cylindrical coordinates) characterizes the efficiency of angular momentum transport. In standard form \citep[e.g.,][]{Ferreira+Pelletier1995, Bai+2016}, one expects
\begin{equation}
    \frac{\dd\dot{M}_{\rm wind}/\dd\ln R}{\dot{M}_{\rm acc}}\approx\frac{1}{2}\frac{1}{(\RA/R_{0w})^2-1}\ ,\label{eq:lever}
\end{equation}
where the left shows the ratio of wind mass loss rate (per logarithmic radius) to the wind-driven accretion rate.

Quantitatively analyzing the wind properties based on the above is, however, no longer straightforward in our simulations. This is for several reasons. We first notice from figure~\ref{fig:Mdot} that the local wind mass loss rate, when measured through the mass flux normal to the surface of ``wind base" at $z=0.35H$ (which is along the $\hat{\theta}$ direction), can become negative. This is also evident from figure~\ref{fig:rzslice}, where we see the negative mass flux is mainly in regions where magnetic flux is sparse. It reflects the limitation of analyzing wind in spherical coordinates when the disk is not very thin, as the wind streamlines can be nearly radial, especially in regions with very little magnetic flux. Second, the wind launching radius $R_\mathrm{0w}$ may not be well characterized, especially in the planet gap region due to its strong magnetization as mentioned earlier. Lastly, we also show the location of the Alfv\'en surface in the top panels of figure~\ref{fig:rzslice}. This surface is meaningful mainly in regions above the wind base as it is derived from the conservation laws under ideal MHD. We see that this surface is only well-defined in regions originating from strong magnetic flux concentration. In regions with magnetic flux deficit, this surface is below the wind base and becomes invalid (not shown).

Here, we seek for qualitative understanding of the outflow properties. The most prominent feature is that the wind is mostly launched from regions with magnetic flux concentration. This can be clearly identified from the streamlines in the bottom panel of figure~\ref{fig:rzslice}, and is also accompanied by super-Keplerian $v_\phi$ in the wind zone as expected. This applies to both the planet gap and the planet-free gaps around $R=3$ in runs Mt1Am3 and Mt3Am3. The wind still shows substantial mass loading, as can be seen in the mass loss rate measured slightly outward of the gap region (as field lines bend radially outward). We may still apply Equation (\ref{eq:lever}) in the wind region for a rough examination of wind properties. From the top panels of figure~\ref{fig:rzslice} which mark the Alfv\'en surface, we may approximately estimate 
the ratio of $R_A/R_\mathrm{0w}$ to be $\lesssim1.3$, $\lesssim1.3$, and $\lesssim1.4$ for field lines originated from the planet gaps in runs with $M_P=1, 3, 5\Mth$, respectively. The small lever arm is consistent with the high value of $\dd\dot{M}_{\rm wind}/\dd\ln R/\Macc$ seen in figure \ref{fig:Mdot}.

There are essentially no poloidal magnetic field lines originate between the gaps, and consequently, no outgoing flow streamlines, in line with the wind mass loss profile shown in figure~\ref{fig:Mdot}. Instead, wind launched from smaller radii essentially flows over such ring-like regions (at $R\sim4\text{--}6$), nearly along the radial direction. In contrast to the strong accretion flow associated with the wind, we notice from figure \ref{fig:rzslice} that the flow structures inward and outward of the planetary gaps in the bulk disk are more stochastic, lacking the characteristic wind-driven accretion flows associated with the vertical gradient of $B_\phi$, but instead exhibit circulation patterns in the meridional plane (which are different from the expected meridional flow seen in hydrodynamic simulations, see next subsection).

\begin{figure*}
    \centering
    \includegraphics[width=\hsize]{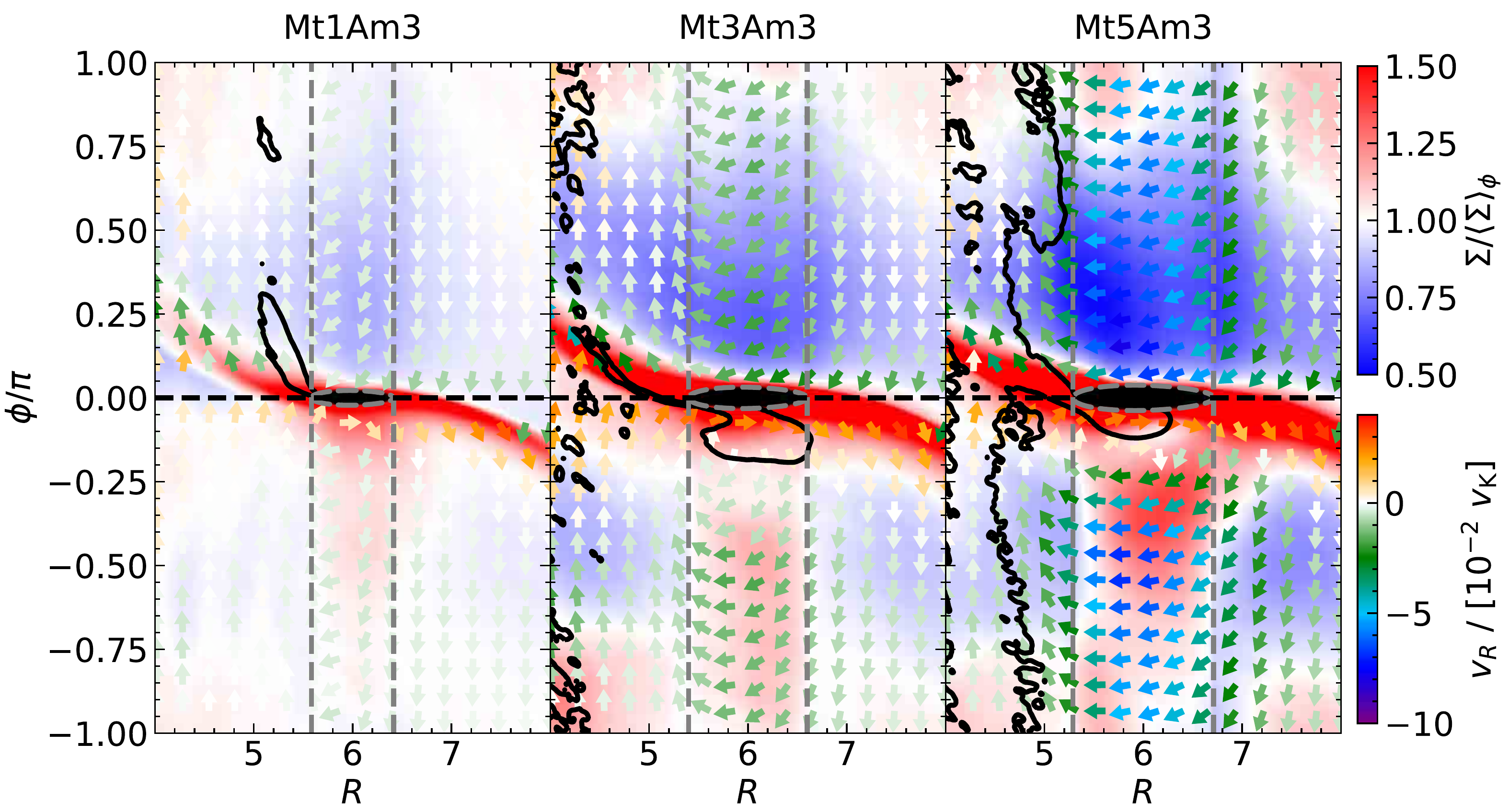}
    \includegraphics[width=\hsize]{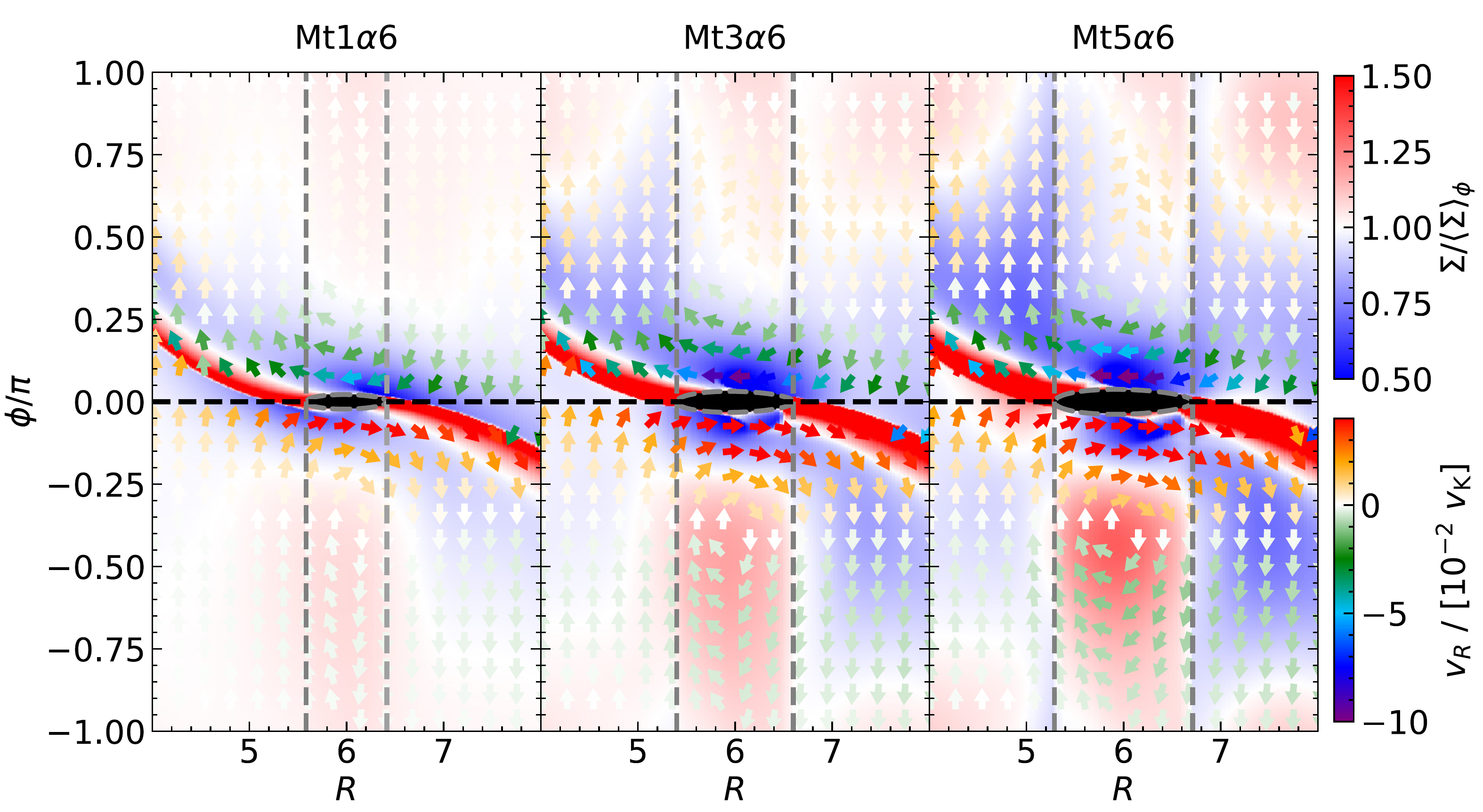}
    \caption{Flow vector (arrow) and relative surface density $\Sigma/\av{\Sigma}_\phi$ (background color) in Mt1Am3 (Mt1\al6), Mt3Am3 (Mt3\al6), and Mt5Am3 (Mt5\al6) from left to right in upper (lower) row.
    The arrow direction shows the planar velocity vector relative to the planet, and the arrow color indicates the radial velocity (NOT the total planar velocity). The velocity is mass-weighted averaged in the vertical ($z$) direction. The black contours in the upper panels enclose the $\Bzm<0$ regions.
    The grey vertical lines and curves correspond to $R=\RP\pm\rH$ and $|\bvec{R} - \bvec{\RP}|=\rH$. The horizontal black dashed line shows $\phi=0$.
    The planet's Hill sphere (black, inside the grey-dashed circle) is masked out when evaluating the $\av{\Sigma}$ average.
    In the windy disk, the fluid supplied form the outer disk passes the corotation region ($|R-\RP|<\rH$) without circulating in the ``horseshoe region''.
    }
    \label{fig:Flow_Hill}
\end{figure*}

\subsection{The meridional flow?}

Previous 3D hydrodynamic simulations found that massive planets trigger meridional circulation in the gap region: the flow gets pushed away from the planet at the midplane, lifts up, then approaches the planet at the high altitudes, and vertically falls onto the planet's orbital region \citep{Morbidelli+2014,Suzulagyi+2014,Fung+Chiang2016}.
There is also observational evidence for such meridional patterns\citep{Teague+2019,Yu+2021}.

In the windy disk, however, we do not see obvious signs of this meridional circulation. When averaged over $\phi$, we have already discussed that the flow pattern around the planet is primarily governed by MHD forces, which drives strong outflow, downflow, and accretion flow, etc.
We do notice that in the case of run Mt3Am3 and Mt5Am3,
there are flows from the gap outer edge towards the planet orbital region at up to $\pm H$, which might disguise as meridional flow if that region happens to be the emission surface. However, at upper altitude at the gap outer edge, gas consistently fly away in the disk outflow. Also note that there is higher-altitude ($z\sim\pm2H$) flows towards the planet orbital region from the inner side of the gap, which are also similar to meridional flow, though they are not present at lower altitudes. 

On the other hand, we identify that such flow pattern occurs only around the planet itself, namely, $\phi\sim0$ (rather than the entire orbital/gap region). At the $\phi=0$ slice where the planet locates, we find that the meridional flow extends to no more than $\pm 2H$, where strong outflow still takes over at higher altitudes ($z\gtrsim 2H$).
As this pattern is more closely related to circumplanetary disks, we will defer for more detailed study in a follow-up paper.

\subsection{Flow within the corotation region}
\label{sec:Flow_Hill}

Typically, in a viscous disk, the fluid in the planetary co-orbital region executes a horseshoe orbit with U-turn at the planet vicinity.
Since the fluid exchanges angular momentum with the planet at this turn, this horseshoe region is crucial for understanding planet migration \citep[e.g.,][]{Ward1991,Masset2001,Paardekooper+Papaloizou2009a}.

In the windy disk, however, this circulation in the corotation region is prevented by the fast radial accretion flow discussed in \S~\ref{sec:Flow_AF_DR}. Figure~\ref{fig:Flow_Hill} shows the horizontal velocity vector in the planet-corotating frame (arrow) and radial velocity (color in the arrows) in $z$-integrated $R$-$\phi$ plane. The velocity vector within the corotation region clearly directs radially inward (upper), while the counterpart in 2D-viscous runs shows the horseshoe circulation (lower). 
This can be understood by comparing the timescales of the horseshoe libration and Hill-region passage.
The libration timescale can be, neglecting the pressure gradient, estimated as 
\citep{Paardekooper+Papaloizou2009a}
\begin{equation}
    \taulib = \frac{8\pi}{3\OmegaP}\frac{R}{\Delta R},
\end{equation}
where $\Delta R$ is the half-width of the horseshoe region.
Since $\Delta R \approx \rH$ when $\MP\gtrsim \Mth$ \citep{Masset+2006,Paardekooper+Papaloizou2009a}, $\taulib \sim 19 ~(\MP/\Mth)^{-1/3}~\tauP$ in our disk model. On the other hand, the mean radial velocity within the corotation region is measured as to be $v_R\sim -0.002$ (Mt1Am3), $-0.01$ (Mt3Am3), and $-0.06~ \vK$ (Mt5A3), resulting in the timescale passing the corotation region to be $\taupass = 16$, $3.2$, and $0.53~\tauP$, all of which are shorter than $\taulib$.
Namely, the fluid passes the corotation region before being able to finish the horseshoe orbit.

In addition, we note that there is outward radial flow at the planetary tailing side ($ -\pi/4<\phi< 0$), which coincides with regions of negative $\Bzm$ (see \S~\ref{sec:MFC}, and the black contours of figure~\ref{fig:Flow_Hill}). We find that in this localized region, $B_\phi$ remains relatively smooth, therefore, a negative $\Bzm$ leads to the reversal of the Lorentz force, and hence this radially-outward gas flow. We highlight that the presence of this localized radial outflow is a robust consequence of the negative $\Bzm$, which itself is a robust phenomenon that we speculate to be associated with magnetic flux concentration. As a result, the flow in the trailing side of the gap exhibits certain localized circulation between $ -\pi/2<\phi< 0$. These patterns differ dramatically from the viscous counterpart that clearly show the horseshoe orbits, and could have important implications to planet migration (see Section \ref{sec:D_mig}).

\section{Gap opening mechanism}
\label{sec:GOM}
The structure of rotating disks is governed by the angular momentum transport, namely, the torque.
In the viscous disk, the planet gravity and viscous torques tend to open and close the gap \citep[e.g.,][]{Lin+Papaloizou1993}, and they balance each other that determine the gap structure upon reaching a steady state.
In this section, we examine how the magnetic fields change the torque balance and, thus, the gap structure.
We discuss the torque balance in the viscous disk and wind-driven disk in \S~\ref{sec:Tor_VD} and \ref{sec:Tor_WD} for the fiducial runs of Mt3\al6 and Mt3Am3, respectively, where the torque components are defined in \S~\ref{sec:M_ana}. Then, we systematically analyze the torques and angular momentum transport in the vicinity of the planet orbital region in \S~\ref{sec:AMT_MHD}.
Lastly, we discuss the dependency on planetary mass in \S~\ref{sec:Tor_MP}.

\begin{figure*}
    \centering
    \includegraphics[width=\hsize]{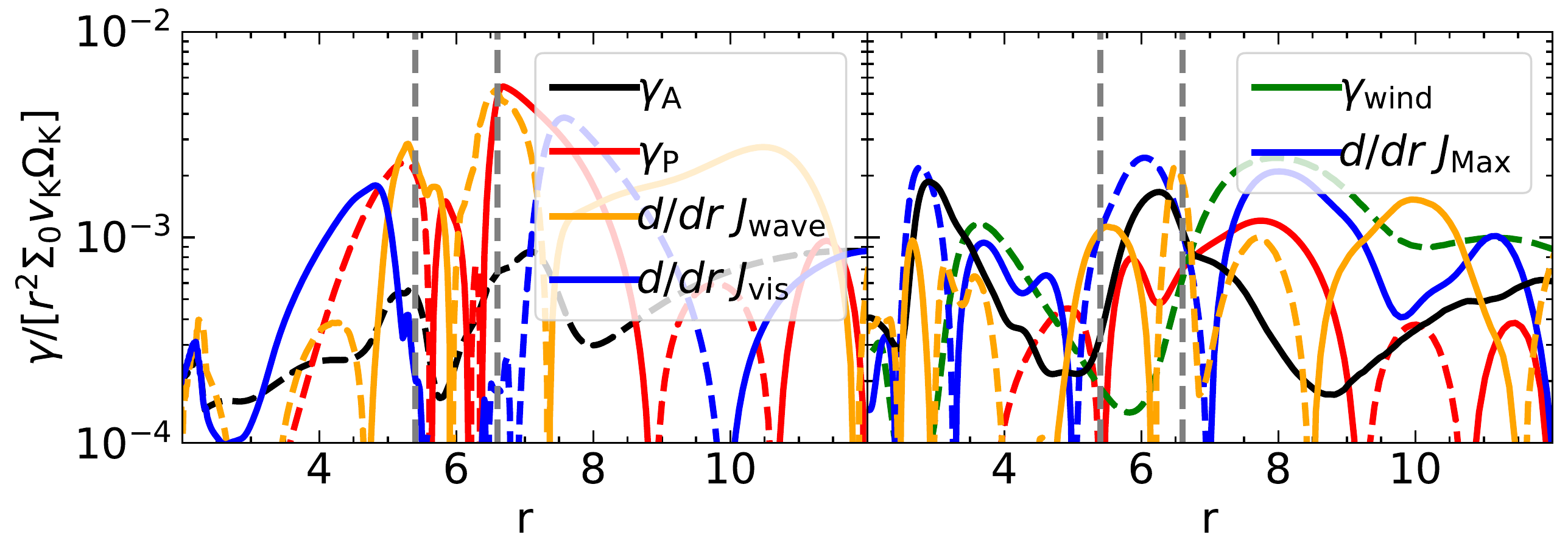}
    \caption{
    Radial torque density due to planet gravity (red), Maxwell stress (blue in the right panel), viscosity (blue in the left panel), disk wind (green), disk accretion flow (black), and density wave/shocks (orange), respectively, for the fiducial run Mt3Am3.
    The left and right panels show the results of the 2D viscous hydrodynamic simulation and of the 3D MHD-wind simulation, respectively.
    The solid and dashed lines correspond to positive and negative torque density in log scale. The grey dashed lines mark $r=\RP\pm\rH$.
    }
    \label{fig:tor}
\end{figure*}

\subsection{Torque balance in a viscous disk}
\label{sec:Tor_VD}
We start from a brief introduction of the torque balance in the viscous accretion disk of the Mt3\al6 run.
The torques in this viscous disk is shown in the left panel of figure~\ref{fig:tor}.
Due to the spiral density waves/shocks, the planet gravity torque is positive and negative at outward and inward of the planet, respectively, indicating that $\GP$ drives gap opening. 
Close to the planet ($5\lesssim R \lesssim 7$), $\Gwave$ has the opposite sign and a similar amplitude to $\GP$, leading to torque balance.
The total planet torque $\GP+\Gwave$ is deposited at further distances \citep[e.g.,][]{Takeuchi+1996, Goodman+Rafikov2001,Crida+2006, Kanagawa+2015,Kanagawa+2017}, in this case at about $3\lesssim R \lesssim 5$ and $7 \lesssim R \lesssim 12$.
The deposited planetary torque is then balanced by the viscous torque $\Gvis$ to fill in the gap. It is the balance of these torques that determine the gap structure \citep[e.g.,][]{Lin+Papaloizou1993}. Note that the torque density associated with accretion $\GA$ is negative here, namely, the disk gas flows outward. This is because of the relatively steep surface density slope adopted here. Nevertheless, this component is negligible in the torque balance in the planetary gap and hardly affects the gap structure.

\subsection{Torques in fiducial run (Mt3Am3)}
\label{sec:Tor_WD}
From now on, we focus on our fiducial run (Mt3Am3), where the relevant torques are shown in the right panel of figure~\ref{fig:tor}. 
Here, the Maxwell stress ($\GMx$) plays the role of viscosity ($\Gvis$). In addition, the MHD-driven wind torque ($\Gwind$) is added to the analysis.

Firstly, we see that $\GP$ and $\Gwave$ share similar shapes compared to the viscous case, but have smaller amplitudes especially in the vicinity of the planet orbit. Therefore, the physics of gap opening by the planet remains similar. As the torque density is proportional to the surface density of the gas, the smaller amplitudes is primarily due to the fact that the gap in windy disks are deeper and wider. As a result, the overall torque balance is largely determined by the MHD torques and the accretion flow, which are strongly affected by magnetic flux concentration.

The wind torque $\Gwind$ mainly dominates in regions outward of the planet gap. As has been discussed in Section \ref{sssec:outflow}, this is the result of field lines bent radially outward at higher altitudes: concentrated poloidal magnetic fields in the gap region emanate from the disk surface from the outer side of the gap (see figure~\ref{fig:rzslice}). Within the gap, the wind torque diminishes as there is a deficit of magnetic flux originating from inward of the planet orbit. We also see the additional bump in the wind torque at $r\sim4$, which results from magnetic flux concentration in the planet-free gap at $r\sim3$.

The radial profile of the torque density resulting from the Maxwell stress is more complex, which is a result of the radial gradient of the angular momentum flux. We can see from the top panels of figure \ref{fig:alp}, that although the $\alphaMx$ profiles peaks within the gap, the unnormalized Maxwell stress itself peaks at outer radii. This leads to a negative torque density in the gap region, and a positive torque density beyond the gap. Although $\dd/\dd r(\JMx)$ is the counterpart of $\dd/\dd r(\Jvis)$, it plays a very different role because the highly inhomogeneous radial distribution of magnetic flux. In particular, in the gap region, the net angular momentum flux is nearly entirely from this component, which drives gas accretion across the gap, leading to gap depletion. Outward of the gap, it is mainly the radial gradient of this $\JMx$ that balances the wind torque.

Piecing together, the aforementioned torques drive the accretion flow through the gap. We note that the torque density contributed from accretion $\GA$ varies substantially across the gap region, in contrast to the relatively smooth variation in the accretion rate (see figure \ref{fig:Mdot}). This can be accounted for by the modulation in disk rotation rate (see Equation (\ref{eq:JA})). For instance, $\dd\langle j\rangle/\dd r$ is steeper in the gap region than the Keplerian case, leading to enhanced $\GA$ in the gap.

The comparison between the left and right panels in figure~\ref{fig:tor} infers why the gap is deeper and wider in the windy disk than in the viscous disk.
In the viscous disk, the gap is carved by the planet-induced torques ($\GP+\Gwave$). Since the planet-induced torque is proportional to density while the viscous torque scales with the density gradient, there is a limit on gap depth once the gap becomes deep enough. In contrast, in the windy disk, gap opening is primarily due to MHD torques boosted by magnetic flux concentration at the planetary orbit. Since the $\dd\JMx/\dd r$ is insensitive to gas density, it keeps depleting the gap regardless the gap depth. We will further show in the next subsection that the role of $\JMx$ and $\Jwind$ are in fact the two sides of the same coin, thus this gap depletion can be equally understood as angular momentum removal from the wind.

\subsection{The nature of the MHD torques}
\label{sec:AMT_MHD}
\begin{figure}
    \centering
    \includegraphics[width=8.55cm]{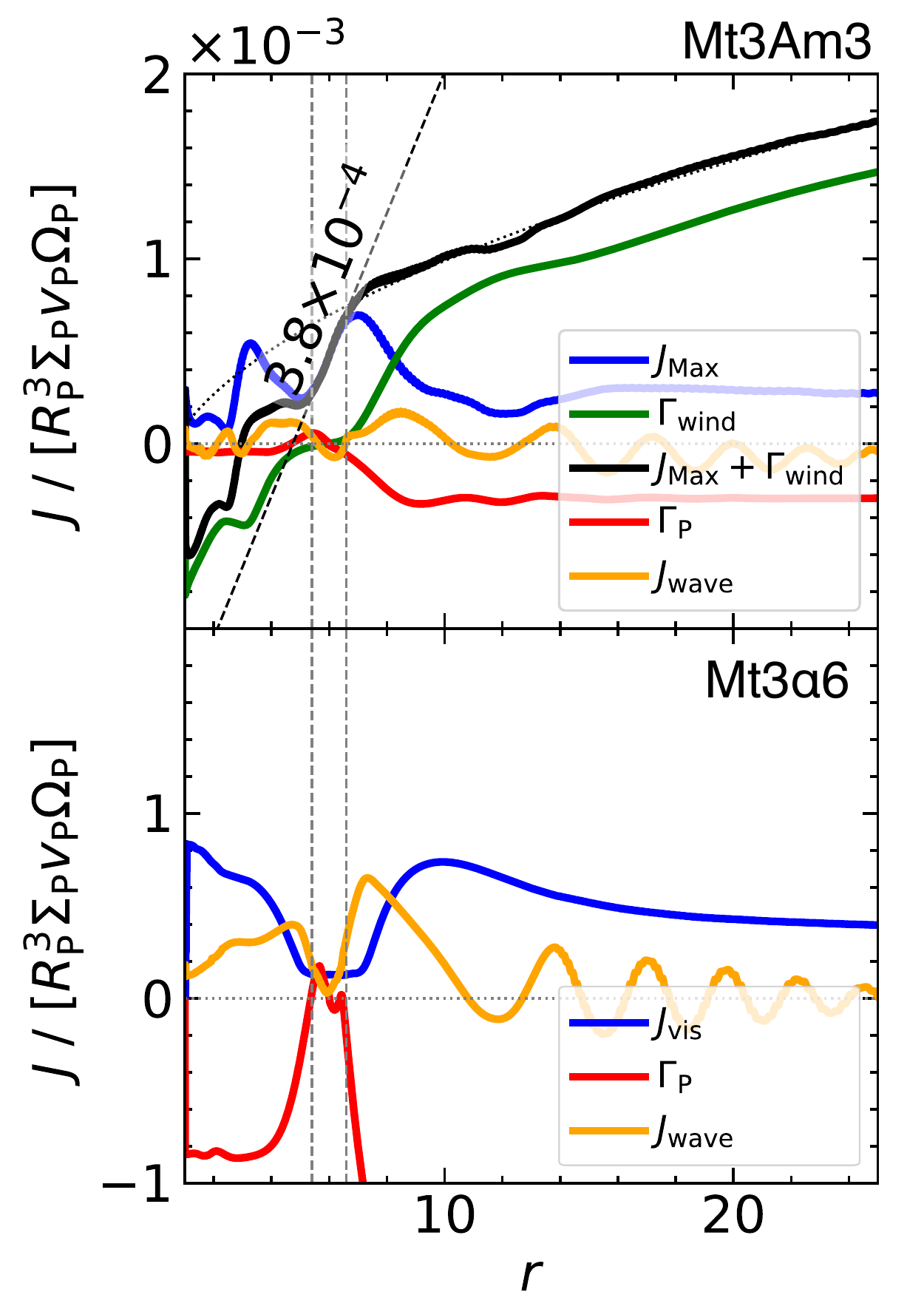}
    \caption{Angular momentum flux $J$ (or cumulative torque $\Gamma$) due to the Maxwell stress ($\JMx$,blue), disk wind ($\Gamma_\mathrm{wind}$, green), planet gravity (red), and density wave/shock (orange) in runs Mt3Am3 (upper, MHD) and Mt\al6 (lower, viscous hydro). 
    Note that the $\Gamma$ is the integral of $-\gamma$, with zero point set at $R=\RP$ (see Eqs.~[\ref{eq:GPt}] and [\ref{eq:Gwt}]). The horizontal and vertical grey dashed lines corresponds to $J=0$ and $R=\RP\pm\rH$, respectively.
    In the viscous simulation, the angular momentum flux due to viscosity ($\Jvis$, blue) is shown instead of $\JMx$, and there is no $\Jwind$.
    The black line shows the effective angular momentum transport due to the MHD effects ($\JMHD \equiv \JMx+\Gamma_\mathrm{wind}$).
    The dashed line has a slope of $3.8\times10^{-4}$ corresponding to the MHD torque in the gap region. The dotted line corresponds to $4.1\sqrt{r}\times10^{-4}+C$ (constant) obtained by fitting $\JMHD$ between $10<R<30$.
    In general, the shape of $\JMHD$ is hardly affected by gap formation except for the corotation region.
    }
    \label{fig:AMF}
\end{figure}

As discussed in the previous subsection, the local torque balance in the windy disk is greatly different from that in the viscous case, primarily due to the MHD torques. However, the inhomogeneity in the torque profiles adds to the complexity against a clear understanding of the underlying physics. In this subsection, we aim to provide a more transparent picture of the torque balance, focusing on the MHD torques.

In figure~\ref{fig:AMF}, we show the {\it cumulative torque} from the two components of the MHD torques, corresponding to the radial angular momentum flux $\JMx$, and the radially-integrated (negative) wind torque $\Gamma_{\rm wind}$. We similarly also compute the cumulative (negative) torques from other components (planetary gravity $\Gamma_P$ and the wave angular momentum flux $\Jwave$). Note that they share the same dimensions, and positive/negative slopes correspond to negative/positive torque density. Here, the cumulative torques $\Gamma$ are set to zero at the planetary orbit $r=6$.

The radial profile of $\JMx$ peaks at about the outer gap edge as a result of magnetic flux concentration, similar to that shown and discussed in the top panels of figure \ref{fig:alp}. As its value falls off beyond the outer gap edge, the wind torque picks up, as a result of winds emanating from included field lines emanating the outer gap edge.
In fact, the increase of the wind torque largely cancels the decrease of the radial angular momentum flux $\JMx$. To see this, we show in black lines the total cumulative MHD torque $\JMHD\equiv\JMx+\Gamma_{\rm wind}$. We find that beyond the planetary gap, $\JMHD$ approximately follows a $r^{1/2}$ scaling, which extends all the way to larger radii. This is consistent with a standard wind torque balancing steady state accretion, which is proportional to $\Macc\dd/\dd r(r^2\OmegaK) \propto \sqrt{r}$. At the planetary gap region, there is little contribution from $\Gwind$ as we discussed in the previous subsection, the MHD torque is dominated by the rising side of $\JMx$, with a steeper slope.

Inward of the planetary gap, we see that the slope of $\JMHD$ first remains similar to that outward of the planetary gap. At around $r=3$ where the planet-free gap is located, there is another bump in $\JMx$ as a result of magnetic flux concentration there. Again, the fall-off of the bump is compensated from the rise in $\Gamma_{\rm wind}$. Overall, $\JMHD$ develops another steeper slope in the planet-free gap region from the rising side of $\JMx$, similar to the planetary gap case.

To summarize the analysis above, we find that despite of the inhomogeneities in individual components of the MHD torque, the combined effect of radial and vertical angular momentum is in fact quite simple: away from the gap region, the total MHD torque behaves as a standard wind torque, whereas in the gap region, the wind torque gets enhanced due to magnetic flux concentration. 
To be more quantitative, when we normalize the torque density (i.e., the slope in figure \ref{fig:AMF})  by $\RP^2\Sigma_{\rm P}\vP \OmegaP$, we find the slope in $\JMHD$ is about $2.3\times 10^{-3}$, as opposed to $5.0\times10^{-4}$ based on the $r^{1/2}$ fit outward of the gap region.\footnote{Note that the $J$ in the figure~\ref{fig:AMF} is normalized by $\RP^3\SigmaP\vP,\OmegaP$, which differs by $\RP=6$ from the dimension of torque-density we use here.} 
There is a factor of $\sim5$ difference, which is largely owing to magnetic flux concentration.
Also note that as the $\Macc$ is largely constant across the gap, this enhancement of torque density is mainly compensated by the steeper $j$ gradient within the gap (see figure~\ref{fig:Mdot} and discussion in the previous subsection).

\begin{figure}
    \centering
    \includegraphics[width=8.55cm]{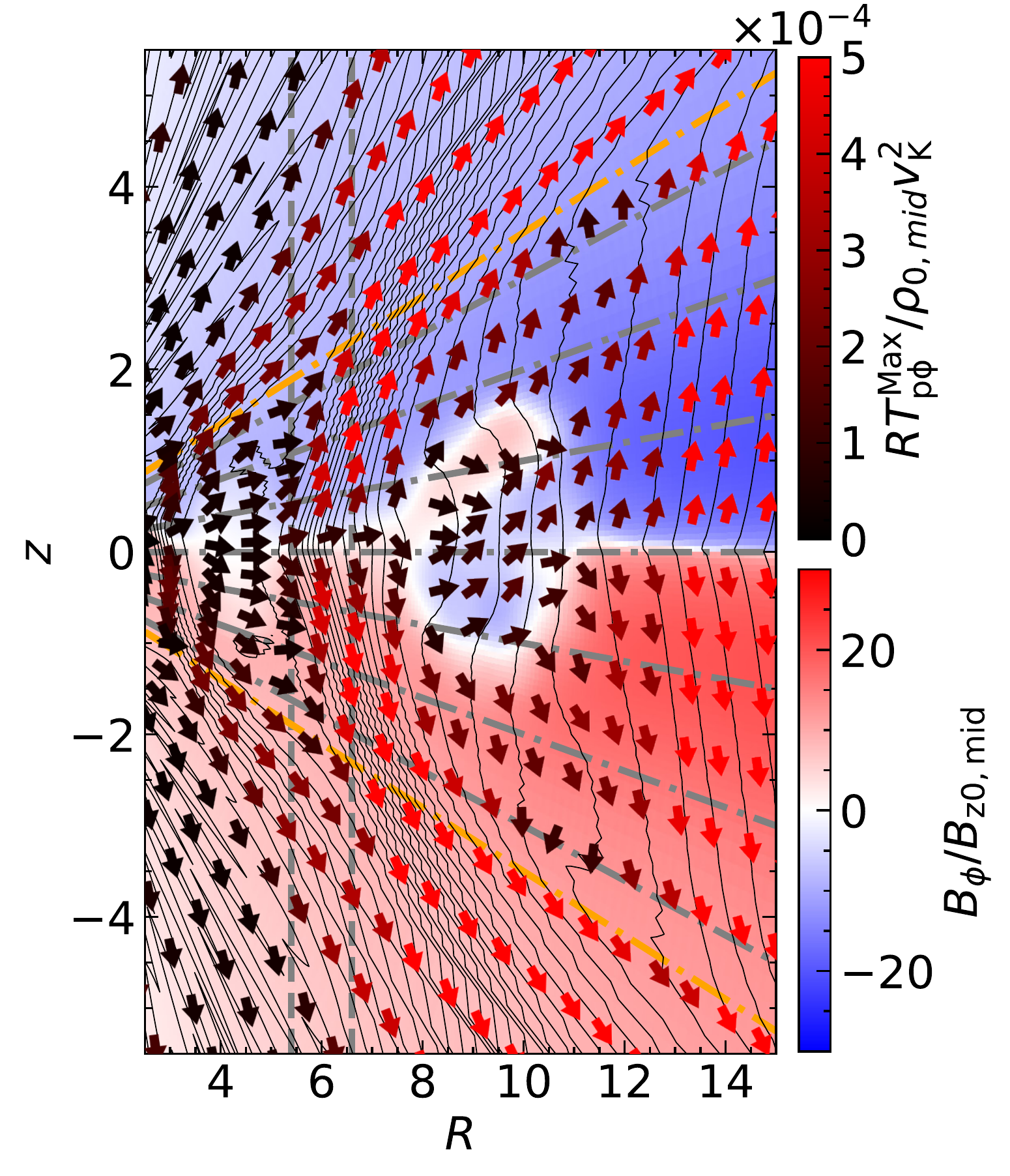}
    \caption{Angular momentum flux due to the Maxwell stress in the poloidal plane ($RT^\mathrm{Max}_{p\phi}$). The background color shows the toroidal magnetic field and the solid black lines show the linearly equally spaced contour of poloidal magnetic flux function $\Phi_B$. The grey dashed lines mark the planetary corotation region $R=\RP\pm\rH$, and the grey dash-dotted lines mark the constant aspect ratio of 0 to $3h$ separated by $1h$. The orange lines correspond to the disk surface ($h=3.5$ in our analysis). Within the planetary gap, the angular momentum flux mainly flows along the magnetic field lines.
    }
    \label{fig:AMflow}
\end{figure}

To better understand why the stresses that correspond to radial and vertical transport of angular momentum are so smoothly connected that leads to the simple picture above, we show in figure~\ref{fig:AMflow} the vectors of the angular momentum flux from the poloidal Maxwell stress $-{\boldsymbol B}_pB_\phi$. We see that in regions with magnetic flux concentration, the angular momentum flows along the poloidal magnetic field lines.
This is characteristic of angular momentum extraction by disk winds. However, the standard diagnostics decomposes this flux into the radial ($-B_rB_\phi$) and vertical ($-B_\theta B_\phi$) components, treating them separately as viscous and wind stresses. This approach artificially separates a single piece of physics into two disparate components, causing certain degree of conceptual confusion. In this figure, we also see that, outward of gap regions, angular momentum transport is largely radially outward in the bulk disk (despite of the relatively small angular momentum flux), conforming to the standard scenario of viscous transport. In the meantime, there is additional wind transport that originate from the disk surface. Therefore, the overall situation in such regions is similar to the planet-free case \cite[e.g.][]{Cui+Bai2021}.

\begin{figure}
    \begin{minipage}[]{8.55cm}
    \centering
    \includegraphics[width=\hsize]{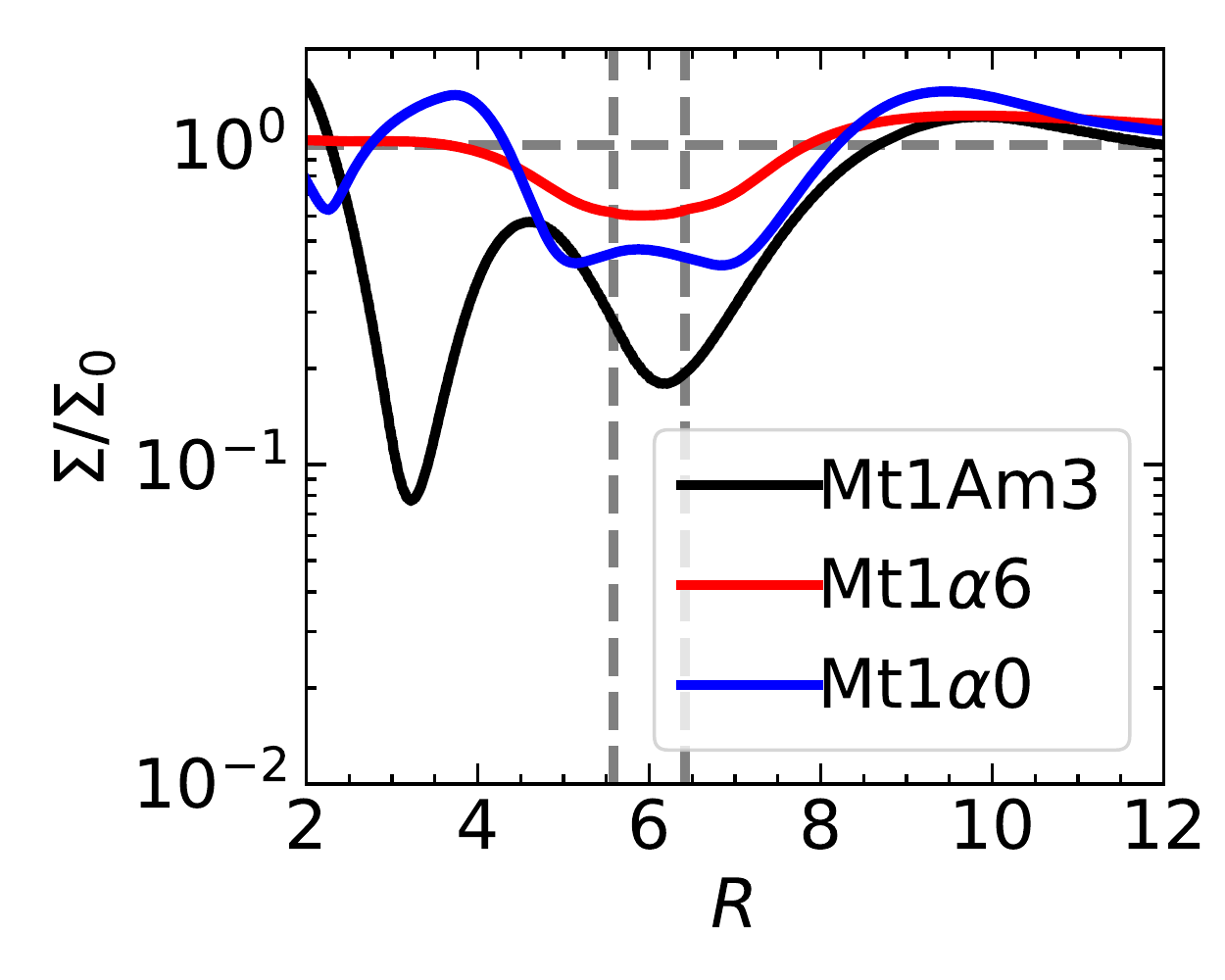}
    \includegraphics[width=\hsize]{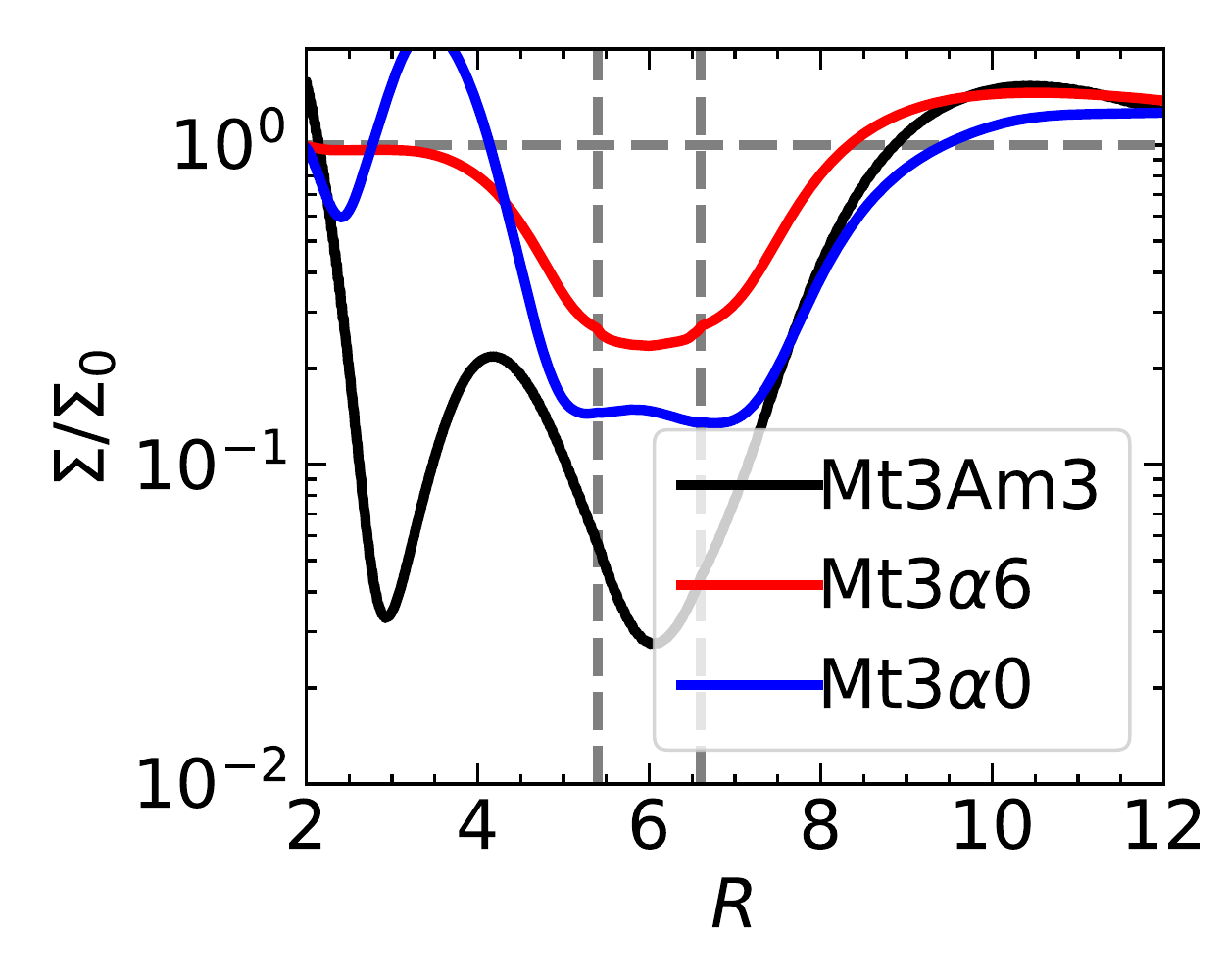}
    \includegraphics[width=\hsize]{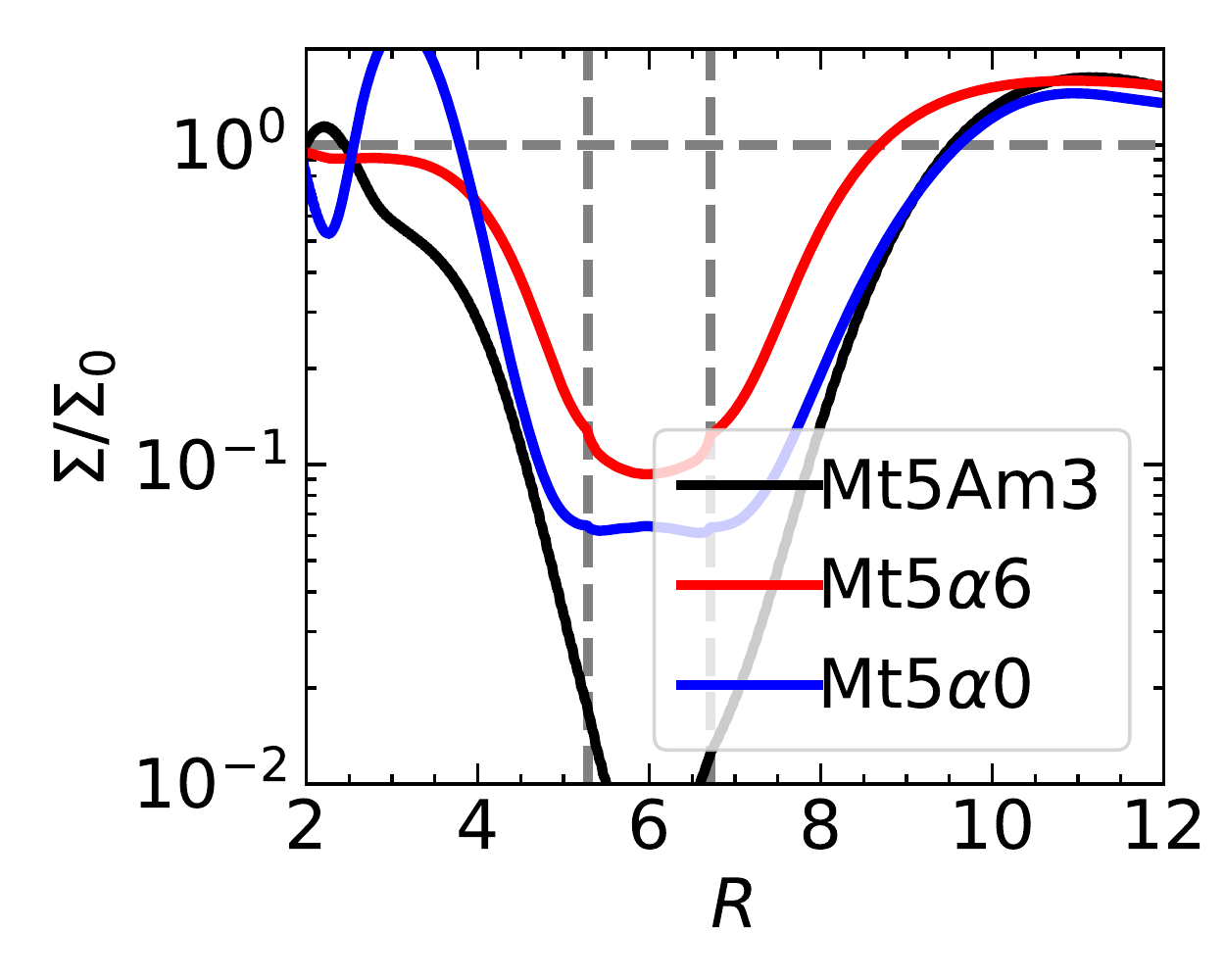}
    \end{minipage}
    \caption{The gap profile in MHD simulations (black) compared with the viscous case ($\alpha=6\times10^{-3}$, red) and the inviscid case (blue) for runs with $\MP=1, 3, 5\Mth$, respectivtly. The vertical and horizontal grey dashed lines mark $\Sigma=\Sigma_0$ and $R=\RP\pm\rH$. 
    The planetary gap in the windy disk is similar in width to the inviscid case, but is deeper.}
    \label{fig:GapProf_invicid}
\end{figure}

\subsubsection{Comparison with inviscid disks}

One aspect of the MHD torques is that as it is mediated by Lorentz force, it is insensitive to changes in the gas density. Moreover, outward of the gap (corotation) region, the MHD torque is largely dominated by disk winds, with very little contribution from viscous stress (as $\JMHD$ can be merged into the wind torque based on earlier discussion). These properties bare certain similarities to the inviscid case, which lacks the density-dependent viscous stress to compensate for the planetary torques. We thus proceed to make such a comparison.

Figure~\ref{fig:GapProf_invicid} shows the radial gap profile in the windy disk (Mt3Am3, black), viscous disk (Mt3\al6, red), and inviscid disk (Mt3\al0, blue). At the gap outer side ($R\gtrsim 7$), the gap profiles are similar between the windy and inviscid disk cases. We also anticipate the similarity to carry over in the inner side of the gap, if the planet-free gap were not present. The gap profiles are not only wider, but also clearly sharper than the viscous case, as there is very little or no viscous diffusion. Also, we find that the gap widths between the inviscid case and the windy case are similar (with the windy case being slightly wider), and the temporal evolution of the gap widths is also similar between the two cases (see figure~\ref{fig:gap_time}). These similarities imply that beyond the gap (Hill) region, the combined effect of MHD torques on the gap profile is similar to the inviscid case. Although MHD/wind-driven accretion is not present in the inviscid case, this accretion component appears to have very limited influence in the gap profile. On the other hand, magnetic flux concentration and wind-driven accretion do have dramatic influence within the corotation region, as discussed earlier.

\subsection{Dependence on planetary mass}
\label{sec:Tor_MP}

\begin{figure}
    \centering
    \includegraphics[width=8.55cm]{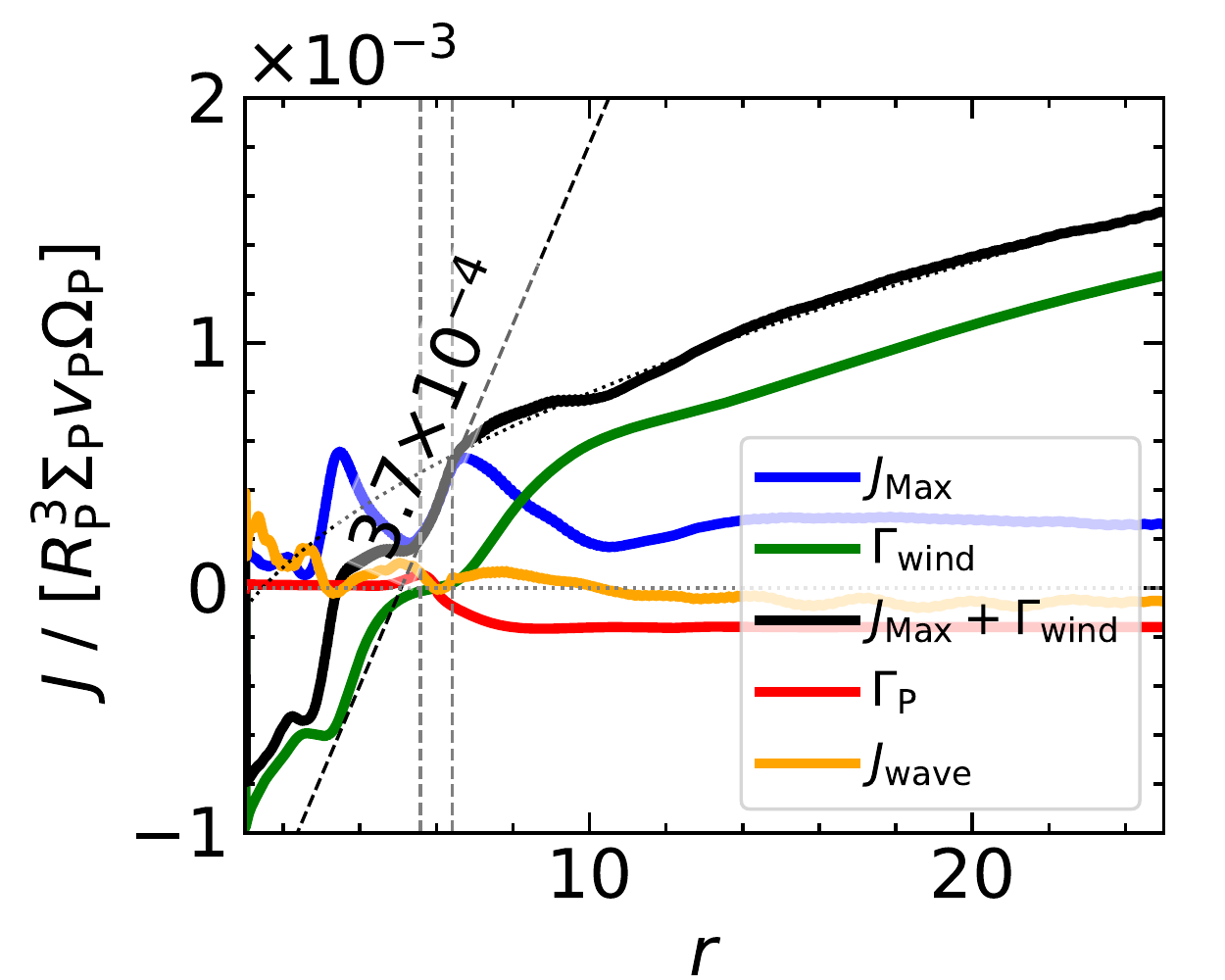}
    \includegraphics[width=8.55cm]{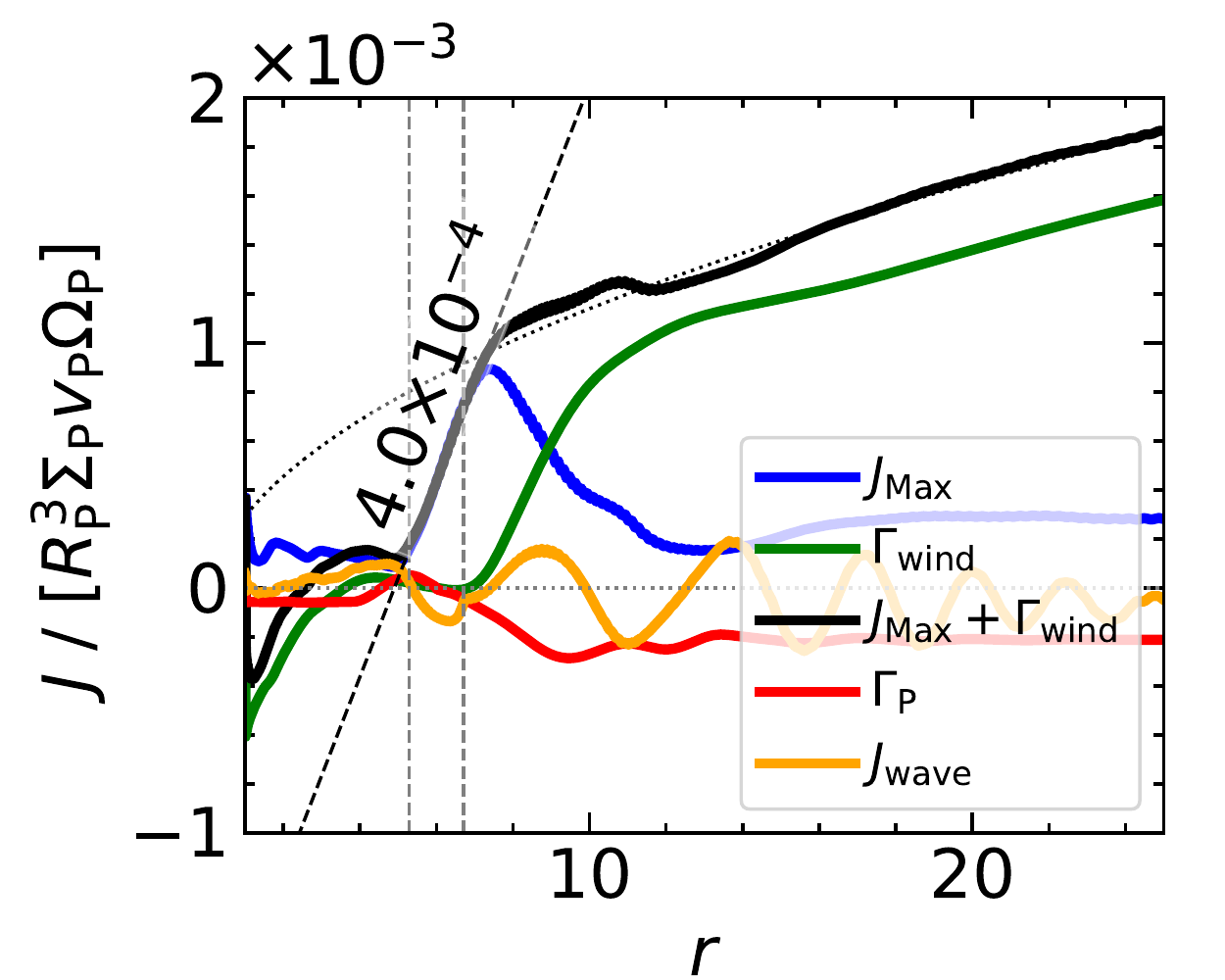}
    \caption{
    Same as figure~\ref{fig:AMF} but for runs Mt1Am3 (upper) and Mt5Am3 (lower), respectively. 
    The fitted slopes in the gap regions are $3.7\times10^{-4}$ and $4.0\times10^{-4}$ for the two runs shown in dashed lines. outward of the gap, the profiles in $\JMHD$ for the two runs correspond to
    $4.0\sqrt{r}\times10^{-4}+C$ and $3.9\sqrt{r}\times10^{-4}+C$ shown in dotted lines.
    }
    \label{fig:AMF_15M}
\end{figure}

In this subsection, we further examine the torque balance in the other two runs Mt1Am3 and Mt5Am3. The radial profiles of the torque densities are shown in figure~\ref{fig:tor_15M} in Appendix~\ref{sec:Am1}, which are qualitatively similar to that of the fiducial run. Here, we mainly focus on the MHD torques, showing the resulting cumulative torques in figure~\ref{fig:AMF_15M}.

The role of the MHD torque on gap formation is largely the same for three planetary masses in our simulations, namely, there is an enhanced torque in the gap region, and the torque becomes similar to a standard wind torque beyond the gap. This is likely the reason why the gap profile is similar to that in the inviscid disk (see figure~\ref{fig:GapProf_invicid}).
Interestingly, the enhanced slope of $\JMHD$ in the gap region is quantitatively similar in different planetary masses ($3.8$, $3.8$, and $4.0 \times$ $10^{-4} \RP^3 \Sigma_0(\RP) v_\mathrm{P}\OmegaP$ for 1, 3, and 5 $\Mth$, respectively). Correspondingly, the negative torque acting on the gap center is same for the three cases within $5\%$.
This is consistent with the fact that the $\Bzm$ has a similar magnitude around the planet in different mass cases (see figure~\ref{fig:GapProf}).
Moreover, figures~\ref{fig:AMF} and \ref{fig:AMF_15M} show that the radial range where the MHD torque is enhanced approximately scales with the Hill length $\rH$, consistent with radial range where $\Bzm$ is enhanced through magnetic flux concentration (see figure~\ref{fig:GapProf}).

For run Mt5Am3, we note that the radial range with enhanced MHD torques extends further outward ($\sim 2\rH$) relative to the other two runs ($\sim\rH$ for runs Mt1Am3 and Mt3Am3). However, we find that such difference appears only after the planet-free gap merges to the planetary gap ($t\gtrsim 110~\tauP$; see \S~\ref{sec:gap_structure}). After the merger, the planetary gap gathers more poloidal magnetic fluxes, which enhances both the $\Bzm$ magnitude and its radial extent (see the red line of the bottom right panel in figure~\ref{fig:GapProf}). Before the merger, both the slope and radial extent are more similar to the other runs.

\section{Planet migration}
\label{sec:D_mig}

\begin{figure*}
    \centering
    \includegraphics[width=\Flh]{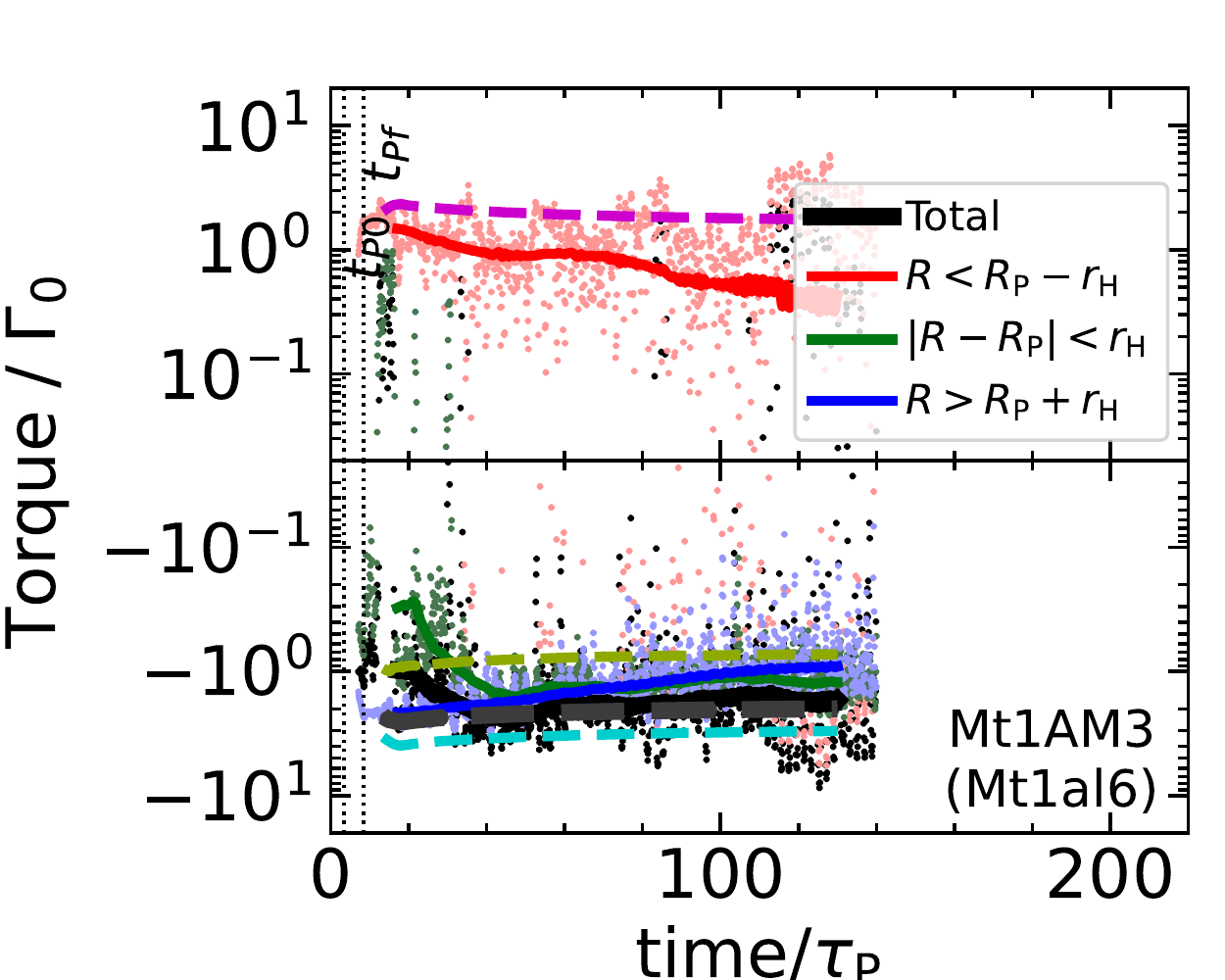}
    \includegraphics[width=\Flh]{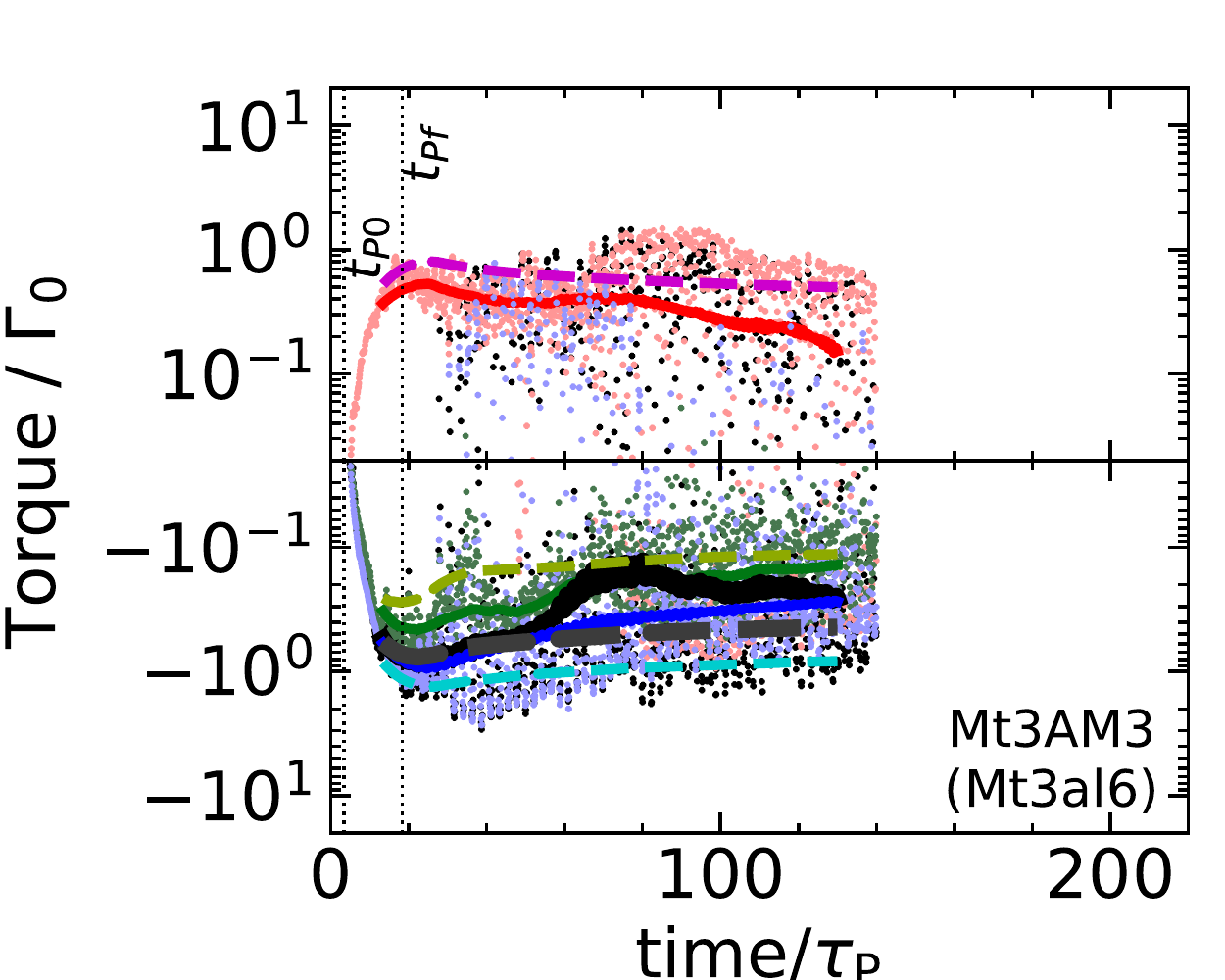}
    \includegraphics[width=\Flh]{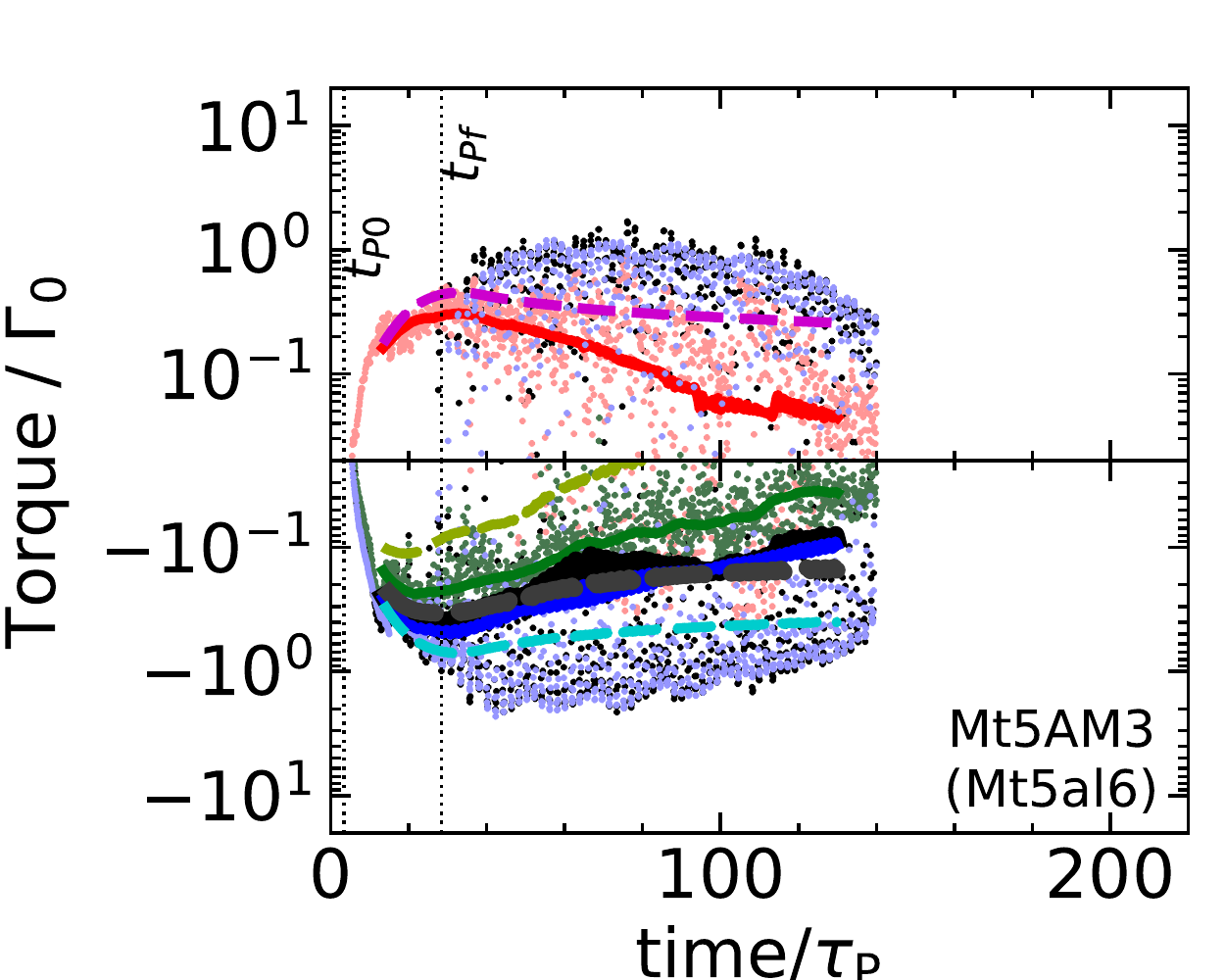}
    \includegraphics[width=\Flh]{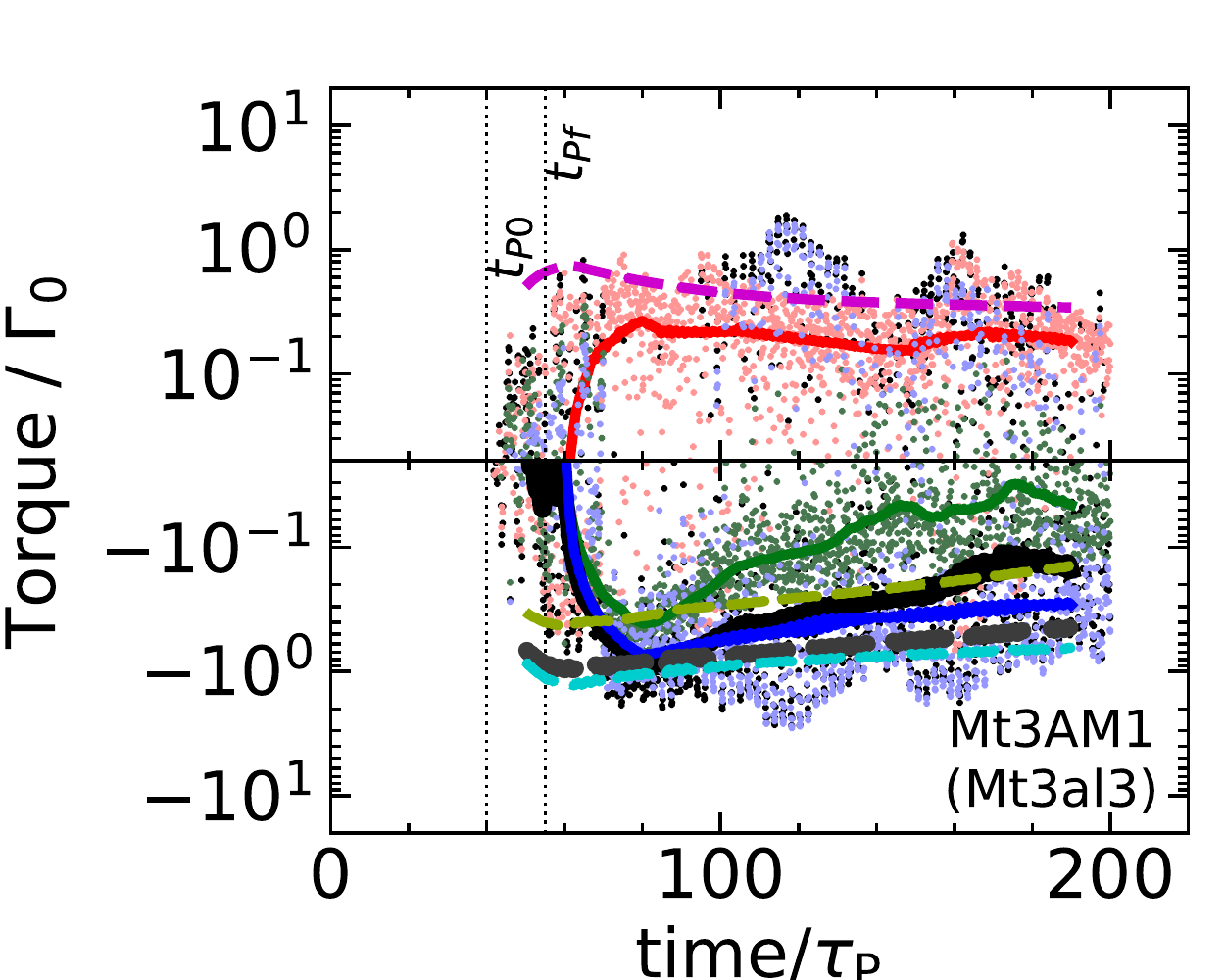}
    \caption{Migration torque acting on the planet for runs Mt1Am3, Mt3Am3, Mt5Am3, and Mt3Am1, respectively. The color corresponds to the torque originating from $R<\RP-\rH$ (red), $|R-\RP|<\rH$ (green), $R>\RP+\rH$ (blue), and whole simulation domain (black). The points and lines correspond to the snapshots and the moving average over $\pm 10~\tauP$, respectively. The averaging takes over a longer time span in order to filter out the temporal fluctuations. 
    The lighter-color dashed lines (grey [total], pink [$R >\RP+\rH$], light green [$|R-\RP|<\rH$], and light blue [$R< \RP-\rH$]) correspond to the 2D viscous simulations of Mt1\al6, Mt3\al6, Mt5\al6, and Mt3\al3, respectively.
    }
    \label{fig:MigrationTorque}
\end{figure*}

Planet migration is one of the most important consequences of the planet-disk interaction.
In the type-II regime, it was conventionally expected that when the planet opens the annular gap, the planet is locked to the gap and migration is associated with viscous disk accretion
\citep[e.g.,][]{Lin+Papaloizou1986,Ward1997}.
However, numerical simulations in the viscous disk scenario showed that the gas can pass through the planetary gap \citep[e.g.,][]{Artymowicz+Lubow1996,Kley1999}, and the direction and speed of the migration depends on the gap structure and disk mass \citep[e.g.,][]{Duffell+2014,Durmann+Kley2015,Durmann+Kley2017,Kanagawa+2018,McNally+2019,Dempsey+2020,Lega+2021,Dempsey+2021}. In our simulations, we have seen that the gas rapidly passes through the planetary gap, primarily due to wind-driven accretion, in stark contrast with the standard type-II migration scenario. We also note that the planet orbit is fixed in our simulations, so as to obtain a first result of migration torques without further complications.

The migration torque in the type II regime is more complex because the planet induces dynamical feedback to the background surface density profile, and in turn alters the migration torque. With disk winds dominate angular momentum transport, it substantially alters the torque balance compared to the viscous case, which is reflected in three aspects:
(1) the planetary gap can be much deeper and modestly wider than the viscous counterpart, with substantial accretion flow across the gap; and (2) there are planet-free gaps whose location can be stochastic; and (3) the gap region can be azimuthally asymmetric.
Figure~\ref{fig:MigrationTorque} shows the migration torque from all our 3D simulations. They are normalized by the characteristic scaling \citep{Goldreich+Tremaine1980, Tanaka+2002,Paardekooper+2010,Kanagawa+2018}
\begin{equation}
    \Gamma_0=(\MP/M_*)^2(H/R)^{-2}\SigmaP R^4\OmegaP^2\ .
\end{equation}
Without more detailed analysis (e.g., see \citealp{Chen+2020}), we simply divide the disk into the inner disk, outer disk, and the corotation region, set by whether the location relative to the planet orbital radius $\RP$ is within the Hill radius $\rH$, and examine the migration torques from these three regions. Note that we soften the torque from the planet nearby to exclude the gas bounded to the planet, by the Fermi-type smoothing function with cut-off length of $0.8~\rH$ \citep[see Eq.~(11) in ][]{Kley+2009}.
The results are also compared to the viscous simulations. It is generally expected that the (Lindblad) torques from the inner/outer regions are positive/negative, as a natural consequence of gravitational interaction between the planet and the trailing density waves/shocks, with a magnitude comparable to $\Gamma_0$ that partially cancel each other (e.g., \citealp{Kley+Nelson2012}). The torques from the corotation region often plays a decisive role on the final sign and magnitude of the migration torque \citep[e.g.,][]{Paardekooper+2011}.

We start by clarifying that our viscous simulations are set to contrast our windy-disk simulations. Specifically, they are not set to achieve steady state configuration as conventional approaches (e.g., \citealp{Duffell+2014,Fung+2014}), and hence the results, especially the net torque, are not to be directly compared with the literature. The viscosity in these simulations are set to $\alpha=6\times10^{-3}$ to compare to rums with $Am=3$, and $\alpha=3\times10^{-3}$ to compare with run Mt3Am1, which are chosen to approximately match the turbulence level (in the absence of the planet) in these runs. Our comparison focuses on the relative difference in migration torques between the viscous and windy-disk simulations, both on the net torque and the torques from the three regions. The main features are listed below.

Firstly, we find that the migration torques in windy-disk simulations at the inner and outer disks are generally smaller than those in 2D viscous simulations. This is mainly because the gaps in windy-disks are generally modestly wider and much deeper, which reduces the Lindblad torques roughly in proportion (e.g., \citealp{Chen+2020}).

Second, the torque from the corotation region is always negative. 
We emphasise that this is not the usual ``corotation torque'' \citep[e.g.,][]{Ward1991,Paardekooper+Papaloizou2009a} because the fluid hardly corotate with the planet but rapidly passes the corotation region without completing the horseshoe orbit. As discussed in Section \ref{sec:Flow_Hill}), the presence of a localized negative $\Bzm$ region on the trailing side of the planet also makes the gas flow radially outward in that region. This flow supplies more gas to the trailing side of the planet, leading to a density asymmetry between the leading and trailing side of the planet (see fig.~\ref{fig:Flow_Hill}), resulting in a negative corotation torque.
Contrary to our findings, when using simple wind prescriptions, \citet{Kimmig+2020} reported that the wind-driven inflow with $\taupass \lesssim 10 \taulib$ enhances the positive corotation torque
(see their figure 11). Our results differ because of faster mean radial flow $\taupass\lesssim\taulib$ that flushes out the horseshoe orbits, together with asymmetric flow structure, thanks to self-consistent simulations.
We note that the corotation torque is negative also in our viscous simulations. 
A negative corotation torque can happen for gap opening planet with low viscosity \citep{Kanagawa+2018}, possibly due to the saturation of the corotation torque (viscous timescale is shorter than $\taulib$; \citealp{Masset+Casoli2010,Paardekooper+2011}).
We also tested that this torque becomes positive in the unsaturated regime (horseshoe drag regime) given the relative steep background density gradient, which can be achieved with a smaller planetary mass or a larger $\alpha$.

Third, while the typical magnitude of the aforementioned torques are of the order $(0.1\text{--}1)\Gamma_0$, their sum always yield a strong negative torque in all our simulations. This is partly due to the planet-free inner gap that leads to the gradual depletion of the inner disk. This fact reduces the outward migration torque, which is then augmented by the negative torque from the corotation region. Consequently, the total torque becomes comparable or even more strongly negative than the outer Lindblad torque (blue).

Finally, we see that the torques from our turbulent windy disk simulations are highly stochastic, with the torques measured at individual snapshots showing orders of magnitude fluctuations. This is in stark contrast with the results from viscous disk simulations, and will likely yield stochastic migration \citep[e.g.,][]{Laughlin+2004}.

In summary, the migration torque in the windy disk tends to be negative while being highly stochastic. We consistently identify a negative torque from the corotation region, while there is additional uncertainty in the planet-free density gap originating from the spontaneous magnetic flux concentration.

\section{Discussion}
\label{sec:Discussion}

\subsection{Implications for observations}
\label{sec:D_obs}

It has become customary to infer the mass of embedded planets by modeling the gap structure in PPDs, where the key is to establish the relationship between the planetary mass and the gap properties (see \citealp{Bae+2022} and \citealp{Paardekooper+2022} for a review). For gas surface density, such relationship has been empirically determined by 2D viscous simulations \citep{Duffell+MacFadyen2013,Fung+2014,Kanagawa+2015,Kanagawa+2016,Kanagawa+2017,Duffell2020} and confirmed by 3D simulations \citep{Fung+Chiang2016, Dong+Fung2017b}. As sub-micron-sized dust is expected to be coupled to the gas, observations at optical and near-infrared wavelengths can be used as a planet mass indicator \citep{Rosotti+2016,Dong+Fung2017b}, and can be combined with upper limits derived from direct imaging to constrain planet formation models \citep[e.g.,][]{Asensio-Torres+2021}. In the meantime, the most-common disk substructures detected in sub-millimeter wavelength reflects the profile millimeter-sized dust that are not well-coupled with the gas. Incorporation of additional dust particles/fluids are needed to interpret the observations \citep[e.g.,][]{Rice+2006, Pinilla+2012, Zhu+2012a, Drazkowska+2016, Taki+2016}, and planet population synthesis \citep{Zhang+2018}.

In this work, we have shown that in the outer disk conditions, the gap in the windy disk is wider and much deeper than the viscous case due to the MHD effects (see \S~\ref{sec:gap_structure}). Therefore, we anticipate that the previous planetary-mass inferred based on modeling the gap structure can be overestimated. 
In other words, smaller-mass planets can potentially open relatively deep gaps to disguise as higher-mass planets. This fact could potentially account for the non-detection of embedded planets by direct imaging \citep[e.g.,][]{Asensio-Torres+2021}.
On the other hand, given that we find the gap width in windy disks is similar to the viscous case, we recommend that without sophisticated simulations, modeling gap structure from inviscid simulations as a reasonable compromise.
Moreover, it is yet to examine whether the same holds for the millimeter-sized dust component, which tends to concentrate towards pressure maxima. Although we measure larger gap width, the separation between the pressure maxima immediately inward and outward of the gap are not necessarily wider by a substantial margin compared with the viscous case (see figures~\ref{fig:GapProf_invicid} and \ref{fig:Mt3AM1}). There are additional complications where the turbulence strength is highly non-uniform across the gap, a situation very different from conventional viscous simulations that adopt constant $\alpha$ viscosity. Further work is needed to address this question.

In addition, the planet-free gap can well co-exist with the planet-induced gap. These gaps could be present at random locations, separated from the planet-induced gap by a few disk scale heights, and can potentially merge with the planet-induced gap when planet mass is large. Given the stochastic nature of the planet-free gaps, we are not in a position to offer a comprehensive comparison between the gap properties, and simply note that more caution should be exercised when interpreting ring-like substructures.

Perhaps the primary way to distinguish planet-free and planet-induced gaps is through the kinematic signatures (see \citealp{Pinte+2022} for a review), particularly velocity kinks \citep{Perez+2015,Perez+2018,Teague+2018,Pinte+2018,Pinte+2019} and meridional flows \citep{Teague+2019,Yu+2021}. We have discussed that the MHD effects change the gas flow structure, and we do not see obvious features of (quasi-axisymmetric) large-scale meridional flows within the parameters space explored here (see \S~\ref{sec:flow}). We anticipate that localized velocity structures are likely the key to distinguish between the two cases, and we will leave more detailed investigations to a follow-up study.

\subsection{The Jupiter barrier in the solar system}
In the solar system context, recent works have established a fundamental isotopic dichotomy between non-carbonaceous (NC) and carbonaceous (CC) meteorite groups. These two groups most likely represent materials from the inner and outer Solar System, respectively, thus implying prolonged spatial separation of inner and outer solar system solid materials \citep{Warren+2011,Kruijer+2017}. While alternative scenarios exist \cite[e.g.][]{Lichtenberg+2021,Liu+2022}, this dichotomy is most easily explained by the early formation and growth of Jupiter (known as the Jupiter barrier): with gap opening, it is anticipated that solids are trapped in pressure bumps on the two sides of its orbits, preventing substantial material exchange \cite[see][for a review]{Kruijer+2020}.

Our work strengthens the scenario of the Jupiter barrier in several aspects. While the formation of a pressure bump outward of the planet orbit is naturally expected, our fiducial simulation shows that, thanks to the presence of the planet-free gap, the system can also form a pressure bumps inward of the planet orbit. Although dust is not included in our simulations, we expect that first, because the planetary gap is much deeper, the outer pressure bump is more effective trapping dust than the viscous and even inviscid counterparts \cite[e.g.,][]{Weber+18}. This could help establish the Jupiter barrier relatively early. Second, the inner pressure bump helps preserve the solids there, preventing them from getting lost through radial drift.

Along the same line, our simulations also share implications to meteorite paleomagnetism, which retrieve the strength of nebular fields from meteorites during their formation \cite[see][for a review]{Weiss+2021}. Very recently, it was discovered that the inferred nebular field strength from the CC meteorites is likely to be stronger than those inferred from the NC meteorite \citep{Borlina+2021,Fu+2021}, whereas in a smooth disk with steady accretion, the mean field strength should rapidly decrease with radius. 
From figure~\ref{fig:rzslice}, we see that the field strength in the midplane region (dominated by the $B_\phi$ component) is relatively low in the inner density bump (right inward of the planetary gap), due to the deficit of magnetic flux.
On the other hand, the field strength outward of the gap can indeed be higher, thus naturally explaining the measurement results. This does not contradict the fact that the disk accretion rate remains smooth across this region, because the field strength in the inner bump is larger at higher altitude to sustain disk accretion.

It should be noted that our earlier study in \citet{Cui+Bai2021} found that without a planet, the field strength in the midplane region is relatively smooth even across the (planet-free) gaps with magnetic flux concentration. Therefore, our results suggest that the presence of a giant planet is likely essential to cause this field strength reversal on the two sides of the gap region. On the other hand, we also caution that our simulations mainly target the outer PPDs in the ambipolar diffusion dominated regime ($\gtrsim10-20$AU). Further works are needed to assess this scenario towards inner regions where the Ohmic resistivity and the Hall effect are equally important \cite[e.g.][]{Bai2017}.

\subsection{A simple prescription of MHD torques for gap opening}
\label{sec:illustration}

\begin{figure}
    \centering
    \includegraphics[width=8.55cm]{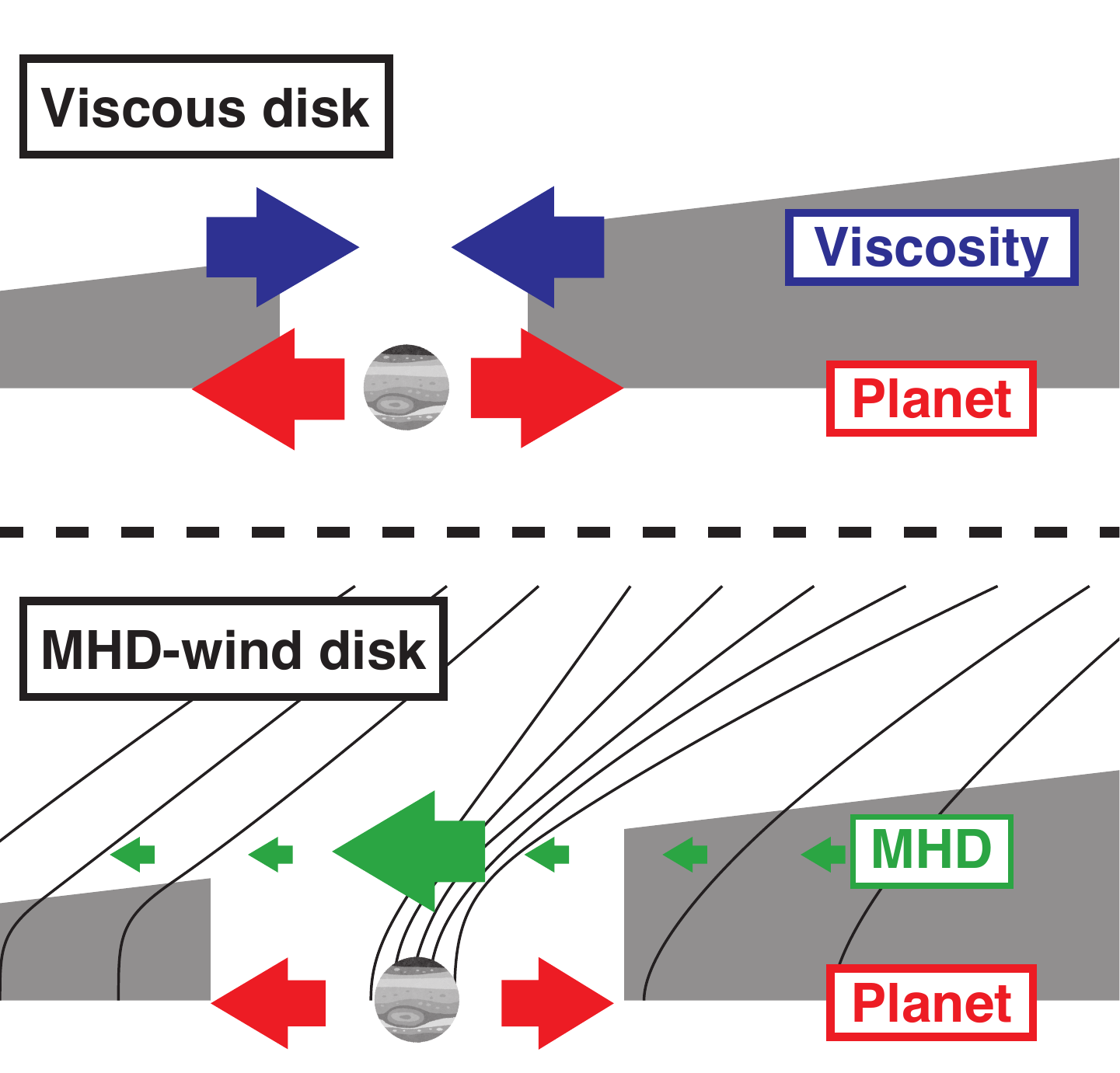}
    \caption{Sketch of the torques around the type-II planetary gap. The planetary torque tends to open the gap in both the viscous and windy disks. In the viscous accretion disk, the viscous torque tends to fill the gap back and balance to the planetary torque. In the windy disk, besides the standard wind torque that drives disk accretion, magnetic flux concentration in the gap region makes both the wind and turbulence inhomogeneous across the planetary gap. On the other hand, the net effect of the MHD torques can be simply represented as an enhanced wind torque in the gap region. This motivates us to provide a simple prescription of MHD torques in Section \ref{sec:illustration}.
    }
    \label{fig:sum_tor}
\end{figure}

Prior to this work, there are a handful studies of planet-disk interaction in wind-driven disks, which all adopt the effect of the disk wind prescribed as an additional wind torque, in both the type-I regime \citep{Ogihara+2015, Ogihara+2018,McNally+2020} and the type-II regime \citep{Kimmig+2020,Elbakyan+2022}. Such prescriptions usually assume that the wind acts uniformly across the entire disk. While it captures the basic effect of disk wind in angular momentum transport, when applied to the type-II regime, it ignores the fact that gap formation would redistribute magnetic flux and hence wind properties. Our simulations provide new insights, from which we can provide an updated prescription of MHD torques in type-II planet-disk interactions.

Figure~\ref{fig:sum_tor} shows the schematic understanding of the planet-disk interaction in the windy disk, compared to the viscous case. 
In the viscous case, it is well known that the planetary torque pushes to open the gap, while the viscous torque attempts to close the gap, leading to a steady gap profile \citep{Lin+Papaloizou1993}. In the windy disk case, despite of the complications from magnetic flux concentration, asymmetric wind-launching, highly enhanced $\alpha$ within the gap, etc., we have seen in figure~\ref{fig:AMF} that the cumulative effect is relatively simple (see more detailed analysis in Section \ref{sec:AMT_MHD}): the cumulative torque generally exhibit one slope outside of the corotation region, representing standard wind-driven accretion, while exhibit a steeper slope in the corotation region, reflecting the effect of magnetic flux concentration with an enhanced torque.
These findings lead to the following simple prescription of the MHD torque densities
\begin{align}
\gamma_\mathrm{MHD} &= \gamma_\mathrm{wind,0}& (|R-\RP|> \rH) \nonumber \\
&= \gamma_\mathrm{E} & (|R-\RP|<\rH),
\end{align}
where $\gamma_\mathrm{wind,0}$ is the unperturbed disk-wind torque density and $\gamma_\mathrm{E}$ is the enhanced MHD torque density.
Here, we can set the radial width over which magnetic flux concentrates to be $2r_H$ (see Section \ref{sec:MFC}).
The value of the torques depends on the parameters, where we can set $\gamma_{\rm wind,0}\approx(j/2r)\Macc$ to be directly related to the unperturbed wind-driven accretion rate, and $\gamma_E\approx3\text{--}5\gamma_{\rm wind,0}$ due to magnetic flux concentration, as found from our simulations.

We emphasize that this prescription is highly approximate in nature, and mainly aims to capture the leading-order effect of the MHD torques in windy disks. It does not account for additional magnetic flux concentration in the planet-free region, and the specific ratio between $\gamma_E$ and $\gamma_{\rm wind,0}$ can be parameter-dependent and random. Nor does this simple prescription account for the asymmetric distribution of the torque in the $\phi-$direction, which could be important for planet migration, though it can in principle be incorporated by more detailed treatments.
In addition, when applying this prescription, certain smoothing between $|R-\RP|<\rH$ and $|R-\RP|>\rH$ regions are needed to avoid abrupt changes.
Finally, away from the planetary region, it is possible to incorporate a background viscosity to account for background turbulence, e.g., with constant $\alpha$. Note that the strong enhancement of $\alpha^{\rm Max}$ in the gap region is already accounted for in the prescription.

\section{Summary and Conclusions}
\label{sec:conclusion}

In this paper, we have conducted 3D global non-ideal MHD (with ambipolar diffusion) simulations of type-II planet-disk interaction relevant in the outer PPDs. The simulations self-consistently captures the launching of magnetized disk winds, together with sufficient resolution to resolve the MRI turbulence. We consider planet mass from $1\text{--}5$ thermal mass $\Mth$, and ambipolar Elsasser number $Am=3$ and $Am=1$. Our main findings are summarized as follows:
\begin{itemize}
    \item The planet triggers the concentration of the poloidal magnetic flux at the corotation region, with mean field stronger than background by a factor of $2\text{--}4$. This process is likely associated with the spiral density shocks at the disk surface and is not axisymmetric. The disk can exhibit additional (likely stochastic) concentration of magnetic flux (which is quasi-axisymmetric) that produces planet-free gaps.
    \item With magnetic flux concentration, the magnetized wind and disk turbulence becomes highly inhomogeneous, with strongly enhanced angular momentum extraction from the gap region along radially inclined poloidal field. Correspondingly, the outflow primarily originates from the disk surface outward of the gap, rather than from the gap region itself.
    \item The inhomogeneous wind strongly alters the torque balance for gap opening, making the gap much deeper than the viscous and inviscid cases. The gap widths are modestly wider than the viscous case and are similar to the inviscid case. The net effect from MHD on gap opening can be understood from a simple wind torque prescription that is enhanced in the corotation region.
    \item There is strong accretion within the corotation region ($|R-\RP|<\rH$), with mean accretion velocity approaching the sound speed, breaking the horseshoe orbit. The radial flow is asymmetric in azimuth, and there is outward radial flow in the trailing side of the planet due to local reversal of vertical field. Large-scale meridional flow is also lacking or weakened.
    \item The migration torque acting on the planet is generally negative and stochastically fluctuating. In particular, the torque from the corotation region is always found to be negative due to the specific aforementioned radial flow structures. As the gap in windy disks is wider, the Lindblad torques on both the inner and outer sides are reduced relative to the viscous case, which also depends on the (less-predictable) presence of neighboring planet-free gaps.
\end{itemize}

The results obtained here imply that caution must be exercised in modeling the gap profile to infer planet mass from disk observations. In particular, due to magnetic flux concentration, gap formation does not necessarily requires the presence of planets. In the presence of planets, they can open deeper and wider gaps than previously thought, thus planetary masses inferred by modeling gap widths tend to overestimate true planet masses. This fact can at least be partially responsible for the deficiency in detecting embedded planets by direct imaging. We anticipate that asymmetric kinematic signatures in the gap region are the key to identifying hidden planets and distinguish them from planet-free gaps.

The deep planetary gap, and the presence of pressure traps on both sides of the planet orbit (thanks to the planet-free magnetic flux concentration), also support the early formation of Jupiter in the solar system as a barrier that leads to the isotopic dichotomy between NC and CC meteorites. Our results are also consistent with the paleomagnetic constraints of nebular field strength where field strength inferred from CC materials appears stronger than that from the NC group.

As a first study of planet-disk interaction in turbulent windy disks, our simulations are simplified in a number of aspects. In particular, while we assumed constant ambipolar Elsasser number in the entire bulk disk, the ionization state in the gap region can be different, and further works are needed to assess the strength of non-ideal MHD effects in the gaps and how it affects magnetic flux concentration and local gap properties. 
We have also fixed disk temperature and assumed a locally isothermal equation of state, which is a reasonable approximation for a full disk, but can be inaccurate when the planet opens a gap due to shadowing \citep[e.g.,][]{Jang-Condell+Turner2013,Hallam+Paardekooper2018,Chrenko+Nesvorny2020}. There is additional heating from the spiral shock and planetary accretion, which can be especially important for understanding circumplanetary disk formation \citep[e.g.,][]{Szulagyi17,Szulagyi+17}.
Finally, planet-disk interaction is expected to leave observable signatures on the dust \cite[e.g.][]{Bi+2021,Szulagyi+2022}, and it is yet to investigate how the presence of magnetized winds affect the dust dynamics and observational signatures by incorporating dust components to our simulations \citep{Huang+2022}.

\software{Athena++(\citet{Athena++2020}),
Matplotlib (\citet{Hunter2007}),
Numpy (\citet{Harris+2020}),
Mayavi (\citet{mayavi}),
Python3 (\citet{Python3})
}

\begin{acknowledgments}
We thank Can Cui for sharing her problem setting and Ruobing Dong for helpful discussions on planet-induced flow structures, and the anonymous referee for a constructive report. This work is supported by the National Science Foundation of China under grant No. 12233004, and the
China Manned Space Project, with No. CMS-CSST-2021-B09. Numerical simulations are conducted on TianHe-1 (A) at National Supercomputer Center in Tianjin, China, and on the Orion Cluster at the Department of Astronomy, Tsinghua University.
\end{acknowledgments}

\appendix
\section{Implementing the rotating frame} \label{sec:rot}

In the frame corotating at the planet orbital frequency $\Omega_P$, the momentum equation is supplemented with centrifugal and Coriolis source terms and becomes
\begin{equation}
\frac{\pa\rho{\mb v}}{\pa t}+\nabla\cdot{\sf M}
=-\rho\nabla\Phi+\rho\Omega_P^2R{\mb e}_R-2\rho\Omega_P{\mb e}_z\times{\mb v}\ ,
\end{equation}
With this formulation, the rotational velocity increase rapidly at large radial distances $v_\phi\sim\Omega_PR$ (at $R\gg\RP$), which leads to large advection error and breaks angular momentum conservation. Here we follow \citet{Kley1998} to reformulate the equations into a conservative form that guarantees angular momentum conservation and reduces advection error.
In doing so, we change variables as
\begin{equation}
v'_\phi\equiv v_\phi + \Omega_P R\ ,
\end{equation}
which is the rotational velocity in non-rotating frame.
For the $\phi-$momentum equation, it is straightforward to rewrite it as an equation for angular momentum as
\begin{equation}\label{eq:am}
\frac{\pa(\rho Rv'_\phi)}{\pa t}+(R\nabla\cdot{\sf M})_\phi
+\nabla\cdot(\rho{\mb v}\Omega_P R^2)=-\rho R(\nabla\Phi)|_\phi\ ,
\end{equation}
where we have used
\begin{equation}
\frac{\pa(\rho\Omega_P R)}{\pa t}=-\Omega_P R\nabla\cdot(\rho{\mb v})\ ,\quad
\nabla\cdot(\rho{\mb v}\Omega_P R^2) = \Omega_P R^2\nabla\cdot(\rho{\mb v})+2\rho\Omega_P Rv_R\ .
\end{equation}

At implementation level, the primitive variables are chosen to be those in the rotating frame, whereas we use the primed quantities (i.e., in the inertial frame) for conserved variables. The reconstruction step, which uses primitive variables, remain unchanged, as well as for the Riemann solver, which yields the momentum and energy fluxes in the corotating frame. 

We note that in Athena++, although it solves the momentum equation in linear momentum not in angular momentum (in both the $\hat{\theta}$ and $\hat{\phi}$ directions), it contains carefully-designed geometric source terms to ensure exact angular momentum conservation (see, e.g., \citealp{Ju2016} in cylindrical coordinates). In spherical polar coordinates, ensuring angular momentum conservation in the $R$ and $\theta$ directions requires the $r$ and $\theta$ factors in $r\rho v_\theta$ and $r\sin\theta\rho v_\phi$ at individual grid points to be interpreted as $\bar{r}_i=(r_{i-1/2}+r_{i+1/2})/2$, and $\overline{\sin\theta}_j=(\sin\theta_{j-1/2}+\sin\theta_{j+1/2})/2$. When extending to the rotating frame, it turns out that in the $\hat{r}$ and $\hat{\theta}$ directions, it suffices implementing the source terms simply requires making a substitution in $M_{\phi\phi}$ in the geometric source terms by replacing $\rho v_\phi^2$ by $\rho(v_\phi+\Omega_P R)^2$.
In the $\hat{\phi}]$ direction, we need to incorporate the additional flux term as shown in Equation (\ref{eq:am}). In doing so, we calculate the additional flux term $\rho{\mb v}\Omega_PR^2$ based on the mass flux $\rho{\mb v}$ obtained from the Riemann solver at cell interfaces, and integrate this flux multiplied by $R^2$ over the area at cell interfaces. The results are then divided by $\bar{R}_{ij}=\bar{r}_i\overline{\sin\theta}_j$ after applying flux divergence.

For completeness, for the energy equation (which is solved although not practically used in this work as we enforce a locally-isothermal equation of state), we can redefine the energy density in the rotating frame $E'$ (where the only difference is in the kinetic energy), and arrive at
\begin{equation}
\frac{\pa E'}{\pa t}-\Omega_P \frac{\pa(\rho Rv'_\phi)}{\pa t}+\nabla\cdot\bigg[\bigg(E+P^*\bigg){\mb v}
-\frac{1}{4\pi}{\mb B}({\mb B}\cdot{\mb v})-\frac{1}{2}\rho{\mb v}\Omega_P^2R^2\bigg]
=-\rho(\nabla\psi)\cdot{\mb v}\ .\label{eq:eng}
\end{equation}
The additional flux term is calculated the same way as in the $\phi-$ momentum equation. We further record the change in $\phi-$momentum in each stage account for the second term above.

\section{Mechanism of magnetic flux concentration}
\label{sec:M_MFC}
\begin{figure*}
    \centering
    \includegraphics[width=\hsize]{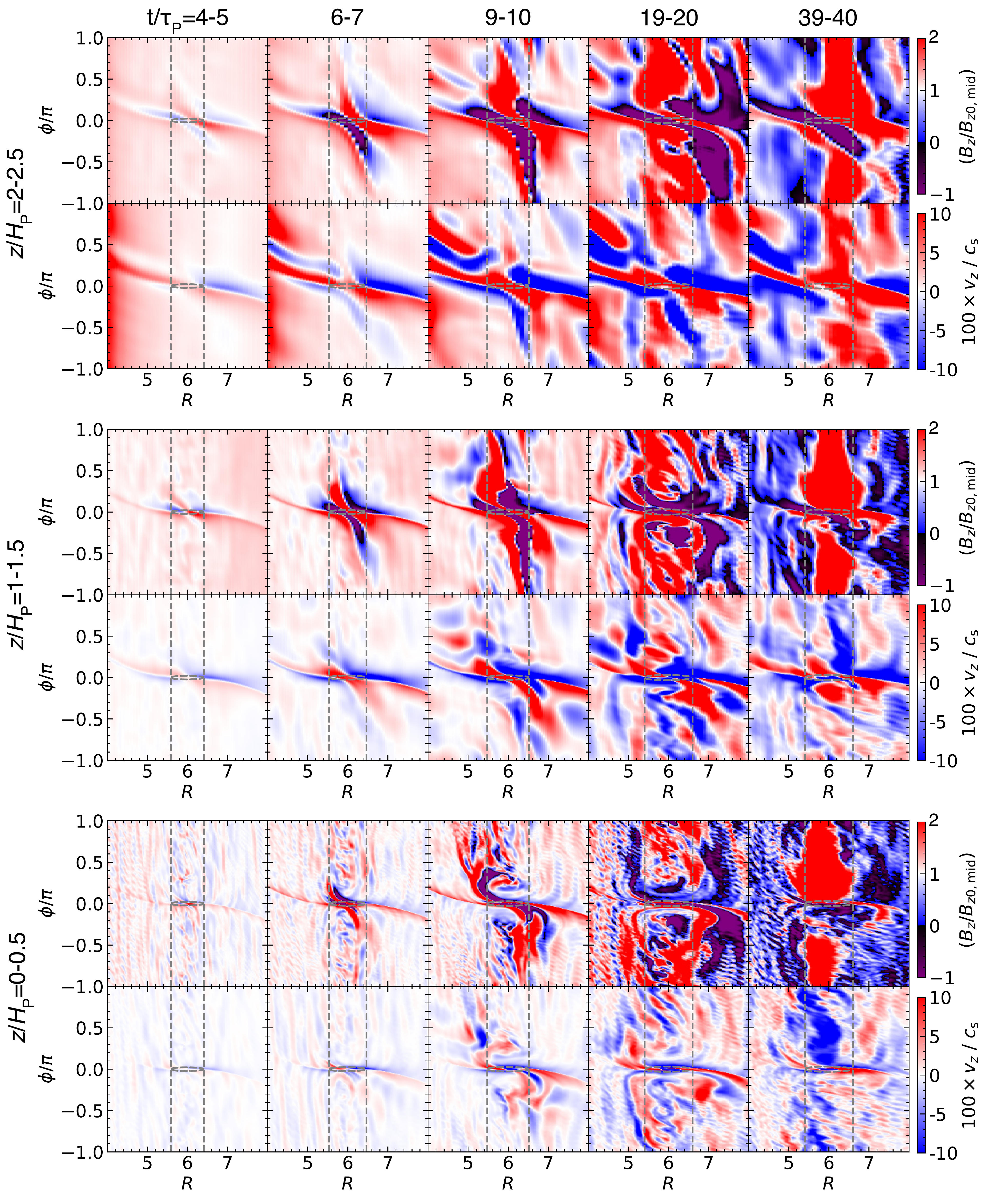}
    \caption{
    Vertical magnetic field ($B_z$, upper) and vertical velocity ($v_z$, lower) averaged over $z/H_\mathrm{P}=0\text{--}0.5$ (bottom), $1\text{--}1.5$ (middle), and $2\text{--}2.5$ (top) from run Mt3Am3 at $t/\tauP=4\text{--}5$, 6--7, and 9--10, 19--20, and 39--40 from left to right, respectively.
    The grey dashed lines and curves shows $R=\RP\pm\rH$ and $|\bvec{R}-\bvec{\RP}|=\rH$ in the $R-\phi$ plane, respectively.
    Here, only the upper hemisphere is shown, and the lower hemisphere largely exhibits similar structures.
    }
    \label{fig:MFC_vertical}
\end{figure*}

In the disk without planet, the spontaneous concentration of magnetic flux has been found in the MHD simulations with net poloidal magnetic flux. Several mechanisms have been proposed to explain this phenomenon, including reconnection of recurrent MRI channel flows \citep{Bai+Stone2014}, reconnection driven by the midplane wind-driven accretion flow \citep{Suriano+2018,Suriano+2019}, and a wind instability \citep{Riols+Lesur2019}. Nevertheless, the actual mechanism responsible for magnetic flux concentration in windy disks with full MRI turbulence remains ambiguous \citep{Cui+Bai2021}, partly due to the stochastic nature of the process.

In the presence of a planet, we have found that magnetic flux concentration in planet-induced gaps is robust rather than being fully stochastic. This suggests that the planet induces additional flow structures that likely responsible for flux concentration. While a detailed understanding is beyond the scope of this work, we identify certain clues that likely play a key role in the flux concentration process in planetary gaps.
We have already seen in Section \ref{sec:MFC} that the azimuthal distribution of magnetic flux is not smooth, and $B_z$ can become negative in the trailing side of the planet, likely associated with planet-induced flows. Here, we show in figure~\ref{fig:MFC_vertical} snapshots of vertical magnetic field ($B_z$, upper) and vertical velocity ($v_z$, lower) at various heights about the midplane, focusing on early stages of magnetic flux concentration.

We see that features of magnetic flux concentration occurs first at the disk upper layer $z\gtrsim H_\mathrm{P}$, associated with the planetary density wave. In particular, there is undulating vertical velocity across the spiral density shock, which is stronger towards the disk surface. As the toroidal field is the dominant field component, this undulating vertical flows then creates the positive and negative $B_z$ regions around the planet (see the upper panels at $t=6\text{--}7\tau_P$). Such undulating flow pattern persists all the way throughout the duration of the simulation, even in the presence of strong turbulence at later time, so does the negative $B_z$ region in the trailing side of the planet.

The magnetic structures formed in the early phases appear to propagate in both the leading and trailing directions in azimuth, which in the meantime gather additional poloidal magnetic flux from the neighborhood. Again, this process proceeds first from the disk upper layer (around $t\sim20\tau_P$), and the flux gets concentrated in the midplane region after about $40\tau_P$. We have identified that this process is robust in all of our simulation runs, including run Mt3Am1 with $Am=1$. Therefore, it is likely that planet-induced flows play a crucial role in gathering magnetic flux into the gap region, which in turn affects the gap dynamics and planet migration.

\section{Vortex formation by the RWI}
\label{sec:vortex}
\begin{figure}
        \includegraphics[width=\Ft]{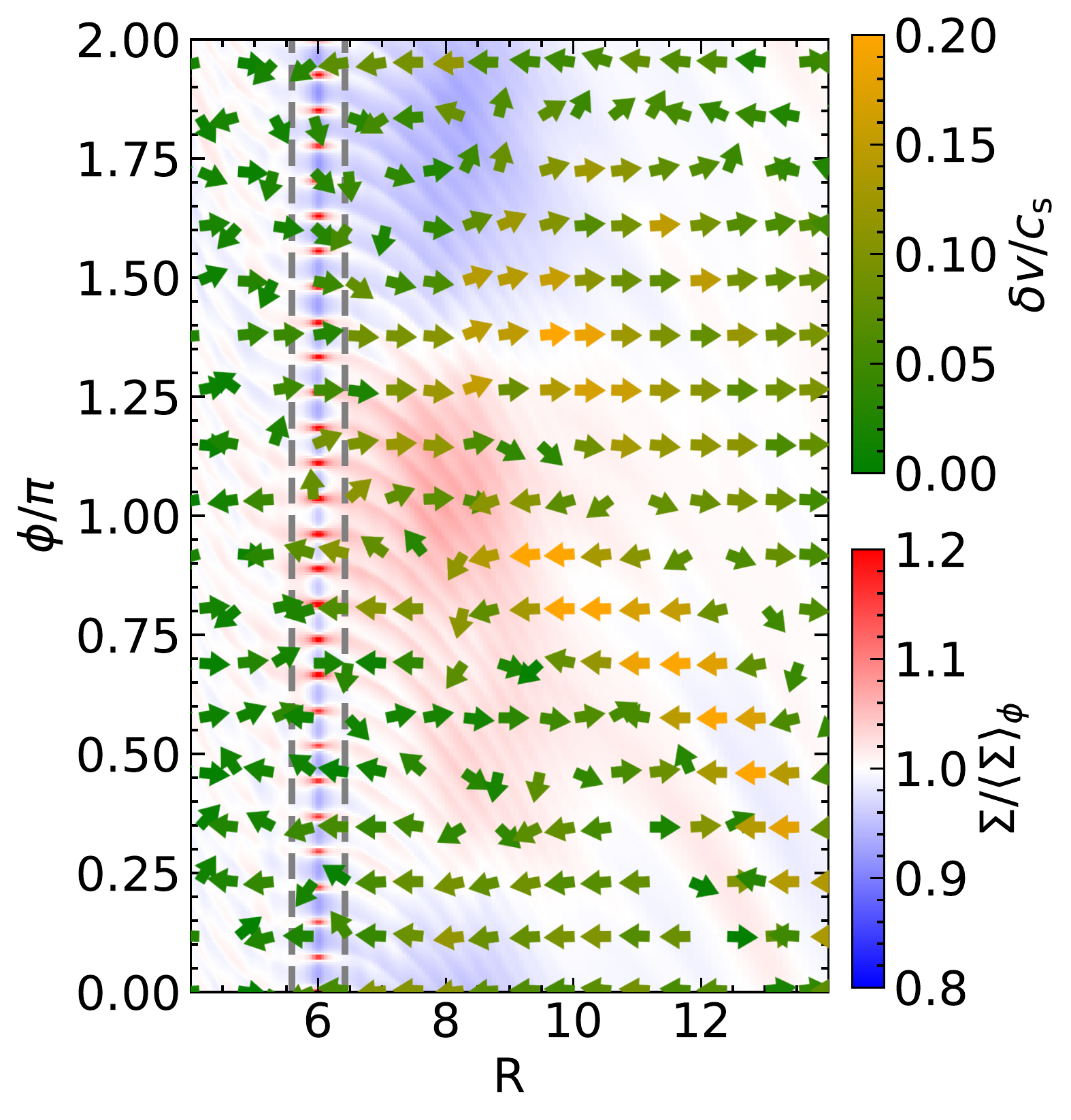}
        \includegraphics[width=\Ft]{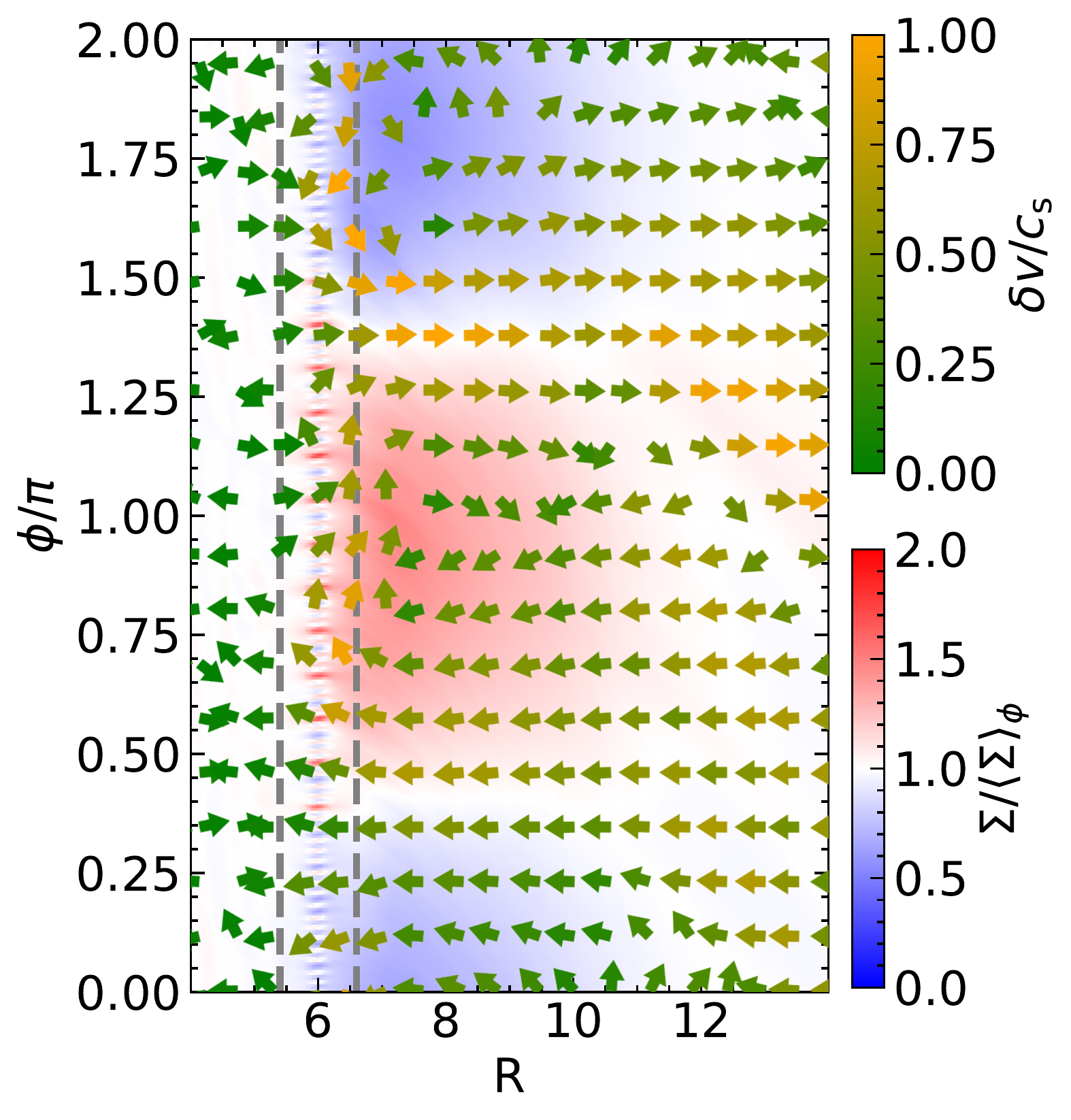}
        \includegraphics[width=\Ft]{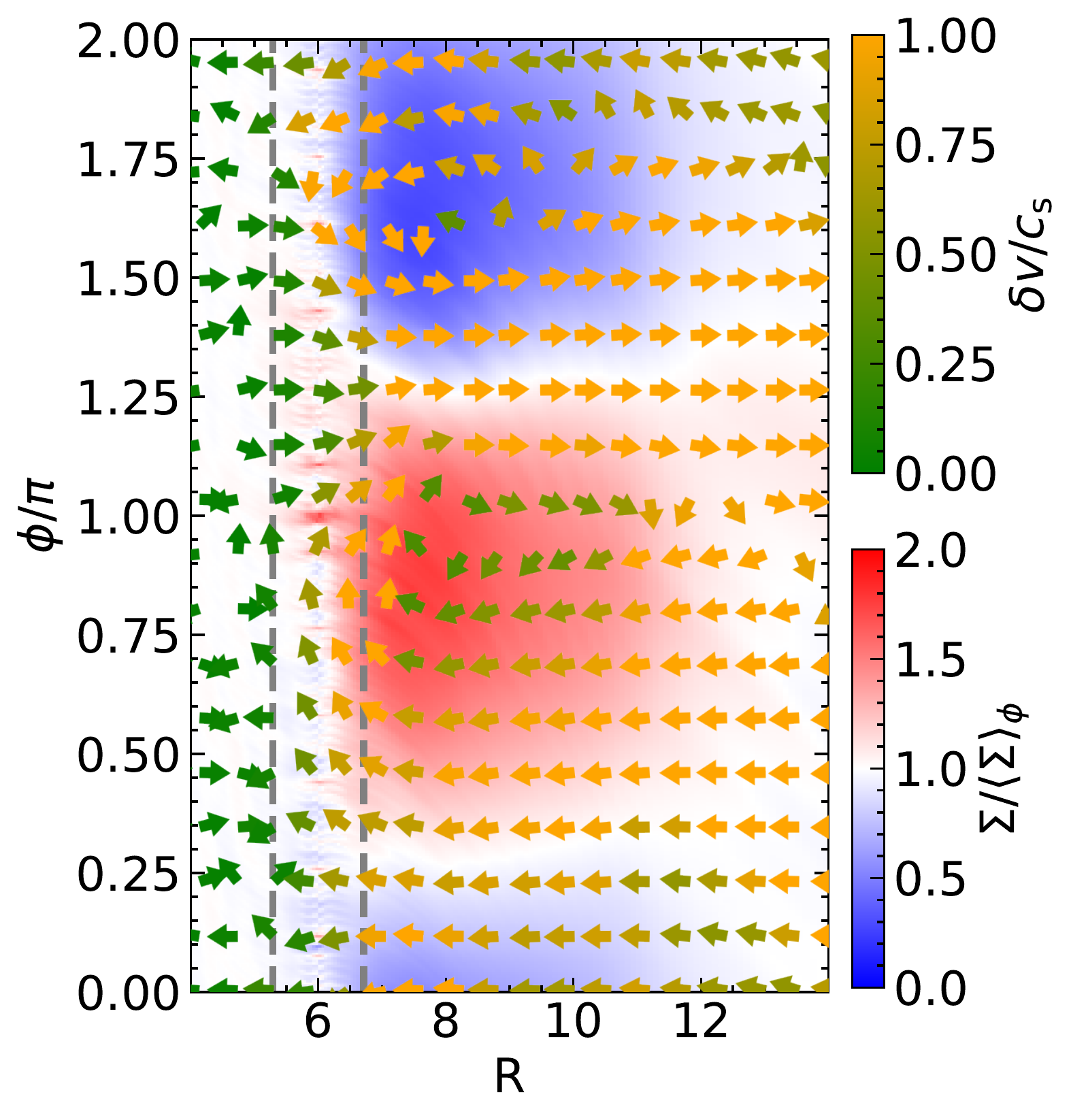}
    \caption{Surface density normalized by the $\phi$-averaged values and cylindrical velocity deviation $\delta \bvec{v} = \bvec{v}_\mathrm{cyl} - \av{\bvec{v}_\mathrm{cyl}}_\phi$ where
    $\bvec{v}_\mathrm{cyl}=\Sigma^{-1}\int dz \rho (\bvec{v}_r \sin{\theta}+ \bvec{v}_\theta \cos{\theta} + \bvec{v}_\phi)$. The time average is taken over $40\text{--}50~\tauP$ with shifting $\phi$ at the vortex phase angular velocity of -0.37, -0.46, and $-0.51 ~\OmegaP$ for runs Mt1Am3, Mt3Am3, and Mt5Am3 in the corotating frame of the plaenet. The grey dashed lines mark $R=\RP\pm\rH$.
    }
    \label{fig:VSD}
\end{figure}

\begin{figure*}
    \centering
    \includegraphics[width=\hsize]{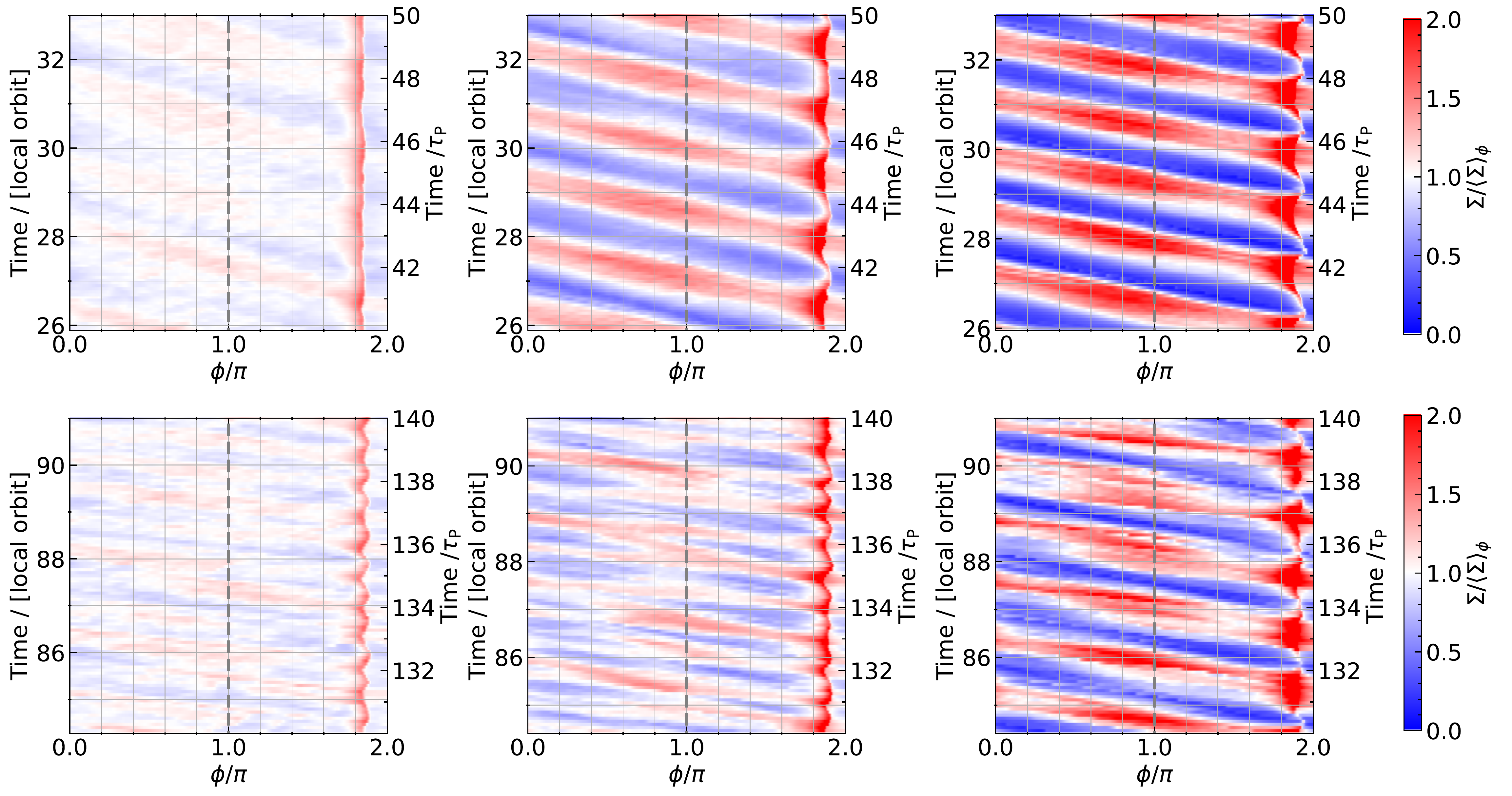}
    \caption{Surface density normalized by the $\phi$-averaged value at $R=8$, shown in $\phi$-time plane. From left to right, each panel shows the case of Mt1Am3, Mt3Am3, and Mt5Am3, respectively. 
    The upper and lower panels show the time from 40 to $50~\tauP$ and from 130 to $140~\tauP$, respectively. 
    }
    \label{fig:tpV50}
\end{figure*}

The planetary density gap generates a pressure bump outward the gap, which is known to be subject to the Rossby wave instability and leads to the formation of large anti-cyclonic vortices outward of the planet orbit \citep[so-called RWI; e.g.][]{Lovelace1999L,Li+2005,Ono+2016}. Such vortices can trap dust, and the RWI is has been widely employed to interpret azimuthal asymmetries observed in PPDs, especially transition disks \citep[e.g.,][]{Casassus+2013,Isella+2013, vanderMarel+2013,Perez+2014}. The RWI has been extensively studied by numerical simulations over the recent years, primarily by means of viscous hydrodynamic simulations with or without planet \citep[e.g.,][]{devalBorro+2007, Meheut+2012a,Fu+2014a,Ono+2018,Huang+2018a}.
Generally, the system first produces multiple vortices which then merge into one large and azimuthally-elongated vortex. The vortex then gradually decay over a finite lifetime depending on the disk conditions \citep[thickness, viscosity, planet growth, etc.,][]{Hammer+2021}.

Studies of the RWI in the MHD context are relatively sparse \citep[e.g.,][in the context of the dead zones]{Lyra+MacLow2012,Flock+2015}. Generally, weak poloidal fields do not strongly affect its linear properites \citep{Yu+Lai2013}. With magnetic field, unstratified simulations by \citet{Zhu+Stone2014} showed that the RWI is suppressed in the ideal MHD simulations while vortex forms when incorporating ambipolar diffusion. Nevertheless, the RWI has not been studied in the presence MHD wind, in addition to turbulence and non-ideal MHD effects.

In our simulations, figure~\ref{fig:GapProf} shows that, other than the density shocks, there are additional azimuthal density variations outward of the gap, which results in a pressure maximum representing an anti-cyclonic vortex. To better visualize the vortex, we further show in figure~\ref{fig:VSD} the surface density normalized by the $\phi$-averaged surface density in $r$-$\phi$ plane. We additionally averaged the data over ten planetary orbits while accounting for the motion of the vortex (based on the phase speed measured from figure~\ref{fig:tpV50}). Due to different phase speeds, the influence of the spiral density waves/shocks is largely suppressed (but not completely, especially in run Mt1Am3 with our finite number of data outputs). We clearly see an $m=1$ azimuthal asymmetry beyond the planet orbit up to $R\sim10$ in all case, with density contrast increasing with increasing planet mass, ranging from $\pm10\%$ for $\MP=\Mth$ to nearly $100\%$ for $\MP=5\Mth$ higher density relative to the average at the orbit. With the time average, it also allows us to clearly visualize the mean flow structure, which clearly shows that the vortex has anti-cyclonic rotation after subtracting background motion.

Figure~\ref{fig:tpV50} shows the temporal evolution of surface density in $t$-$\phi$ plane at $R=8$. 
The density is normalized by the $\phi$-averaged value $\av{\Sigma}_\phi$ at each time, so as to compensate the evolution of background density profile during gap opening.
Note that the high density band at $\phi\sim1.9\pi$ corresponds to the planetary spiral shock.
The angular velocity of the vortex is measured as $0.63$, $0.54$, and $0.49~\OmegaP$ for Mt1Am3, Mt3Am3, and Mt5Am3, respectively, in the non-rotating frame. Note that the $\phi$-coordinate co-rotates with the planet, and the vortex has a retrograde rotation in this co-rotating frame.
The corotation radius is measured as $8.1$ (Mt1Am3), $9.0$ (Mt3Am3), and $9.6$ (Mt5Am3), which are consistent with the local minimum of the background vortensity, as expected \citep[e.g.,][]{Ono+2016}. This corotation radius is often located close to the local density maximum of the vortex \citep[e.g.,][]{Ono+2018}. In figure~\ref{fig:VSD}, we can identify the vortex center as the location where the radial density profile peaks, which are at $R\sim7.5$ and $8$ for runs Mt3Am3 and Mt5Am3, respectively.
We speculate that the difference with the corotation radii likely reflects the influence from additional physics, especially the launching of disk winds.

In our simulations, we note that the value of $\alpha$ at the vortex center is of the order of $10^{-2}$ (see figure~\ref{fig:alp}), corresponding to modestly strong turbulence. The level of turbulence is even larger towards the planet. In conventional viscous simulations, the formation and survival of vortices becomes significantly degraded when $\alpha\gtrsim10^{-3}$, with lifetime less than 100 local orbits \citep[e.g.,][]{Rometsch+2021}.
Therefore, the threshold of turbulence level to trigger and sustain the RWI in our simulations is higher. The exact threshold could depend on multiple factors, where with $\MP=1\Mth$, the presence of an RWI vortex is only marginal. We also see that the amplitude of the vortices in all three runs generally becomes smaller at later time, suggestive of finite lifetime while the lifetime is already much longer than the viscous case (if the RWI exists). Generally, we expect that these results are related to the fact that the density gradient of the planetary gap in our simulations is steeper than that in the viscous disk counterpart due to magnetic flux concentration, making it more prone to the RWI.

\section{Simulation with $\AM=1$}
\label{sec:Am1}

\begin{figure*}
    \centering
    \includegraphics[width=\hsize]{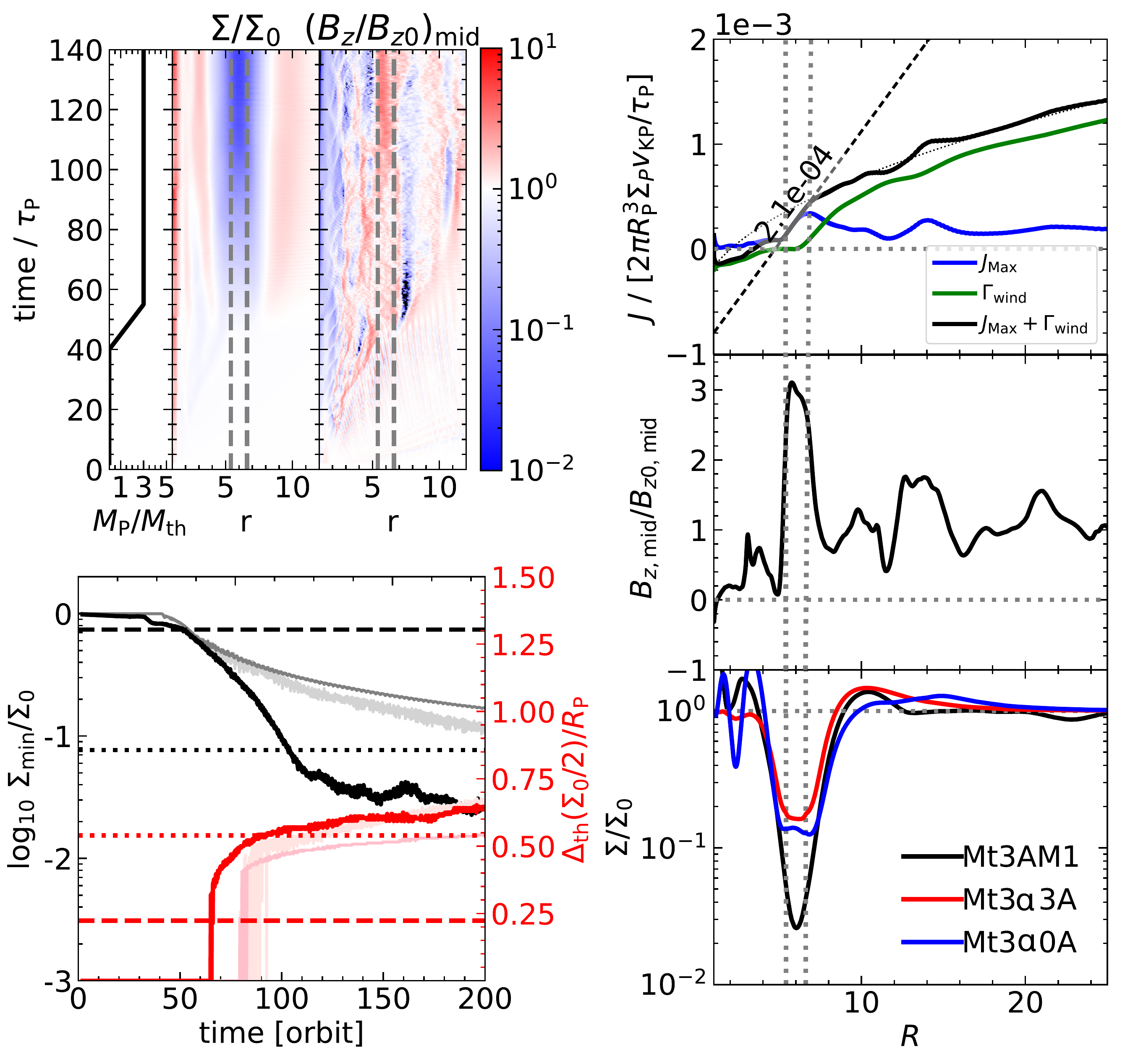}
    \caption{Same as figures~\ref{fig:trSB} (upper left), \ref{fig:gap_time} (lower left), \ref{fig:AMF} (top right), and \ref{fig:GapProf_invicid} (bottom right), respectively, but for Mt3Am1. The right middle panel shows the time-averaged $\Bzm$. The right panels are averaged over $190<t/\tauP<200$.
    In the lower left panel, the dotted and dashed lines correspond to the predictions with $\alpha=3\times10^{-3}$ and $1.0\times10^{-1}$, respectively.
    In the top right panel, the $\JMHD$ profile outward of the gap corresponds to
    $3.9\sqrt{r}\times10^{-4}+C$ shown in dotted lines.
    }
    \label{fig:Mt3AM1}
\end{figure*}

\begin{figure*}
    \centering 
    \includegraphics[width=\hsize]{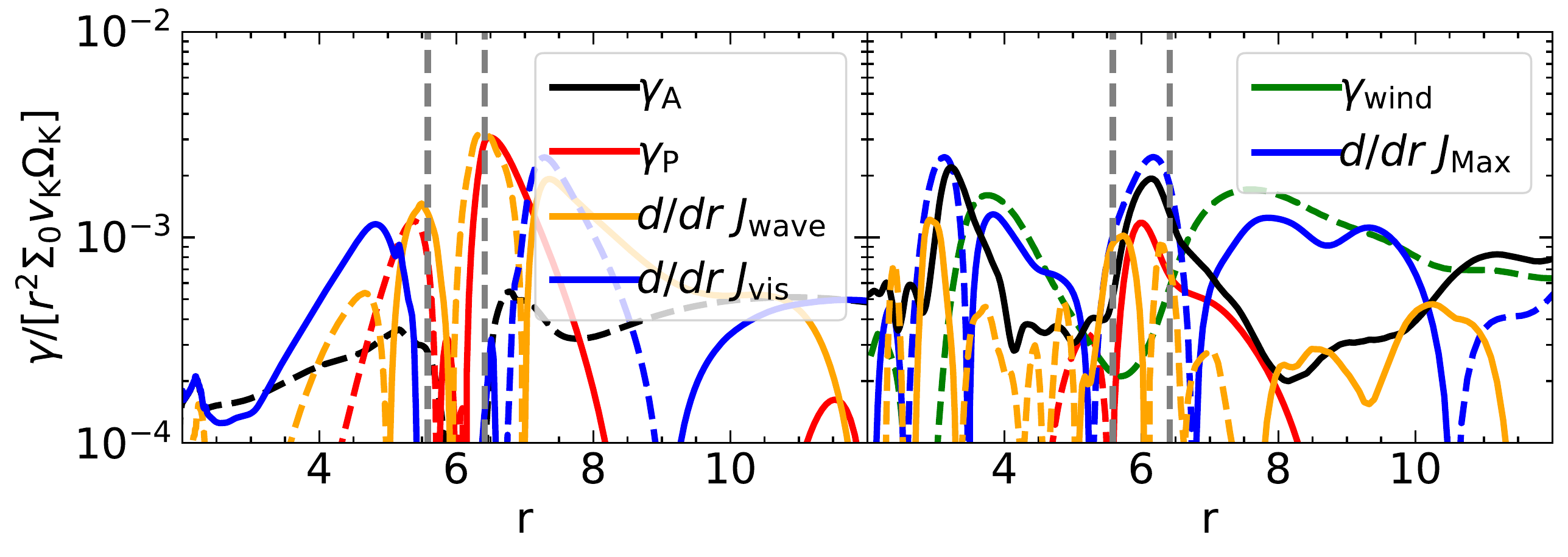}
     \includegraphics[width=\hsize]{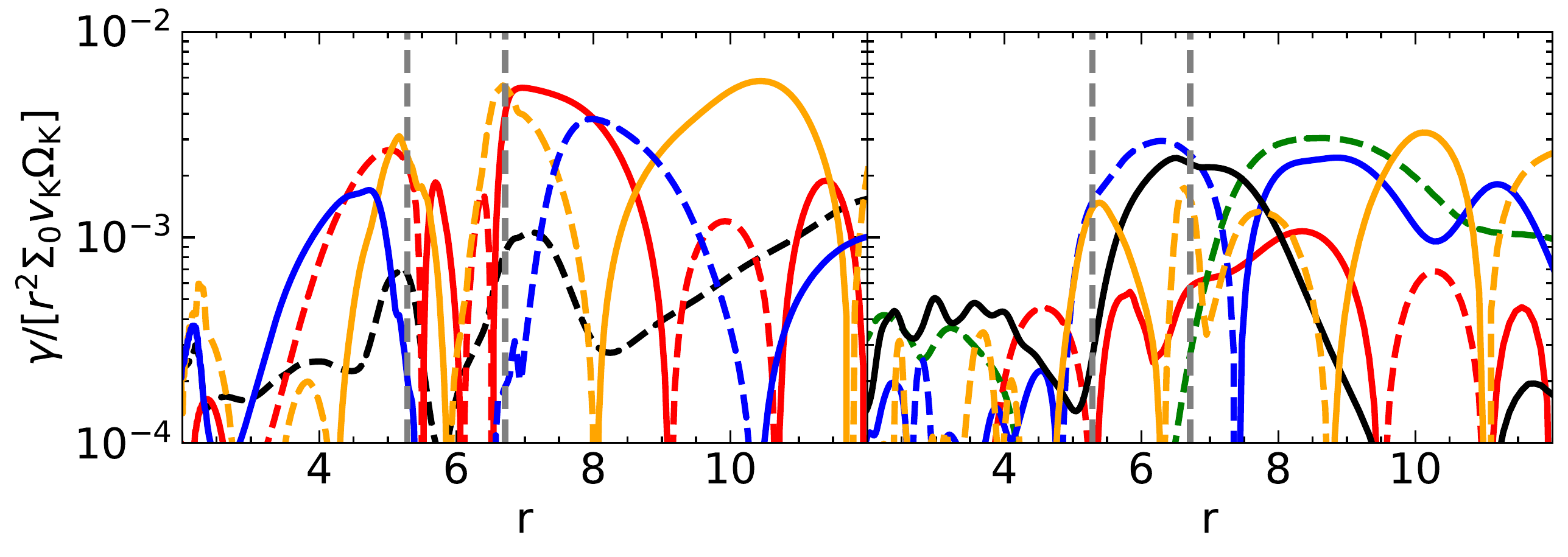}
    \includegraphics[width=\hsize]{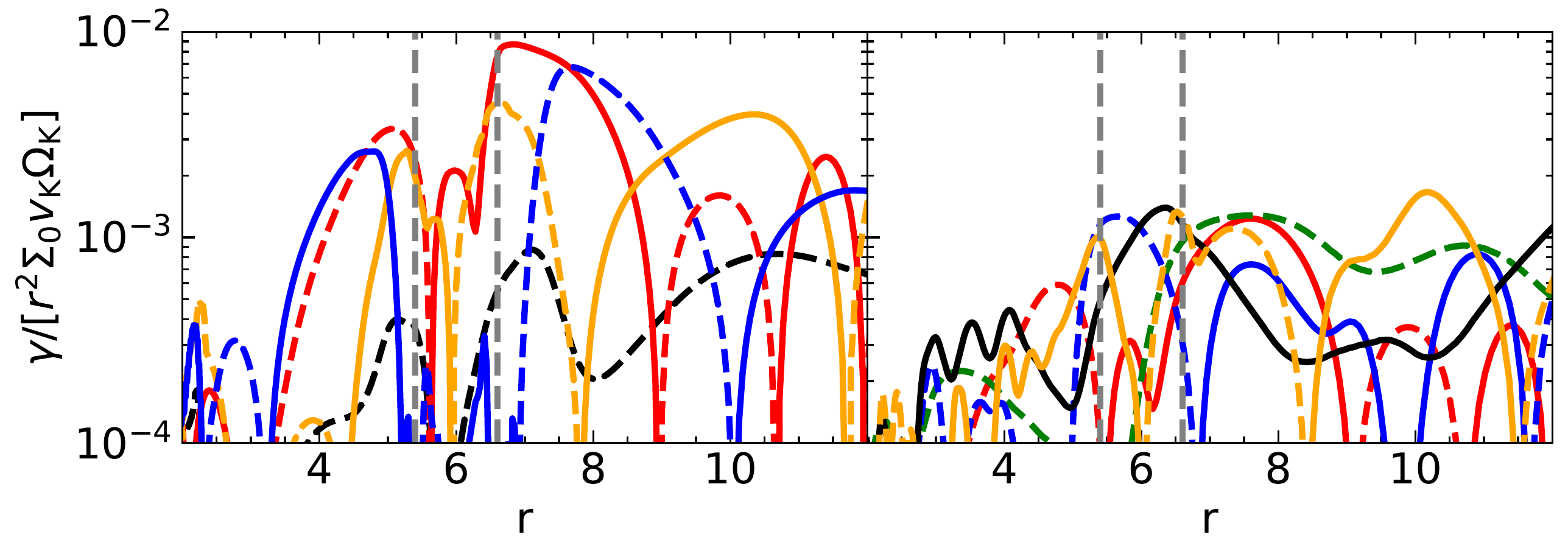}
    \caption{
    Same as figure~\ref{fig:tor} but for Mt1Am3 (upper),  Mt5Am3 (middle), and Mt3Am1 (lower), respectively.
    }
    \label{fig:tor_15M}
\end{figure*}

The three simulations of Mt1Am3, Mt3Am3, and Mt5Am3, which we discussed in \S\S~\ref{sec:Disk} and \ref{sec:GOM}, share the same initial condition and evolution path before inserting the planets.
To validate our simulation results and to expand the parameter space, we perform one additional simulation of Mt3Am1. The differences from the fiducial run of Mt3Am3 include a different $\AMzero=1$, later planet mass insertion $t_\mathrm{P0}=40~\tauP$, and a flatter surface density profile $q_\rho=2$ (see Table~\ref{tab:param}).
The most important difference is the $\AMzero$, where the smaller $\AM$ means the stronger ambipolar diffusion, which will weaken the MRI turbulence \citep{Bai+Stone2011,Cui+Bai2021}.
Next, the larger $t_\mathrm{P0}$ allows the MRI turbulence to fully develop before introducing the planet, and it also allows for the initial development of magnetic flux concentration of $\Bzm$ to perturb the surface density at planetary orbit. 
Finally, a slightly flatter surface density profile also slightly alters the initial distribution of magnetic flux, with more magnetic flux at the disk outer region. The main results are summarized in figure~\ref{fig:Mt3AM1}.

With $\AMzero=1$, the planet still gathers poloidal magnetic flux to its orbital region (upper left panel). The planet-free magnetic flux concentration appears as well, but is much weaker than the $\AMzero=3$ cases, and such flux sheets migrate stochastically. Consequently, these planet-free concentrations of magnetic flux hardly open deep density gaps, and we can thus focus on the pure-planetary gap. The level of magnetic flux concentration in the planetary gap is similar but slightly weaker than the $Am=3$ case, with typical $\Bzm/B_{z0}\sim3$. It also leads to slightly weaker $\alpha$ values in the gap region of $\sim0.1$, as opposed to $\sim0.3$ in run Mt3Am3.

Also similar to the $\AM=3$ cases, the planet opens a much deeper, and modestly wider gaps than the viscous counterpart, as seen from the bottom left of the figure. The gap center in the windy disk is deeper by a factor of 4 and 6 than the inviscid and viscous disks, respectively, due to the enhanced MHD torque. On the other hand, beyond the corotation region, the gap profile is again compatible with the inviscid case, as seen from the bottom right panel of the figure.

The top right panel of figure \ref{fig:Mt3AM1} shows the cumulative MHD torques. Again, we see that the cumulative MHD torque exhibits a steeper slope within $|R-\RP| \lesssim \rH $, while it returns to the normal slope corresponding to the planet-free wind-torques beyond the gap region.
The enhanced slope (dashed line), which corresponds to the local torque density, is measured as $1.2\times 10^{-3}~[ \RP^2 \Sigma_0(\RP)~ \vP\OmegaP]$, which is $2.6$ times steeper than the normal slope (black dotted curve at the planetary orbit), measured as $4.7\times10^{-4}$ in the same units. The enhanced slope here is less strong compared to the $Am=3$ simulations, where the contrast is a factor of $\sim5$. This is likely related to the slightly weaker level of flux concentration and the Maxwell stresses in this simulation.

In brief, while run Mt3Am1 exhibits weaker turbulence and weaker level of magnetic flux concentration, the main conclusions in the $Am=3$ simulations are equally applicable.

\bibliography{MHD}
\bibliographystyle{aasjournal}

\end{document}